\documentclass[a4paper,11pt]{article}
\pdfoutput=1 

\usepackage{jcappub} 
\usepackage{aas_macros}
\usepackage{hyperref}
\usepackage{graphicx}
\usepackage{subcaption}
\usepackage{float}
\usepackage{color}
\usepackage{cancel}
\usepackage{array}
\usepackage{relsize}
\usepackage{multirow}
\usepackage{orcidlink}
\usepackage{colortbl}
\usepackage{ragged2e}	
\usepackage{etoolbox}  

\apptocmd{\thebibliography}{\justifying}{}{}

\newcommand{\bea}{\begin{eqnarray}}
	\newcommand{\eea}{\end{eqnarray}}

\title{\boldmath Estimators for the cross-pairwise kSZ effect: forecasts for the dark energy and modified gravity parameters with Simons Observatory and CMB-S4}

\author[]{Aritra Kumar Gon\,\orcidlink{0000-0002-0004-9563},}
\author[]{ Rishi Khatri}

\affiliation[]{ Department of Theoretical Physics, Tata Institute of Fundamental Research, \\
	Mumbai, India}

\emailAdd{aritra.gon@theory.tifr.res.in}
\emailAdd{khatri@theory.tifr.res.in}
\abstract{We present and study a new cross-pairwise estimator to extract the kinetic Sunyaev Zeldovich (kSZ) signal from galaxy clusters. The existing pairwise kSZ method involves pairing clusters with other clusters and stacking them. In the cross-pairwise method, we propose to pair clusters with galaxies from a spectroscopic survey and then do the stacking. Cross-pairing decreases the measurement, instrumentation, and statistical noise, thus boosting the signal-to-noise ratio. However, we also need data from a galaxy survey in addition to the CMB temperature maps and a cluster catalog in order to use this method. We do a Fisher matrix analysis for the optimised pairwise and cross-pairwise estimators and forecast the ability of future Cosmic Microwave Background (CMB) experiments and galaxy surveys to measure cosmological parameters with the kSZ effect when combined with primary CMB and Baryon Acoustic Oscillation (BAO) data. { \color{black}We show that using the cross-pairwise kSZ estimator  leads to a factor of 2(3) improvement in the $1\sigma$ error of the dark energy parameters $w_0$ and $w_a$ and a factor of 4(6)  improvement for the growth rate index $\gamma$ compared to the pairwise estimator, when we pair the clusters from the Simons Observatory (CMB-S4) with the galaxies from the DESI survey.  }}

\begin{document}
	\maketitle
	\flushbottom
	\section{Introduction \label{sec_intro}}	
	Galaxy clusters are the largest collapsed objects in the observable universe. The intra cluster medium (ICM) of galaxy clusters consists of hot ionised gas which acts as a source of free electrons. As the Cosmic Microwave Background (CMB) radiation traverses through the galaxy cluster, it interacts with the free electrons via Thomson scattering producing different types of secondary (produced after recombination) effects in the CMB. One such phenomenon is the Doppler shift of the CMB photons due to the bulk peculiar motion of the clusters relative to the CMB rest frame. This is known as the kinetic Sunyaev Zeldovich (kSZ) effect \cite{1972_Sunyaev_Zeldovich,SZ_80}. Since this is a linear order effect in the peculiar velocity of clusters, this does not result in a distortion of the CMB spectrum. The Doppler shift due to the peculiar motion of electrons results in a temperature shift in the blackbody distribution of the photons.  The kSZ effect is proportional to the product of the Thomson scattering optical depth and the line of sight (LoS) peculiar velocity of the cluster. Alongside the kSZ effect, the thermal motion of hot electrons inside the clusters can result in a net up-scattering of CMB photons from the Rayleigh-Jeans side to the Wien tail. This is called the thermal SZ effect (tSZ) \cite{zeldovich1969interaction}. The tSZ effect results in a distortion of the CMB blackbody spectrum and can be distinguished from the other CMB anisotropies because of its characteristic spectrum. The thermal velocities of the electrons in the cluster is many orders of magnitude larger than the electron bulk peculiar
	velocity.  Consequently, at the frequency with the greatest distortion, the tSZ signal can reach approximately $100\, \mu \mathrm{K}$, while at the same frequency, the kSZ signal will be around $5\, \mu \mathrm{K}$ \cite{2002_Carlstorm}, even though the thermal SZ effect is second order in velocity of electrons while the kSZ effect is linear in velocity. Moreover, having a blackbody spectrum makes the detection of the kSZ effect and separation from primary CMB anisotropy, which has the same spectrum, very challenging.\\\\
	We note that the only way to directly measure the peculiar velocity of a galaxy cluster is through its effect on the CMB, e.g. from Thomson scattering of CMB by electrons in a cluster (kSZ and polarised kSZ effects) \cite{1972_Sunyaev_Zeldovich,SZ_80,2002_Carlstorm,2014_renaux,2022JCAP_Hotinli,2022JCAP_Gon} or lensing of the CMB by the moving cluster (moving lens effect) \cite{2019_movinglens_hotinli,2019_movlens_Elena}. Other probes such as the redshift space distortions \cite{1987_Kaiser_RSD,2009_Percival,2013_RSD_Fabian,2017_sdss12} give statistical measurement of the correlations of the velocity field of a tracer of large scale structure such as the galaxies. However, measurement of the peculiar velocity of individual clusters directly using the kSZ effect is difficult and only possible in rare cases \cite{2013_Sayers}. This has inspired the development of alternative methods for statistical measurement of the kSZ signal collectively from a large number of clusters. One such method is known as the pairwise kSZ effect. On small scales ($\sim 50\,\mathrm{Mpc}$), the mutual gravitational attraction between galaxy clusters causes them to move towards each other on average. Thus, on average, when observed from Earth, the LoS component of the peculiar velocity of the cluster in a pair further from(closer to) us will be directed towards(away from) us. The kSZ effect will create a pattern in the CMB, which consists of hot and cold spots at the locations of clusters moving towards and away from the observer respectively. Therefore, if we subtract the intensity (or equivalently temperature) in a CMB map at locations of two different clusters and average over many cluster pairs, the pairwise kSZ signal adds up. We thus boost signal-to-noise ratio as the random noise, the primary CMB anisotropies, and other foregrounds uncorrelated with the position of the clusters average out. This pairwise motion was indeed the way by which the kSZ signal was first detected \cite{2012_Hand}. More recently, the measurement of the pairwise kSZ effect has been improved since its first detection \cite{2016_DESYear1andSPT,2018_PlanckandBOSSdata,2021_PRD_AtacamaCosmologyTelescope,2022MNRAS_DESIgalaxyclustersandPlanck,2023_SPT_3GandDESCollaborations}. The average pairwise signal is a function of the separation between the clusters and a measurement of the kSZ effect as a function of separation can be used to measure cosmological parameters.
	The pairwise signal is also proportional to the mean pairwise relative velocity and the Thomson optical depth of the clusters. The mean pairwise velocity was initially introduced in the context of the \textcolor{black}{Bogoliubov–Born–Green–Kirkwood–Yvon (BBGKY)} theory applied to cosmology \cite{1994ApJ_PerturbativeGrowth,1977ApJ_Davis,1980_peebles_lss} and has since been studied by several authors \cite{1992ApJ_Fry,1999ApJ_DynamicalEstimator,2016PRL_ProjectedFields:ANovelProbe}. The kSZ effect is a probe of the cosmological velocity field which in turn probes the large-scale gravitational potential around the cluster.  In addition, the Thomson optical depth of the clusters depends on their mass, and the distribution of cluster masses depends on the cosmological parameters. The pairwise kSZ signal, kSZ power spectrum, and kSZ cross-power spectrum and bispectrum with galaxies as a probe of dark energy, modifications to gravity, radial inhomogeneity, and neutrino masses has been studied previously \cite{2005_Spergel,2008_Bhattacharya,2012_Philip,2015ApJ_DarkEnergy,2018_Madhavacheril,2015PRD_massiveneutrinos,2021_kszgalaxy_Kusiak,2016PRL_ProjectedFields:ANovelProbe}. However, the cross-correlation in a pairwise estimator has not been explored before.\\\\
	We propose a new estimator for the kSZ effect by pairing clusters with galaxies from a spectroscopic survey such as the DESI survey, similar to the estimator we developed for the polarised kSZ effect \cite{2023_Gon}.  In a cluster-cluster pair, the second cluster acts as an indicator of the direction in which the first cluster moves and vice-versa. The main principle behind the pairwise estimator is that the correlation of the peculiar velocity of a cluster with the nearby clusters leads to a coherent addition of the kSZ signal during the stacking process. Instead of pairing clusters with other clusters, we can pair up a cluster with any other tracer of the large-scale gravitational potential responsible for its peculiar velocity. This we call the cross-pairwise kSZ effect. \textcolor{black}{For a cluster-galaxy pair, the kSZ signal from the galaxy is negligible and we are only measuring the motion of the cluster through its kSZ effect.} The cross-pairwise effect has two advantages. Although galaxies contribute negligibly to the kSZ signal, resulting in some loss of signal, there are significantly more galaxy-cluster pairs than cluster-cluster pairs. This higher number of cluster-galaxy pairs reduces the measurement and instrument noise when we stack them. Secondly, in terms of the statistical noise, the Poisson noise is suppressed as the clusters and galaxies are two independent tracers of the underlying peculiar velocity field \cite{2009_poisson_smith,2017_poisson_noise}. The suppression of the Poisson noise and the decrease in the measurement noise increases the signal-to-noise ratio. We note that the forecasts in the previous works were done for the un-optimised pairwise estimator \cite{2008_Bhattacharya,2015ApJ_DarkEnergy}. We also revisit the forecast for the pairwise kSZ effect using the optimised estimator. The reduction of both the statistical and measurement noise and the resultant improvement in the signal-to-noise ratio translates into a more precise measurement of the cosmological parameters. To compare the precision with which we can measure the cosmological parameters using optimised pairwise and cross-pairwise estimators, we do a Fisher matrix analysis. We use a 10 parameter cosmological model, containing a Chevallier-Polarski-Linder (CPL) parameterization of the Dark Energy equation of state \cite{2001_CP,2003_Linder}. Further information about the cosmological parameters and their fiducial values are given in section \ref{constraints}.  { \color{black}For our forecasts, we assume the tSZ cluster catalogs expected from the upcoming Simons Observatory (SO) \cite{2019JCAP_Simons_Obs} large aperture telescope (LAT) and the CMB-S4 wide \cite{2016_cmbs4} surveys}. The kSZ signal in the CMB temperature maps from the same survey is measured at the position of these clusters. For the galaxies we take the early data release of the DESI \cite{2024_DESI,2023_DESI_EDR} survey and the LSST \cite{2009_LSST} survey forecasts assuming there will be a spectroscopic follow up for LSST galaxies. We note that instead of tSZ cluster catalog we can use any other cluster catalog covering the same region of the sky as the CMB survey, e.g. X-ray cluster catalog from eRosita survey \cite{2012_erosita}.
	\section{Theoretical formalism \label{formalism}}
	{\color{black}The kSZ effect from a particular galaxy cluster results in a change in the intensity of the primary CMB radiation, while the spectrum remains a blackbody spectrum at first order}. We can express the fractional change in the observed CMB blackbody intensity as a fractional change in the CMB temperature, ${\Delta T}/{T_{\mathrm{CMB}}}$. The change is proportional to the peculiar velocity of the cluster along the LoS direction and the Thomson optical depth.
	\begin{align}
		\label{ksz_starting}
		\frac{\Delta T}{T_{\mathrm{CMB}}}(\mathbf{\hat{n}})=-\sigma_{\mathrm{T}}\int d\chi\,e^{-\tau(\chi)}\, n_e({\mathbf{r}})\,a(\chi)\;\left(\mathbf{v}({\mathbf{x}})\cdot\mathbf{\hat{n}}^\mathrm{cl}\right),
	\end{align}
	where \textcolor{black}{$T_{\mathrm{CMB}}=2.725\,$K, $\sigma_{\mathrm{T}}$ is the Thomson scattering cross section, $\mathbf{\hat{n}}^\mathrm{cl}$ is the LoS direction, $\mathbf{r}$ is the position vector of a point along the LoS direction, $\mathbf{v}$ is the cluster's peculiar velocity, $\mathbf{x}$ is the position vector to the centre of the cluster, $\chi$ is the comoving distance, a is the scale factor, and $n_e$ is the electron number density}. {\color{black} The negative sign in Eq.(\ref{ksz_starting}) ensures that the kSZ signal is positive if the cluster is moving towards us.} A galaxy cluster is an object of finite angular size in the sky, so there are multiple lines of sight which pass through the cluster. Averaging the optical depth over the cluster defines an effective optical depth ($\tau_{\mathrm{eff}}$) for each cluster \cite{2017ApJ_Fender}. Since the velocity and the LoS direction is approximately constant over the cluster, we can bring it out of the integral and after averaging over the cluster, rewrite Eq.(\ref{ksz_starting}) as,
	\begin{align}
		\label{ksz_2}
		\frac{\Delta T}{T_{\mathrm{CMB}}}(\mathbf{\hat{n}}^\mathrm{cl})=-\tau_{\mathrm{eff}}\;\left(\mathbf{v}({\mathbf{x}})\cdot\mathbf{\hat{n}}^\mathrm{cl}\right)\equiv \mathcal{T}(\mathbf{\hat{n}}^\mathrm{cl}),
	\end{align}
	where $\mathbf{\hat{n}}^\mathrm{cl}$ now represents the LoS direction towards the cluster centre. This expression assumes not more than a single scattering per photon,
	which is a good assumption at the low optical depth. For most clusters, $\tau_{\mathrm{eff}}\sim\mathcal{O}(10^{-3})$. The typical peculiar velocity of galaxy clusters is of $\mathcal{O}(10^{-3})$ in units of the speed of light (c). Thus, from Eq.(\ref{ksz_2}), it is evident that the kSZ effect is a few $\mu K$ in amplitude.
	\subsection{Pairwise  kSZ estimator \label{pairwise} }
	Two galaxy clusters falling towards each other due to their mutual gravitational attraction will have their LoS velocity component in the opposite direction. In our reference frame one cluster will have its LoS velocity towards us and the other cluster will have its LoS velocity away from us. Consequently, from Eq.(\ref{ksz_2}), we can see that the kSZ effect induced by the two clusters will be of opposite sign. As a result, this will leave a dipole pattern in the CMB temperature anisotropy. This simple example, with just two clusters, suggest an estimator for the kSZ effect. In general, when stacking all possible cluster pairs, the direction of the velocity of one cluster will be correlated with the position of the second cluster in a pair on average. Therefore to detect this effect, we formally define the pairwise kSZ estimator as,
	\begin{align}
		\label{esti_defn}
		\hat{T}_{\mathrm{pairwise}}=\sum_i \hat{T}^i_{\mathrm{pairwise}}=\sum_{i}w_i\left(\mathcal{T}_{i1}-\mathcal{T}_{i2}\right),
	\end{align}
	where the sum is over all distinct cluster pairs, and $\mathcal{T}_{i1}$ and $\mathcal{T}_{i2}$ denote the kSZ signal from the two clusters in a pair.  The normalised weights are denoted by $w_i$ which we want to use to create the optimal estimator. In the case where the two clusters are moving towards each other, if cluster labeled $1$ is closer to us it is moving away from us and the kSZ signal from cluster 1 is negative. Similarly, the kSZ signal from the cluster 2 is positive and in the Eq.(\ref{esti_defn}) the signal from the two clusters add up to give a negative pairwise kSZ signal. In this case we choose the sign of the weight, $w_i$ to be positive. On the other hand, if cluster 1(2) is the further(closer) cluster, the  kSZ signal from it is positive(negative). In this case we choose the sign of the weights to be negative to again get a negative final pairwise signal.\footnote{The sign of weights, $w_i$, effects a rotation in 1-D, analogous to the rotation in 2-D we needed in case of the pkSZ effect \cite{2023_Gon}. This is required so that the kSZ signal from all pairs add up coherently irrespective of which cluster in a pair is labeled $1$.}  We also want to choose the magnitudes of the weights to be proportional to the signal to maximize the signal-to-noise ratio. The expectation value of the estimator is given by,
	\begin{align}
		\Big\langle\hat{T}_{\mathrm{pairwise}}\Big\rangle_2=\sum_{\mathrm{i}}w_i	\Big\langle(\mathcal{T}_{i1}-\mathcal{T}_{i2})\Big\rangle_2,
	\end{align}
	where the subscript 2 on the angular brackets $\langle \cdots \rangle_2 $ indicates that the ensemble average is over the two-particle (or two-cluster) distribution function. {\color{black}The two-particle distribution function signifies that the correlations between the two particles are included while doing the ensemble average \cite{1980_peebles_lss}.}  The expectation value of the pairwise estimator has been previously derived using the BBGKY formalism \cite{1980_peebles_lss,1977ApJ_Davis,1994ApJ_PerturbativeGrowth}. It has also been derived using density-weighted velocity moments \cite{1992ApJ_Fry,2002PhR_Bernardeau,1999ApJ_DynamicalEstimator}. In this work, we employ the same method as in our previous study \cite{2023_Gon} to derive both the expectation value of the optimised pairwise estimator and the associated covariance matrix. Our method, based on \cite{2004ApJ_Ma_CosmologicalKineticTheory}, demonstrates the connection between the BBGKY approach and the density-weighted velocity moments. Considering a cluster pair at a given average comoving distance $\chi$ and a given average  cluster mass $m$, we can write the expectation value as,
	{\color{black}\begin{align}
			\label{expect_val_1}
			\Big\langle(\mathcal{T}_{i1}-\mathcal{T}_{i2})\Big\rangle_2 =  -\tau_{\mathrm{eff}}(m_i,\chi_i)	\Big\langle \mathbf{v_{i1}}({\mathbf{x_{i1}}})\cdot\mathbf{\hat{n}^\mathrm{cl}_{i1}}-\mathbf{v_{i2}}({\mathbf{x_{i2}}})\cdot\mathbf{\hat{n}^\mathrm{cl}_{i2}}\Big\rangle_2.
	\end{align}}The two clusters in the pair are separated by a comoving distance given by $r=|\mathbf{ x_1}-\mathbf{ x_2}|$ in 3-D space. We further assume that the angular separation between the clusters is small such that, $\mathbf{\hat{n}^\mathrm{cl}_{i1}}\simeq \mathbf{\hat{n}^\mathrm{cl}_{i2}}\simeq(\mathbf{\hat{n}^\mathrm{cl}_{i1}}+ \mathbf{\hat{n}^\mathrm{cl}_{i2}})/2=\mathbf{\hat{n}_{i}}$. Henceforth, for brevity, we will suppress the subscript `i'. From Eq.(\ref{expect_val_1}) we get,
	{\color{black}\begin{align}
			\label{expect_val_2}
			\Big\langle(\mathcal{T}_{1}-\mathcal{T}_{2})\Big\rangle_2 = - \tau_{\mathrm{eff}}(m,\chi)	\Big\langle \left(\mathbf{v_1}({\mathbf{x_1}})-\mathbf{v_2}({\mathbf{x_2}})\right)\Big\rangle_2\cdot\mathbf{\hat{n}}.
	\end{align}}
	The expectation value, $	\Big\langle \left(\mathbf{v_1}({\mathbf{x_1}})-\mathbf{v_2}({\mathbf{x_2}})\right)\Big\rangle_2$, is defined as, 
	\begin{align}
		\label{expect_val_3}
		\Big\langle \left(\mathbf{v_1}-\mathbf{v_2}\right) \Big\rangle_2(\mathbf{x_1},\mathbf{x_2}|m,\chi)\equiv\frac{\int\int  \left(\mathbf{v_1}-\mathbf{v_2}\right) f_2(\mathbf{x_1},\mathbf{x_2},\mathbf{v_1},\mathbf{v_2}|\chi,m)d^3\mathbf{v_1}d^3\mathbf{v_2}}{\int\int f_2(\mathbf{x_1},\mathbf{x_2},\mathbf{v_1},\mathbf{v_2}|\chi,m)d^3\mathbf{v_1}d^3\mathbf{v_2}},
	\end{align}
	where $f_2(\mathbf{x_1},\mathbf{x_2},\mathbf{v_1},\mathbf{v_2}|\chi,m)$ is the two-particle distribution function of the halos in the position-velocity phase space for a given mass m at a comoving distance $\chi$. Note that we can replace momentum with velocity as one of the phase space variables in the non-relativistic limit. Following \cite{2023_Gon} we can show that,
	\begin{align}
		\label{expect_val_4}
		\Big\langle \left(\mathbf{v_1}-\mathbf{v_2}\right)\Big\rangle_2=\frac{\Big\langle \left(\mathbf{v_1}-\mathbf{v_2}\right)(1+\delta^h_1)(1+\delta^h_2)\Big\rangle}{\Big\langle (1+\delta^h_1)(1+\delta^h_2)\Big\rangle},
	\end{align}
	where $\delta^h$ is the local halo overdensity field. The angular brackets on the right hand side of the equation indicate the usual ensemble average over the density and velocity fluctuations. The local halo overdensity field is related to the local dark matter overdensity as,
	\begin{align}
		\label{halo_dm_rel}
		\delta^h(\mathbf{x},\chi;m,z)=\sum_{p}\frac{b_p(m,z)}{p!}\delta^p(\mathbf{x},\chi),
	\end{align}
	where $\delta(\mathbf{x},\chi)$ is the dark matter density field and $b_p(m,z)$ are the halo bias parameters. We assume the halo velocities to be unbiased tracers of the matter velocity field. The dark matter overdensity field $\delta$ can be expanded in perturbation theory as, 
	\begin{align}
		\label{dm_pert_exp}
		\delta(\mathbf{x},\chi)=\sum_{q}\delta^{(q)}(\mathbf{x},\chi),
	\end{align}
	where $\delta^{(q)}$ is the $q^{th}$ order density field. We can use Eq.(\ref{halo_dm_rel}) and Eq.(\ref{dm_pert_exp}) in Eq.(\ref{expect_val_4}) to find the expectation value of the pairwise velocity. Let us first consider the numerator in Eq.(\ref{expect_val_4}). Since the two-point functions give non-zero contributions, we only include terms to 1st order in perturbation theory. Therefore we get,
	\begin{align}
		\Big\langle \left(\mathbf{v_1}-\mathbf{v_2}\right)(1+\delta^h_1)(1+\delta^h_2)\Big\rangle = b_1(m,\chi)\left[\Big\langle\mathbf{v_1}^{(1)}\;\delta_2^{(1)}\Big\rangle- \Big\langle\mathbf{v_2}^{(1)}\;\delta_1^{(1)}\Big\rangle\right]+\mathrm{higher\,order\,terms}.
	\end{align}
	Similarly the denominator in Eq.(\ref{expect_val_4})  is given by,
	\begin{align}
		\Big\langle (1+\delta^h_1)(1+\delta^h_2)\Big\rangle = 1+\left(b_1(m,\chi)\right)^2\Big\langle\delta_1^{(1)}\;\delta_2^{(1)}\Big\rangle.
	\end{align}
	To proceed further we move to Fourier space. In Fourier space, the density field and the velocity field at first order in perturbation theory are given by \cite{2002PhR_Bernardeau,1994ApJ_Bhuvnesh},
	\begin{align}
		\label{delta_fourier}
		&\delta^{(1)}(\mathbf{x},\chi)=D(\chi)\int\frac{d^3\mathbf{k}}{(2\pi)^3}\exp\left(i \mathbf{k}\cdot\mathbf{x}\right)\delta^{(1)}(\mathbf{k}),
	\end{align}
	\begin{align}
		\label{vel_fourier}
		&\mathbf{v}^{(1)}(\mathbf{x},\chi)=D(\chi)\,[afH](\chi)\int\frac{d^3\mathbf{k}}{(2\pi)^3}\exp\left(i \mathbf{k}\cdot\mathbf{x}\right) \frac{i\mathbf{k}}{k^2}\delta^{(1)}(\mathbf{k}),
	\end{align}
	where $\delta^{(1)}(\mathbf{k})=\mathcal{R}(\mathbf{k})T(k)$, $\mathcal{R}(\mathbf{k})$ is the primordial curvature perturbation, $T(k)$  is the transfer function, $D(\chi)$ is the growth factor normalised as D(0)=1, $f={d\ln D}/{d \ln a}$ is the growth rate, a is the scale factor, and $H(\chi)$ is the Hubble parameter. In $\Lambda$CDM cosmology we have $f=\Omega^{0.55}_m(\chi)$ \cite{2007_growthrate_Linder}. After performing a few more steps of calculations as described in Appendix \ref{App:Review_kSZ}, we can write Eq.(\ref{expect_val_4}) as,
	{\color{black}\begin{align}
			\label{expect_fin1}
			\Big\langle \mathbf{v_{1}}-\mathbf{v_{2}}\Big\rangle_2(\mathbf{r}|m,\chi)=\frac{2}{3}[afH](\chi)\frac{\,b_1(\chi,m)\,	\bar{\xi}(r)}{1+\left(b_1(m,\chi)\right)^2\,\xi(r)}\;\mathbf{r},
	\end{align}}where {\color{black}$\mathbf{r}=\mathbf{ x_2}-\mathbf{ x_1}$.} The two-point correlation function of the matter overdensity field is $\xi(x)$ and $\bar{\xi}$ is its volume average over a spherical volume of radius $r$. We note that the ensemble average of the velocities over the two-particle distribution function is just a function of the inter-cluster separation vector as expected from the assumption of homogeneity of space.  Therefore, Using Eq.(\ref{expect_val_2}) and Eq (\ref{expect_fin1}) in Eq.(\ref{expect_val_1}) we can write the expectation value of the pairwise estimator as,
	\begin{align}
		\label{pair_ksz_final}
		T_{\mathrm{pairwise}}(r)\equiv	\Big\langle\hat{T}_{\mathrm{pairwise}}\Big\rangle_2&=-\frac{2}{3}\sum_{i}w_i\,[afH](\chi_i)\,\tau_{\mathrm{eff}}(m_i,\chi_i)\frac{r\,b_1(\chi_i,m_i)	\bar{\xi}(r)}{1+\left(b_1(m_i,\chi_i)\right)^2\,\xi(r)}\;\cos\theta_i\nonumber\\
		&=\sum_i T^i_{\mathrm{pairwise}}(r),
	\end{align}
	where $\theta_i=\cos^{-1}\left(\mathbf{\hat{r}_i}\cdot\mathbf{\hat{n}_{i}}\right)$. The average cluster mass and the average comoving distance of the i$^\mathrm{th}$ cluster pair are $m_i$ and $\chi_i$ respectively. The sum is over all cluster pairs in the cluster catalog. {\color{black} We note that the choice of separation vector to point from cluster 1 to cluster 2, $\mathbf{ r}=\mathbf{ x_2}-\mathbf{ x_1}$, ensures that Eq.(\ref{pair_ksz_final}) is consistent with our choice of sign convention in Eq.(\ref{esti_defn}).} In the case of pairwise kSZ signal, for cluster pairs with the same average mass and spatial separation, the signal-to-noise ratio will be higher for pairs for which the separation vector is aligned along our LoS direction. Also, cluster pairs with higher average mass will have higher optical depth, resulting in a higher kSZ signal. Following the optimization procedure given in \cite{2022MNRAS_DESIgalaxyclustersandPlanck} we find that the optimal choice of weight is,
	\begin{align}
		\label{weights}
		w_i= \frac{m_i|\cos \theta_i|}{\sum_i m_i|\cos \theta_i|}\mathrm{sign}(\cos \theta_i)= \frac{m_i\cos \theta_i}{\sum_i m_i|\cos \theta_i|}.
	\end{align}
	We note that the sign of $\cos \theta_i$ gives the correct sign to the weights in agreement with our sign convention. We can see directly from Eq.(\ref{pair_ksz_final}) that with weights $\propto \cos\theta_i$, the pairwise signal is $\propto \cos^2\theta_i$ and thus always negative as long as the cosmological correlations are positive. {\color{black}We note that we need to know the mass and orientation of the cluster a priori in order to calculate the weights.  We assume that the mass of the cluster would be known from tSZ effect and their positions from the spectroscopic redshifts.} To estimate the pairwise kSZ signal, we also need to estimate the mean optical depth for a cluster pair. The optical depth depends on the electron density profile of the ICM, which we model using a spherically symmetric $\beta$ profile. To calculate the effective optical depth, we average the optical depth over the angular area of the cluster following \cite{2017ApJ_Fender,2023_Gon}. 
	\subsection{Cross pairing clusters with galaxies \label{cross-pairwise} }
	{ \color{black} If we had the knowledge of the underlying peculiar velocity field (for example by reconstructing the velocity field from a galaxy survey), we could just stack individual clusters by aligning the direction of their motion. In this case, we could use the kSZ effect as a probe of the optical depth of clusters and learn about their astrophysics. We are, however, interested in using the kSZ effect as a probe of the velocity fields.} In fact, one of the main advantages of the kSZ effect is that we can probe the statistical properties of the cosmic velocity fields without having to reconstruct the velocity fields themselves. One of the ways is to directly measure the power spectrum of the kSZ effect. However, this is made difficult because the signal is so small and there is confusion with the primary CMB anisotropies which have the same frequency spectrum. The pairwise effect discussed in the previous section has proved to be more successful, where clusters themselves are also the indicator of the peculiar velocity fields. However, clusters are rare objects and inefficient/noisy tracers of the large-scale gravitational potential and the peculiar velocity field. We showed in \cite{2023_Gon}, that we can improve the signal-to-noise ratio by cross-pairing clusters with a better tracer, i.e. galaxies. Thus, instead of pairing up two clusters, we can replace one of the clusters with a galaxy. We should expect improvement in signal-to-noise ratio similar to the cross-power spectrum of kSZ effect with galaxies \cite{2016PRL_ProjectedFields:ANovelProbe, 2021_kszgalaxy_Kusiak}. In the case of cross-pairing clusters with galaxies, there are two competing effects. The Thomson optical depth of galaxies is very small and, therefore, the kSZ signal from a galaxy is negligible. Also, since galaxies are less biased than clusters, the pairwise signal from an individual pair decreases. Nevertheless, there are many more cluster-galaxy pairs than there are cluster pairs. While CMB-S4 wide survey will detect $\sim10^5$ clusters, surveys like DESI \cite{2024_DESI,2023_DESI_EDR} will detect $\sim30$ million galaxies with spectroscopic information in its entire operational time. 
	\begin{figure}
		\hspace{-0.5cm}
		\begin{subfigure}{0.5\textwidth}
			\centering
			\includegraphics[width=1.05\linewidth]{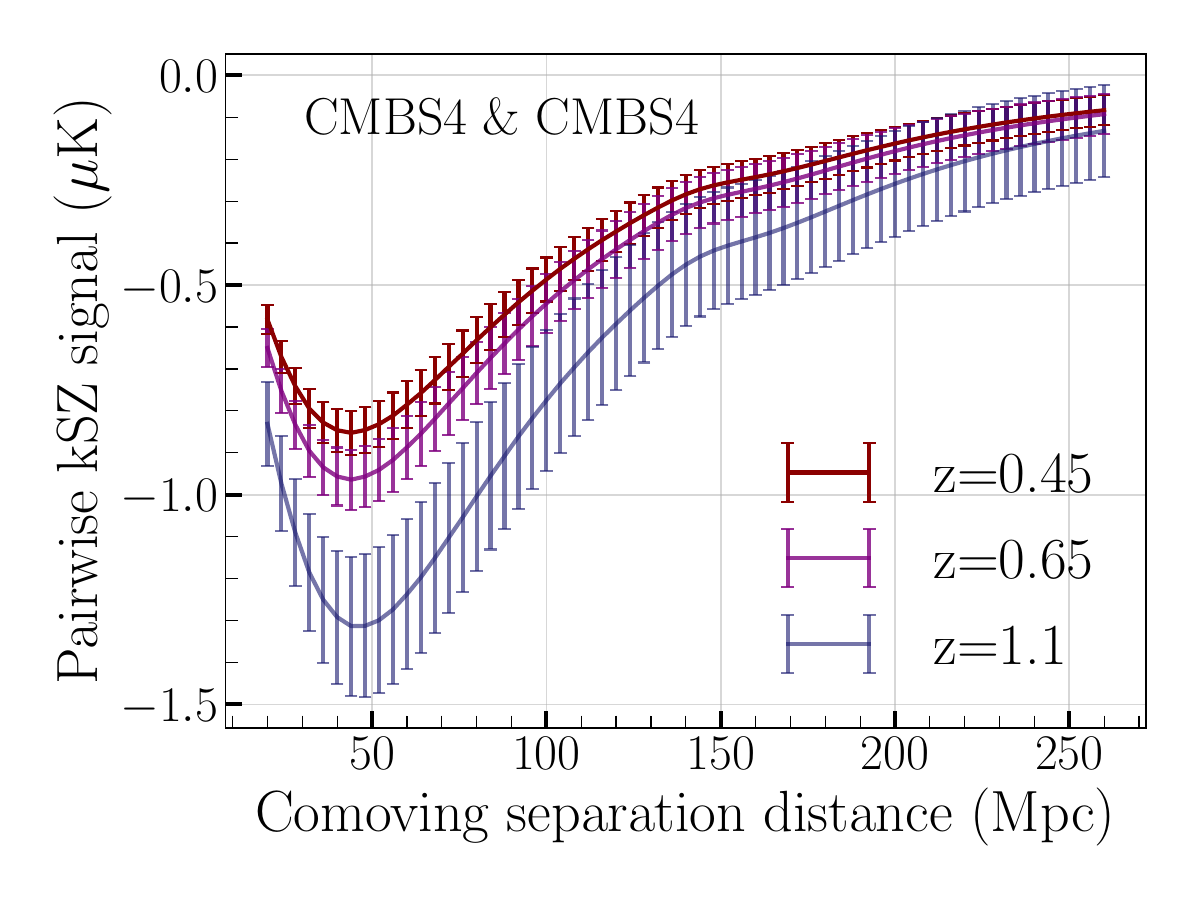}
			\caption{Pairwise signal}
			\label{fig:pairwisesignal}
		\end{subfigure}%
		\hspace{0.1cm}
		\begin{subfigure}{0.5\textwidth}
			\centering
			\includegraphics[width=1.05\linewidth]{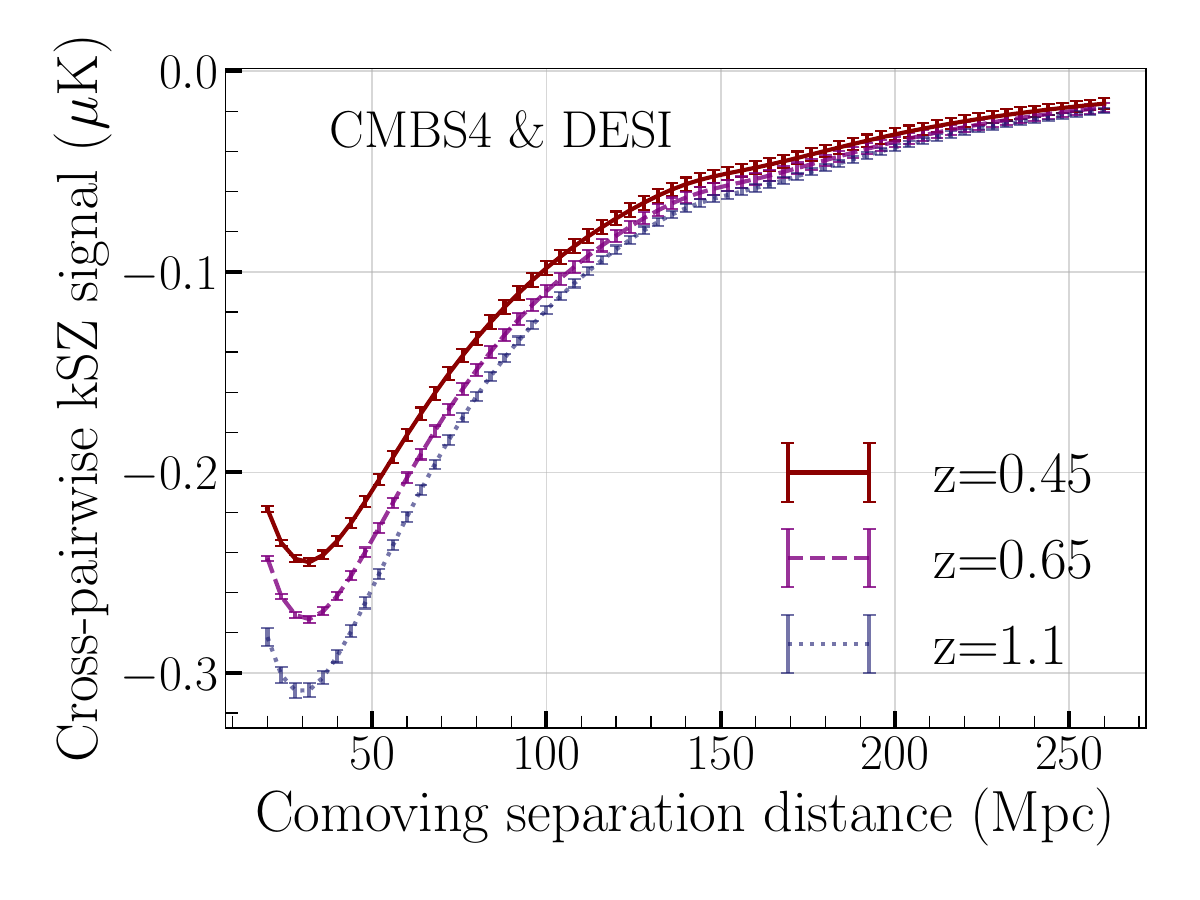}
			\caption{Cross-pairwise signal}
			\label{fig:crosspairwisesignal}
		\end{subfigure}
		\caption{The pairwise and cross-pairwise signal at different redshifts. For the cross-pairwise case we show the signal for luminous red galaxy sample of the DESI survey.}
		\label{fig:pairwisecrosspairwisesignal}
	\end{figure}
	Future surveys like LSST will detect even more galaxies, $\sim 20$ billion in total. For the cross-pairwise kSZ effect, each pair has one cluster paired with one galaxy and we stack the CMB temperature maps only at the position of the clusters. We thus define the cross-pairwise estimator as,
	\begin{align}
		\hat{T}_{\mathrm{cross-pairwise}}=\sum_{i}w_i\left(\mathcal{T}_{i \,\mathrm{cluster}}-\mathcal{T}_{i\,\mathrm{galaxy}}\right)\simeq\sum_{i}w_i\,\mathcal{T}_{i \,\mathrm{cluster}},
	\end{align}
	and {\color{black}the separation vector points from the cluster to the galaxy, i.e. $\mathbf{ r}=\mathbf{ x_\mathrm{galaxy}}-\mathbf{ x_\mathrm{cluster}}$. }
	We can modify Eq.(\ref{pair_ksz_final}) to write the expectation value of the estimator for the cluster-galaxy cross pairwise kSZ effect, 
	\begin{align}
		\label{cross_pair_ksz}
		T_{\mathrm{cl-gal}}(r)&=-\frac{1}{3}\sum_{i}w_i[afH](\chi_i)\tau_{\mathrm{eff}}(m_i,\chi_i)\frac{r\,b^{\mathrm{gal}}_1\;	\bar{\xi}(r)}{1+b^{\mathrm{gal}}_1\,b^{\mathrm{cl}}_1(m_i,\chi_i)\,\xi(r)}\;\cos\theta_i\nonumber\\
		&=\sum_i 	T^i_{\mathrm{cl-gal}}(r),
	\end{align}
	where $b^{\mathrm{gal}}_1$ is the linear bias factor of the galaxy. Here, we have specifically written the bias factor coming from clusters as $b^{\mathrm{cl}}_1$ to avoid confusion. {\color{black} In figure \ref{fig:pairwisecrosspairwisesignal}, we show the expectation value of pairwise and cross-pairwise kSZ estimators at different redshifts. We also show the diagonoal components of the covariance matrix as
		error bars. We note that the neighbouring points are in general correlated  since the covariance matrix has non-zero off-diagonal components. We show the full covariance matrix for these two case in figure \ref{fig:pairwisecrosspairwisecovmat} in Appendix \ref{App:Cov_mat}. The Fisher forecasts use the full covariance matrix, which is discussed in the next section.} We see that the average cross-pairwise signal from a cluster-galaxy pair is smaller by a factor of $\sim4-6$ compared to the cluster-cluster pairwise signal. We want to forecast the precision with which we can measure the cosmological parameters with future CMB missions and galaxy surveys using these estimators. Unlike the polarised kSZ case, the unpolarised kSZ signal has a much larger amplitude, so the instrument or the measurement uncertainty is not the dominant source of error. We also need to take into account the statistical error from Cosmic variance and Poisson noise in the computation of the covariance matrix.
	\section{The covariance matrix for the optimised pairwise and cross-pairwise estimators\label{fish_mat}}
	Although the optimal pairwise estimator has been used previously to detect the pairwise kSZ signal \cite{2022MNRAS_DESIgalaxyclustersandPlanck}, the more recent forecasts for future experiments have used the un-optimised estimator \cite{2008_Bhattacharya,2015ApJ_DarkEnergy,2015PRD_massiveneutrinos}. Measurement of the pairwise and cross-pairwise kSZ effect is associated with both statistical uncertainties and measurement errors. In statistical uncertainties, we include both cosmic variance and shot noise. Measurement errors include the instrumentation noise and uncertainties in measuring the mass and hence the effective Thomson optical depth of galaxy clusters. We will first calculate the cosmic variance and shot noise. Cosmic variance occurs because, in observations, we can only sample clusters and galaxies within a finite survey volume. The shot noise arises because we are sampling density and peculiar velocity fields using clusters and galaxies which are discrete objects. The statistical uncertainty for the pairwise kSZ effect can be quantified by the following equation for the covariance $C_{T_{\mathrm{pairwise}}}(r,r')$,
	\begin{align}
		\label{cov_start}
		C_{T_{\mathrm{pairwise}}}(r,r')=&\Big\langle\int \frac{d^3\mathbf{x}d^3\mathbf{x}'}{V_s^2}\int \frac{ d\Omega_{r} d\Omega_{r'}}{(4\pi)^2}\hat{T}_{\mathrm{pairwise}}\;\hat{T'}_{\mathrm{pairwise}}\Big\rangle_4\nonumber\\
		&\hspace{6cm}-T_{\mathrm{pairwise}}(r)T'_{\mathrm{pairwise}}(r'),
	\end{align}
	where
	\begin{align}
		&\hat{T}_{\mathrm{pairwise}}=\sum_{i}\;w_i\;\tau_{\mathrm{eff}}(m_i,\chi_i) \left(\mathbf{v_{i1}}({\mathbf{x}})-\mathbf{v_{i2}}({\mathbf{x+r}})\right)\cdot\mathbf{\hat{n}_{i}}\hspace{1cm}\mathrm{and}\\
		&\hat{T'}_{\mathrm{pairwise}}=\sum_{j}\;w_j\;\tau_{\mathrm{eff}}(m_j,\chi_j) \left(\mathbf{v_{j1}}({\mathbf{x'}})-\mathbf{v_{j2}}({\mathbf{x'+r'}})\right)\cdot\mathbf{\hat{n}_{j}}.
	\end{align}
	The angular brackets $\langle\cdots\rangle_4$ indicates that the ensemble average
	is over the four-particle distribution function, $T_{\mathrm{pairwise}}(r)$ and $T'_{\mathrm{pairwise}}(r')$ are defined in Eq.(\ref{pair_ksz_final}). We refer to appendix \ref{pairwise_cosmic_variance} for more details.
	The separation vectors of the two cluster pairs are $\mathbf{r}$ and $\mathbf{r'}$ and they are at an average comoving distance of $\chi$ and $\chi'$ respectively. It should be noted that we should perform similar volume and angular averaging in the definition of our estimator itself in Eq.(\ref{esti_defn}). Since the ensemble average is dependent only on the separation between the two clusters in a pair and not their orientation or absolute position, the volume averaging and angular averaging on the estimators can be performed trivially and would just give a factor of unity. Starting from Eq.(\ref{cov_start}), we can do the integrals and ensemble averages analytically as shown in Appendix \ref{App:Noise_case}. The result is the following final expression for the statistical part of the covariance matrix:
	\begin{align}
		\label{C_pair}
		&C_{T_{\mathrm{pairwise}}}(r,r')=\frac{1}{V_s}\; \frac{4}{15\pi^2}\sum_{i,j}\;m_i\;m_j\,\tau_{\mathrm{eff}}(m_i,\chi_i)\tau_{\mathrm{eff}}(m_j,\chi_j)\,\bigg[
		1+2(\mathbf{\hat{n}_{i}}\cdot\mathbf{\hat{n}'_{j}})^2\bigg]\, [afH](\chi_i)\nonumber\\
		&\,[afH](\chi_j)\int dk\,\left(b_1(m_i,\chi_i)P(k,\chi_i)+\frac{1}{n_{\mathrm{cl}}}\right)\left(b_1(m_j,\chi_j)P(k,\chi_j)+\frac{1}{n_{\mathrm{cl}}}\right)\;j_{1}(kr)\;j_{1}(kr')\nonumber\\
		&\left[\left(1+b^2_1(m_i,\chi_i)\xi(x,\chi_i)\right)\left(1+b^2_1(m_j,\chi_j)\xi(x',\chi_j)\right)\right]^{-1},
	\end{align}
	where $V_s$ is the survey volume and $n_{\mathrm{cl}}$ is the cluster number density. Eq.(\ref{C_pair}) includes the effect of both the cosmic variance and the Poisson noise. Besides the statistical uncertainty, there are the instrument and the measurement noise which contribute to the diagonal elements of the covariance matrix.  
	The sensitivity $\kappa$ of CMB experiments is usually specified in units of $\mu \mathrm{K\,arcmin}$. If a cluster is resolved and has an angular area $\sigma_{\mathrm{cluster}}$ in $\mathrm{arcmin}^2$, the average noise variance for a cluster is given by \cite{2023_Gon},
	\begin{align}
		\label{noise_cl}
		\left(	\mathcal{N}_\mathrm{cl}\right)^2= \frac{1}{(T_{\mathrm{CMB}})^2}\left(\frac{\kappa^2}{\sigma_{\mathrm{cluster}}}\right).
	\end{align}
	Since, in the pairwise estimator, we are measuring the kSZ signal from the two clusters in the pair, the noise variance for the cluster pair is $2\,\mathcal{N}^2_\mathrm{cl}$. The  resultant noise for our measurement after performing a weighted sum over $\mathrm{N_{pairs}}$ cluster pairs is,
	\begin{align}
		\label{noise}
		\mathcal{N}^{\mathrm{inst}}_{\mathrm{pairwise}}(r)= \sqrt{2\,\sum_{i}\,w^2_i\left(\mathcal{N}^{\,i}_{\mathrm{cl}}\right)^2},
	\end{align} 
	where the sum is over all distinct cluster pairs. We find the total number of pairs expected in a cosmological survey using the Peebles-Hauser (natural) estimator \cite{1974ApJ_pH_estimator,2013A&A_estimator},
	\begin{align}
		\label{DD}
		DD(r)=\left[1+(\bar{b}_1)^2\,\xi(r)\right]RR(r),
	\end{align}
	where $\bar{b}_1$ is the mass-averaged linear bias factor of clusters at a given redshift (z), $RR(r)$ denotes the total number of cluster pairs at a given separation in a randomly distributed sample, and $DD(r)$ denotes the total number of cluster pairs in the presence of clustering at a fixed separation $r$. We bin the separation distance in finite bin sizes and also in redshift bins. We refer to \cite{2023_Gon} for detailed calculation of estimating the total number of pairs. The uncertainty in the measurement of the optical depth of the clusters is a significant source of error.  Following \cite{2015ApJ_DarkEnergy} we have assumed, 
	$\left(\Delta \tau_{\mathrm{eff}}/\tau_{\mathrm{eff}}\right)^2=0.15. $
	Therefore, the error due to the uncertainty in measurement of the optical depth becomes,
	\begin{align}
		\mathcal{N}^{\mathrm{tau}}_{\mathrm{pairwise}}(r)=  \left(\Delta \tau_{\mathrm{eff}}/\tau_{\mathrm{eff}}\right)\sqrt{2\,\sum_{i}\,\left(	T^i_{\mathrm{pairwise}}(r)\right)^2}.
	\end{align}
	The sum is again over all cluster pairs. Therefore, the total measurement noise contribution to the covariance matrix is,
	\begin{align}
		\mathcal{N}^{2}_{\mathrm{pairwise}}(r) = \left(	\mathcal{N}^{\mathrm{inst}}_{\mathrm{pairwise}}(r)\right)^2+ \left(\mathcal{N}^{\mathrm{tau}}_{\mathrm{pairwise}}(r)\right)^2.
	\end{align}
	In our mock catalog, once we have a fair sample of the clusters expected in a survey, we can scale the results for the total number of expected cluster pairs as $\mathcal{N}^{2}_{\mathrm{pairwise}}\propto1/	\mathrm{N_{pairs}}(r)$, where $\mathrm{N_{pairs}}(r)$ is the total number of cluster-cluster pairs at a separation r.
	\subsection{Covariance for the cross-pairwise estimator \label{cov_cross_pair}}
	The Poisson noise is absent in the case of the cross-pairwise estimator \cite{1980_peebles_lss,2009_poisson_smith,2017_poisson_noise}, where we cross-correlate clusters with galaxies. This is because these are two independent tracers of the large-scale structure as long as we are looking at separation distances which are much larger than the cluster sizes.\footnote{We note that in reality there will be some correlation between the Poisson terms from galaxies and clusters \cite{2009_poisson_smith,2017_poisson_noise}. To explore the influence of these results on our forecasts, we consider the extreme case of fully correlated Poisson noise in appendix \ref{App:Noise_case}. } Thus, for the cross-pairwise effect, we only have the cosmic variance contribution to the statistical uncertainty,
	\begin{align}
		&C_{T_{\mathrm{cl-gal}}}(r,r')=\frac{1}{V_s}\; \frac{1}{15\pi^2}\sum_{i,j}\;m_i\;m_j\,\tau_{\mathrm{eff}}(m_i,\chi_i)\tau_{\mathrm{eff}}(m_j,\chi_j)\,\bigg[
		1+2(\mathbf{\hat{n}_{i}}\cdot\mathbf{\hat{n}'_{j}})^2\bigg]\, [afH](\chi_i)\nonumber\\
		&\hspace{1cm}\,[afH](\chi_j)\int dk\,\left(b^{\mathrm{gal}}_1(\chi_i)P(k,\chi_i)\right)\left(b^{\mathrm{gal}}_1(\chi_j)P(k,\chi_j)\right)\;j_{1}(kr)\;j_{1}(kr')\nonumber\\
		&\hspace{1cm}\left[\left(1+b^{\mathrm{gal}}_1(\chi_i)b^{\mathrm{cl}}_1(m_i,\chi_i)\xi(x,\chi_i)\right)\left(1+b^{\mathrm{gal}}_1(\chi_j)b^{\mathrm{cl}}_1(m_j,\chi_j)\xi(x',\chi_j)\right)\right]^{-1}.
	\end{align}
	Furthermore, since we are just stacking clusters, the instrument and measurement noise variance also get reduced by a factor of $1/2$ compared to the pairwise estimator covariance. Therefore,
	\begin{align}
		\mathcal{N}_{\mathrm{cl-gal}}(r)= \sqrt{\,\sum_{i}\,w^2_i\left(\mathcal{N}^{\,i}_{\mathrm{cl}}\right)^2}.
	\end{align} 
	This is also true for the noise due to measurement error in the optical depth.
	\begin{align}
		\mathcal{N}^{\mathrm{tau}}_{\mathrm{cl-gal}}(r)=  \left(\Delta \tau_{\mathrm{eff}}/\tau_{\mathrm{eff}}\right)\sqrt{\sum_{i}\,\left(	T^i_{\mathrm{cl-gal}}(r)\right)^2},
	\end{align}
	where the sum is over all cluster-galaxy pairs.
	The absence of the Poisson noise diminishes the covariance significantly. The number of cluster-galaxy pairs is also significantly larger  than the number of cluster pairs. Therefore, the measurement noise also decreases in the cross-pairwise method. The covariance matrix provides information regarding the precision with which we can measure the pairwise and cross-pairwise kSZ signal. {\color{black}We show the covariance matrices for the pairwise and cross -pairwise kSZ estimators in Appendix \ref{App:Cov_mat}. }We can now use this information to forecasts how precisely we can measure cosmological parameters within the context of a particular cosmological model using the pairwise and cross-pairwise kSZ measurements. 
	\begin{table}[]
		\hspace{-0.0cm}
		\centering
		\begin{tabular}{|c|clccccc|}
			\hline
			\multicolumn{1}{|l|}{}                                      & \multicolumn{7}{c|}{Number of clusters/galaxies}                                                                                                                                                                                                                                                                                                                                                                                                                                                                                                                                                                                                          \\ \hline
			\begin{tabular}[c]{@{}c@{}}Redshift (z)\\ bins\end{tabular} & \multicolumn{1}{c|}{\begin{tabular}[c]{@{}c@{}}CMB-S4 \\ clusters\\ ($\times10^3$)\end{tabular}} & \multicolumn{1}{c|}{\begin{tabular}[c]{@{}c@{}}SO\\ clusters\\ ($\times10^2$)\end{tabular}} & \multicolumn{1}{c|}{\begin{tabular}[c]{@{}c@{}}LSST \\ galaxies\\ ($\times10^8$)\end{tabular}} & \multicolumn{1}{c|}{\begin{tabular}[c]{@{}c@{}}DESI \\ BGS\\ ($\times10^5$)\end{tabular}} & \multicolumn{1}{c|}{\begin{tabular}[c]{@{}c@{}}DESI \\ LRG\\ ($\times10^5$)\end{tabular}} & \multicolumn{1}{c|}{\begin{tabular}[c]{@{}c@{}}DESI\\ ELG\\ ($\times10^5$)\end{tabular}} & \begin{tabular}[c]{@{}c@{}}DESI \\ QSO\\ ($\times10^4$)\end{tabular} \\ \hline
			0.0 - 0.3                                                   & \multicolumn{1}{c|}{8.207}                                                                       & \multicolumn{1}{l|}{37.4}                                                                   & \multicolumn{1}{c|}{3.032}                                                                     & \multicolumn{1}{c|}{115.0}                                                                & \multicolumn{1}{c|}{-}                                                                    & \multicolumn{1}{c|}{-}                                                                   & -                                                                    \\ \hline
			0.3 - 0.4                                                   & \multicolumn{1}{c|}{8.811}                                                                       & \multicolumn{1}{l|}{32.5}                                                                   & \multicolumn{1}{c|}{2.699}                                                                     & \multicolumn{1}{c|}{30.63}                                                                & \multicolumn{1}{c|}{-}                                                                    & \multicolumn{1}{c|}{-}                                                                   & -                                                                    \\ \hline
			0.4 - 0.5                                                   & \multicolumn{1}{c|}{10.318}                                                                      & \multicolumn{1}{l|}{36.6}                                                                   & \multicolumn{1}{c|}{3.237}                                                                     & \multicolumn{1}{c|}{11.91}                                                                & \multicolumn{1}{c|}{5.665}                                                                & \multicolumn{1}{c|}{-}                                                                   & -                                                                    \\ \hline
			0.5 - 0.6                                                   & \multicolumn{1}{c|}{10.648}                                                                      & \multicolumn{1}{l|}{36.1}                                                                   & \multicolumn{1}{c|}{3.509}                                                                     & \multicolumn{1}{c|}{3.195}                                                                & \multicolumn{1}{c|}{9.371}                                                                & \multicolumn{1}{c|}{-}                                                                   & -                                                                    \\ \hline
			0.6 - 0.7                                                   & \multicolumn{1}{c|}{11.245}                                                                      & \multicolumn{1}{l|}{32.7}                                                                   & \multicolumn{1}{c|}{3.556}                                                                     & \multicolumn{1}{c|}{-}                                                                    & \multicolumn{1}{c|}{12.68}                                                                & \multicolumn{1}{c|}{1.893}                                                               & 3.438                                                                \\ \hline
			0.7 - 0.8                                                   & \multicolumn{1}{c|}{10.262}                                                                      & \multicolumn{1}{l|}{29.5}                                                                   & \multicolumn{1}{c|}{3.435}                                                                     & \multicolumn{1}{c|}{-}                                                                    & \multicolumn{1}{c|}{13.68}                                                                & \multicolumn{1}{c|}{11.72}                                                               & 5.332                                                                \\ \hline
			0.8 - 1.0                                                   & \multicolumn{1}{c|}{15.787}                                                                      & \multicolumn{1}{l|}{14.2}                                                                   & \multicolumn{1}{c|}{6.101}                                                                     & \multicolumn{1}{c|}{-}                                                                    & \multicolumn{1}{c|}{25.71}                                                                & \multicolumn{1}{c|}{57.65}                                                               & 14.36                                                                \\ \hline
			1.0 - 1.2                                                   & \multicolumn{1}{c|}{12.789}                                                                      & \multicolumn{1}{l|}{6.38}                                                                   & \multicolumn{1}{c|}{4.805}                                                                     & \multicolumn{1}{c|}{-}                                                                    & \multicolumn{1}{c|}{4.287}                                                                & \multicolumn{1}{c|}{47.37}                                                               & 18.84                                                                \\ \hline
			1.2 - 3.0                                                   & \multicolumn{1}{c|}{29.802}                                                                      & \multicolumn{1}{l|}{3.50}                                                                   & \multicolumn{1}{c|}{10.44}                                                                     & \multicolumn{1}{c|}{-}                                                                    & \multicolumn{1}{c|}{-}                                                                    & \multicolumn{1}{c|}{66.39}                                                               & 161.2                                                                \\ \hline
		\end{tabular}
		\caption{Number of clusters and galaxies used in each redshift bin for calculation of the signal and noise for the pairwise and cross-pairwise estimator. These are expected numbers of clusters and galaxies that will be detected by these future and ongoing surveys. CMB-S4 has a sky coverage of $50\%$. In the case of cross-pairing CMB-S4 and LSST, the sky overlap is $45\%$ of the sky. Hence, the number of clusters used to calculate cluster-galaxy pairs is reduced by 9/10. For cross-pairing CMB-S4 and DESI galaxies, the sky overlap is $25\%$ of the sky, so the cluster number is reduced by 1/2. {\color{black}Similarly, the sky coverage for SO is $40\%$ and it has a complete overlap with DESI.} }
		\label{tab:cl_gal_no}
	\end{table}
	\section{Forecast for cosmological parameters: Fisher matrix analysis \label{constraints}}
	We can use the Fisher matrix formalism to forecast the precision with which a future experimental setup can measure the cosmological parameters, if they are not too far from the fiducial values. The Fisher matrix is a local Gaussian approximation to the distribution of the parameters in the vicinity of their fiducial values.  We use 10 free parameters in our extension to $\Lambda$CDM cosmological model as shown in table \ref{tab:central_val},
	\begin{table}
		\hspace{-0.2cm}
		\begin{tabular}{|c|c|l|l|l|l|l|l|l|l|l|}
			\hline
			\multirow{2}{*}{Parameter}           & \multirow{2}{*}{$\Omega_bh^2$} & \multirow{2}{*}{$\Omega_m$} & \multirow{2}{*}{$\Omega_k$} & \multirow{2}{*}{$H_0$}  & \multirow{2}{*}{$\ln(10^{10}A_s)$} & \multirow{2}{*}{$n_s$}    & \multirow{2}{*}{$w_0$}     & \multirow{2}{*}{$w_a$}    & \multirow{2}{*}{$\tau_{\mathrm{reion}}$} & \multirow{2}{*}{$\gamma$} \\
			&                                &                             &                             &                         &                                    &                           &                            &                           &                                          &                           \\ \hline
			\multicolumn{1}{|l|}{Fiducial Value} & 0.0224                         & \multicolumn{1}{c|}{0.3111} & \multicolumn{1}{c|}{0.0}    & \multicolumn{1}{c|}{67} & \multicolumn{1}{c|}{3.195}         & \multicolumn{1}{c|}{0.96} & \multicolumn{1}{c|}{-0.95} & \multicolumn{1}{c|}{0.0} & \multicolumn{1}{c|}{0.06}                & \multicolumn{1}{c|}{0.55} \\ \hline
		\end{tabular}
		\caption{The 10 parameter cosmological model and the corresponding fiducial values used in our analysis}
		\label{tab:central_val}
	\end{table}
	where $\Omega_b, \Omega_m,$ and $\Omega_k$ are the dimensionless baryon, matter, and curvature density parameters respectively,
	$H_0$ is the Hubble constant, and $h = H_0/100$. The dark energy equation of state parameters are $w_0$ and $w_a$ with the equation of state, $w(a) = w_0 + (1 - a) w_a$. The growth rate exponent is  $\gamma$, such that the growth rate $f ={d\ln D}/{d \ln a}= \left(\Omega_m(a)\right)^{\gamma}$,
	and $n_s$ and $A_s$ are the spectral index and normalisation of the primordial spectrum of curvature perturbations respectively.  Deviation of the $w(a)$ and $\gamma$ from their $\Lambda$CDM values would tell us about the properties of dark energy and gravity on large scales. The Fisher matrix for two parameters $p_\mu$ and $p_\nu$ is given by \cite{1999_Tegmark,2009_Coe,2015ApJ_DarkEnergy},
	
	\begin{align}
		F_{\mu\nu}=\sum^{N_\chi}_{\chi}\sum^{N_r}_{a,b}\frac{\partial T(r_a,\chi)}{\partial p_\mu}C^{-1}_T(r_a,r'_b,\chi)\frac{\partial T(r'_b,\chi)}{\partial p_\nu},
	\end{align}
	\begin{figure}
		\hspace{-0.5cm}
		\centering
		\begin{subfigure}[t]{0.51\textwidth}
			\centering
			\includegraphics[width=\linewidth]{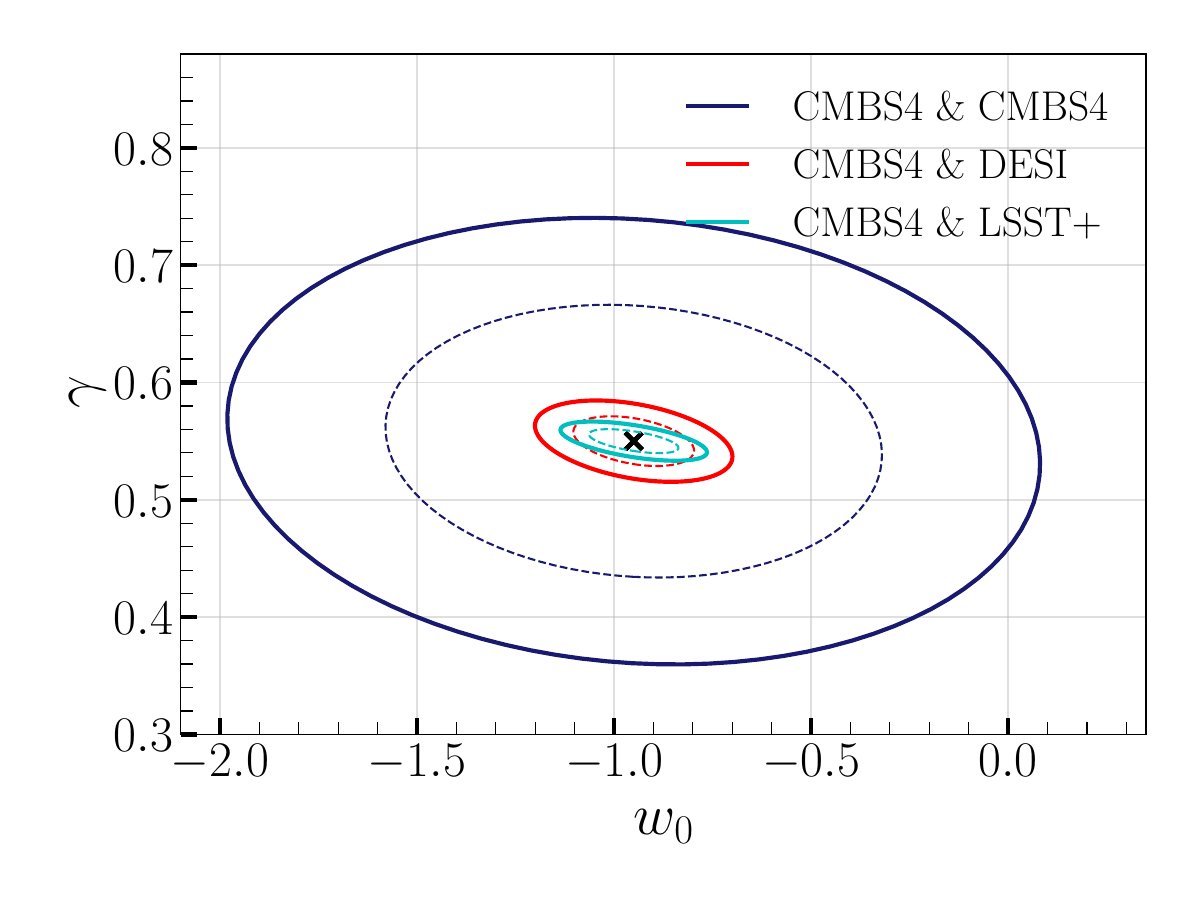}
			\caption{Planck (CMB)+CMB lensing prior}
			\label{fig:wo_gamma_planck}
		\end{subfigure}%
		\hfill
		\begin{subfigure}[t]{0.51\textwidth}
			\centering
			\includegraphics[width=\linewidth]{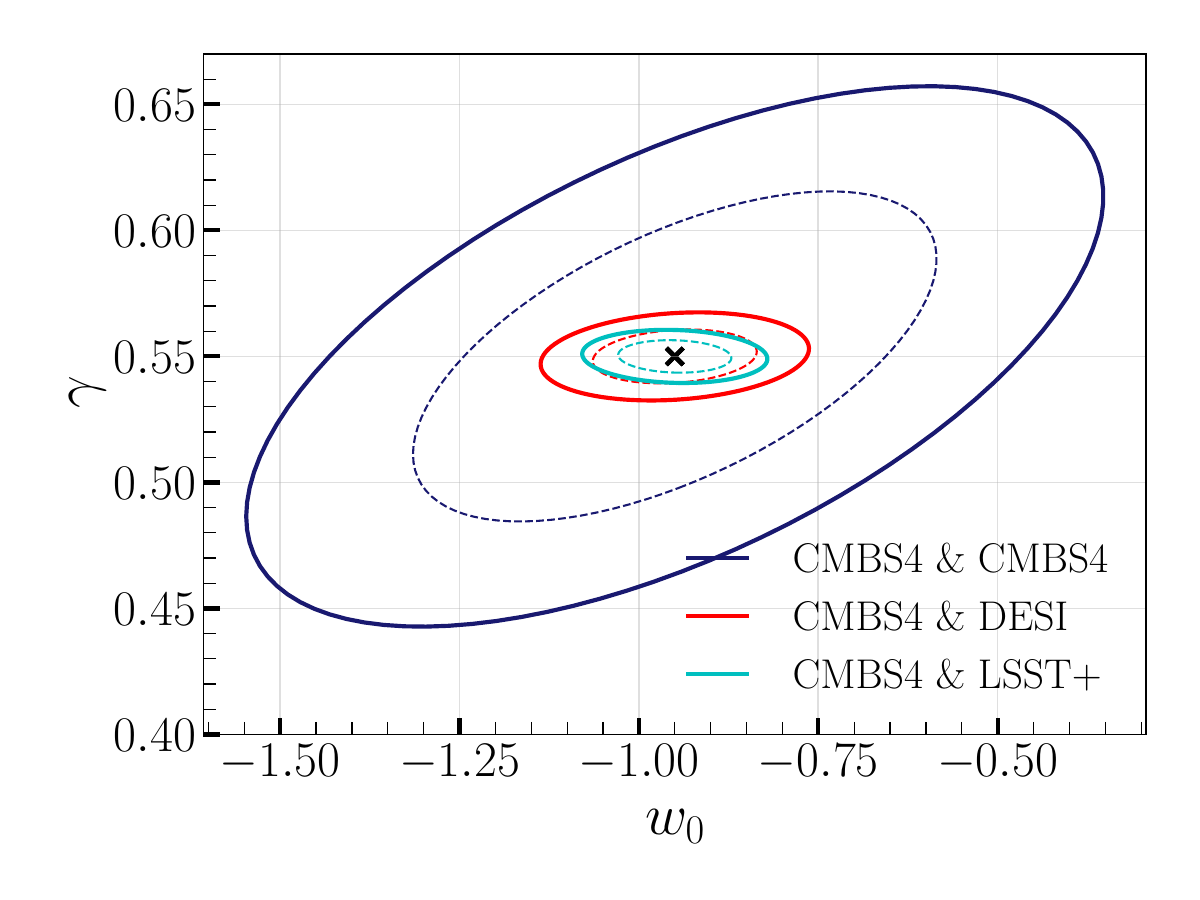}
			\caption{Planck (CMB)+CMB lensing+DESI BAO prior}
			\label{fig:wo_gamma_planck_BAO}
		\end{subfigure}
		
		\begin{subfigure}[t]{0.51\textwidth}
			\hspace{-0.7cm}
			\centering
			\includegraphics[width=\linewidth]{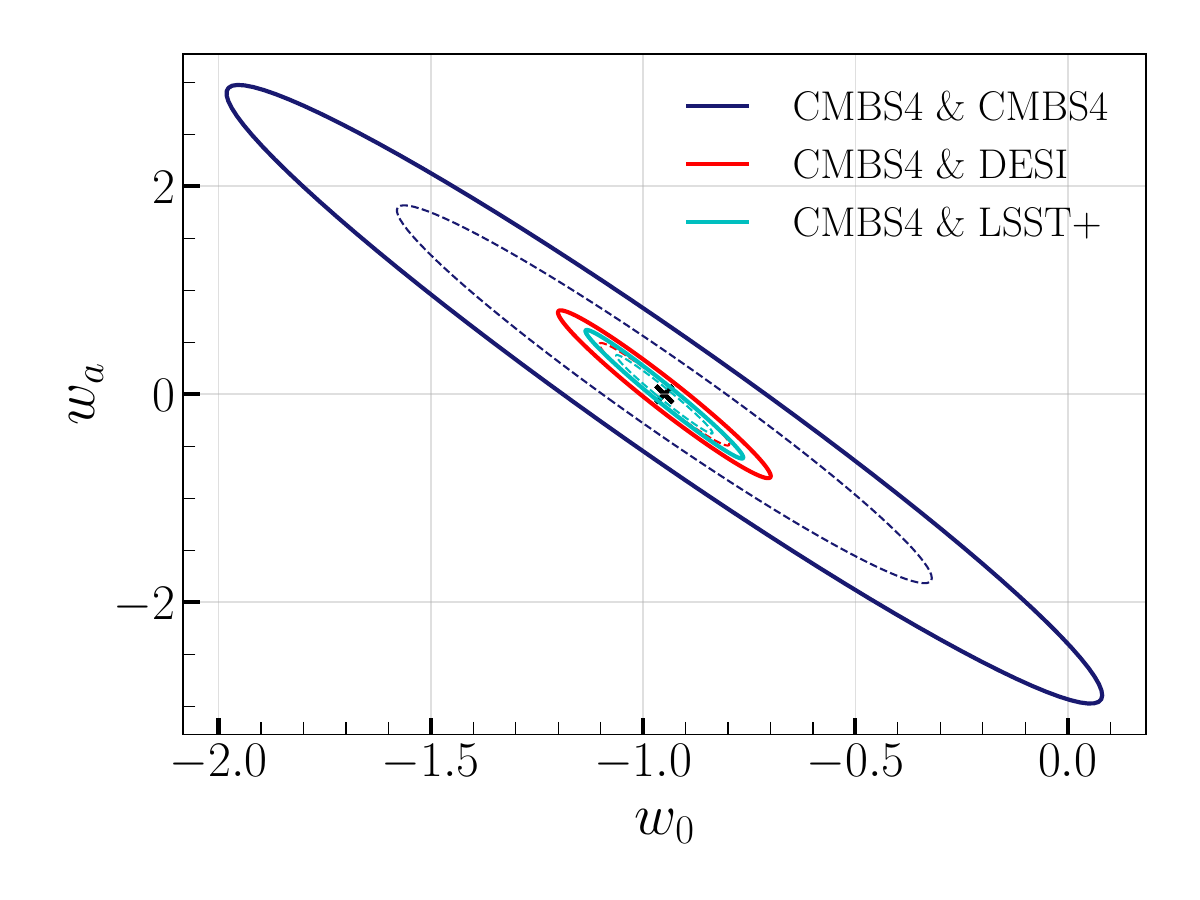}
			\caption{Planck (CMB)+CMB lensing prior}
			\label{fig:wo_wa_planck}
		\end{subfigure}%
		\begin{subfigure}[t]{0.51\textwidth}
			\hspace{-0.7cm}
			\centering
			\includegraphics[width=\linewidth]{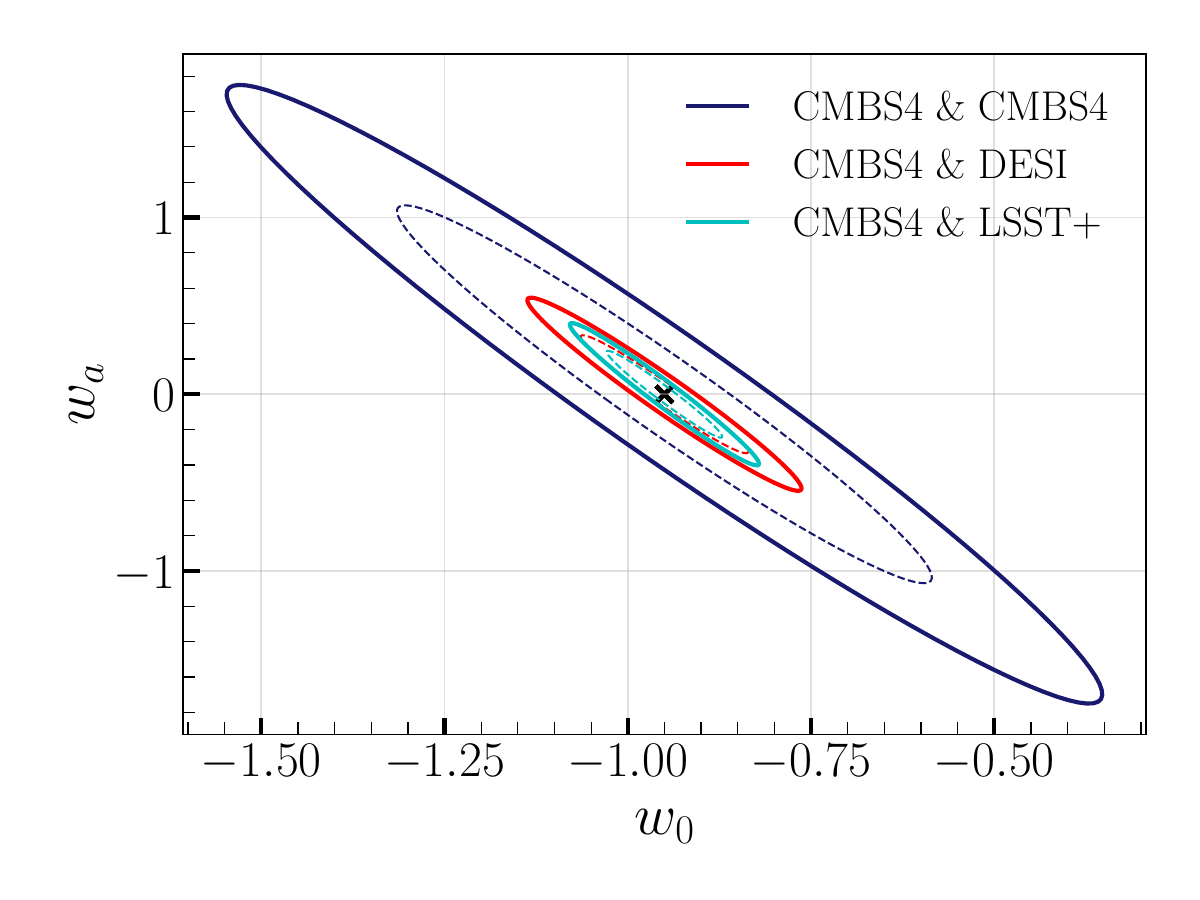}
			\caption{Planck (CMB)+CMB lensing+DESI BAO prior}
			\label{fig:wo_wa_planck_BAO}
		\end{subfigure}
		
		\caption{The dashed and solid lines represent the $1\sigma$ and $2\sigma$ confidence ellipses respectively. In dark blue, we show the result for the pairwise estimator. In red, we give the result for the cross-pairwise estimator where we cross-correlate clusters with galaxies from DESI, and in cyan, we show the cross-pairwise results using LSST galaxies. The upper panel provides the constraints in the $w_0-\gamma$ plane from two different prior choices, on the left with only Planck (CMB)+CMB lensing prior and on the right with Planck (CMB)+CMB lensing+DESI BAO prior. The bottom panel gives the constraints for the $w_0-w_a$ plane.}
		\label{fig:Conf_ellipse}
	\end{figure}

	\begin{figure}
		\centering
		\hspace{-0.5cm}
		\begin{subfigure}[t]{0.51\textwidth}
			\centering
			\includegraphics[width=\linewidth]{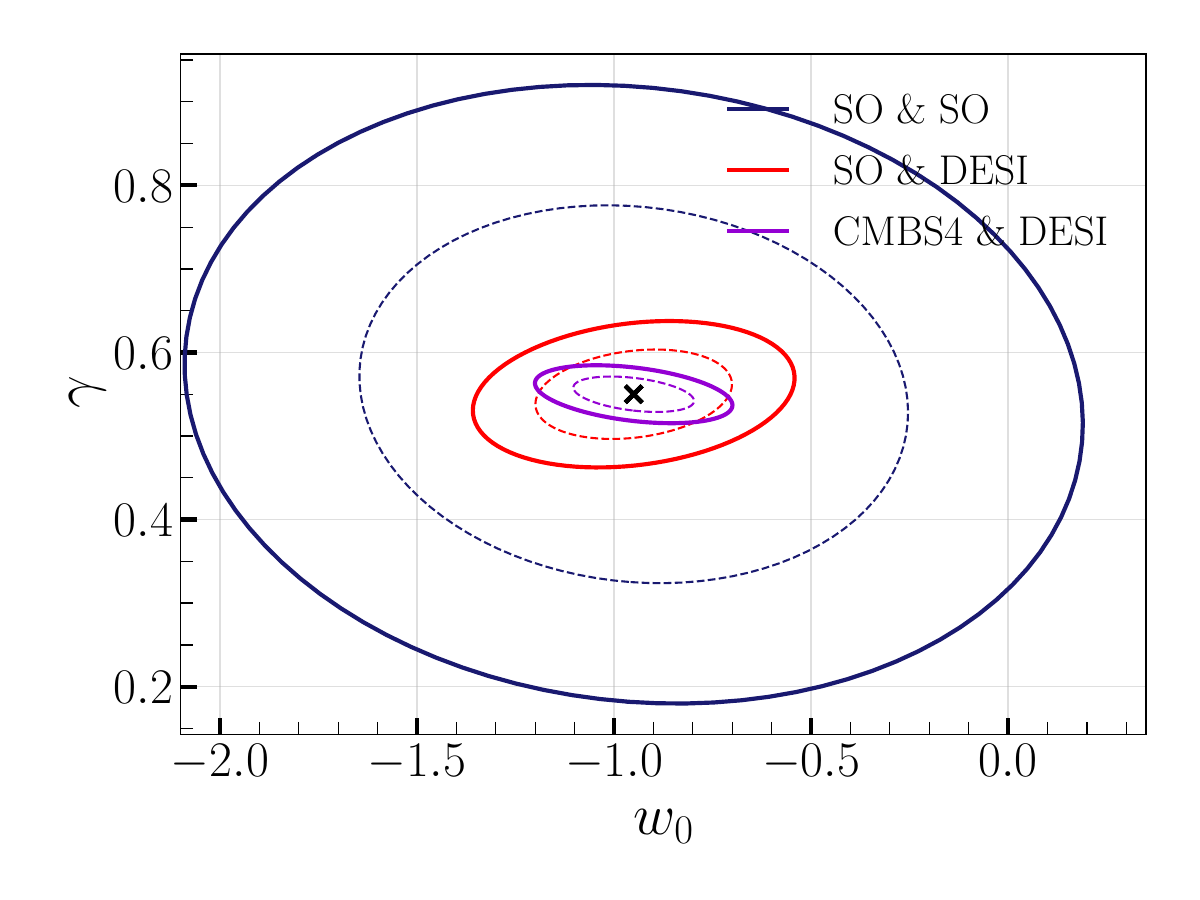}
			\caption{Planck (CMB)+CMB lensing prior}
			\label{fig:wo_gamma_planck_for_SO}
		\end{subfigure}%
		\hfill
		\begin{subfigure}[t]{0.51\textwidth}
			\centering
			\includegraphics[width=\linewidth]{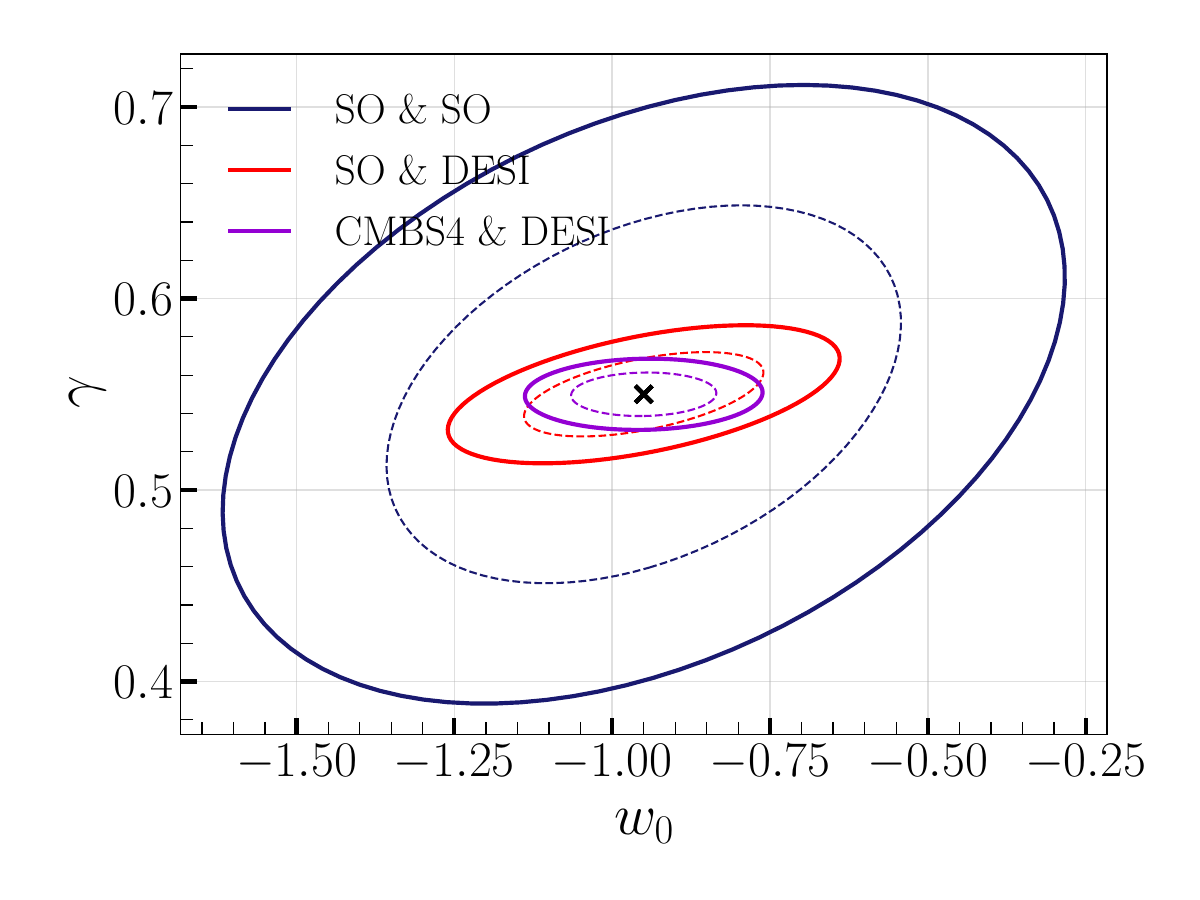}
			\caption{Planck (CMB)+CMB lensing+DESI BAO prior}
			\label{fig:wo_gamma_planck_BAO_for_SO}
		\end{subfigure}
		
		
		\begin{subfigure}[t]{0.51\textwidth}
			\centering
			\hspace{-0.7cm}
			\includegraphics[width=\linewidth]{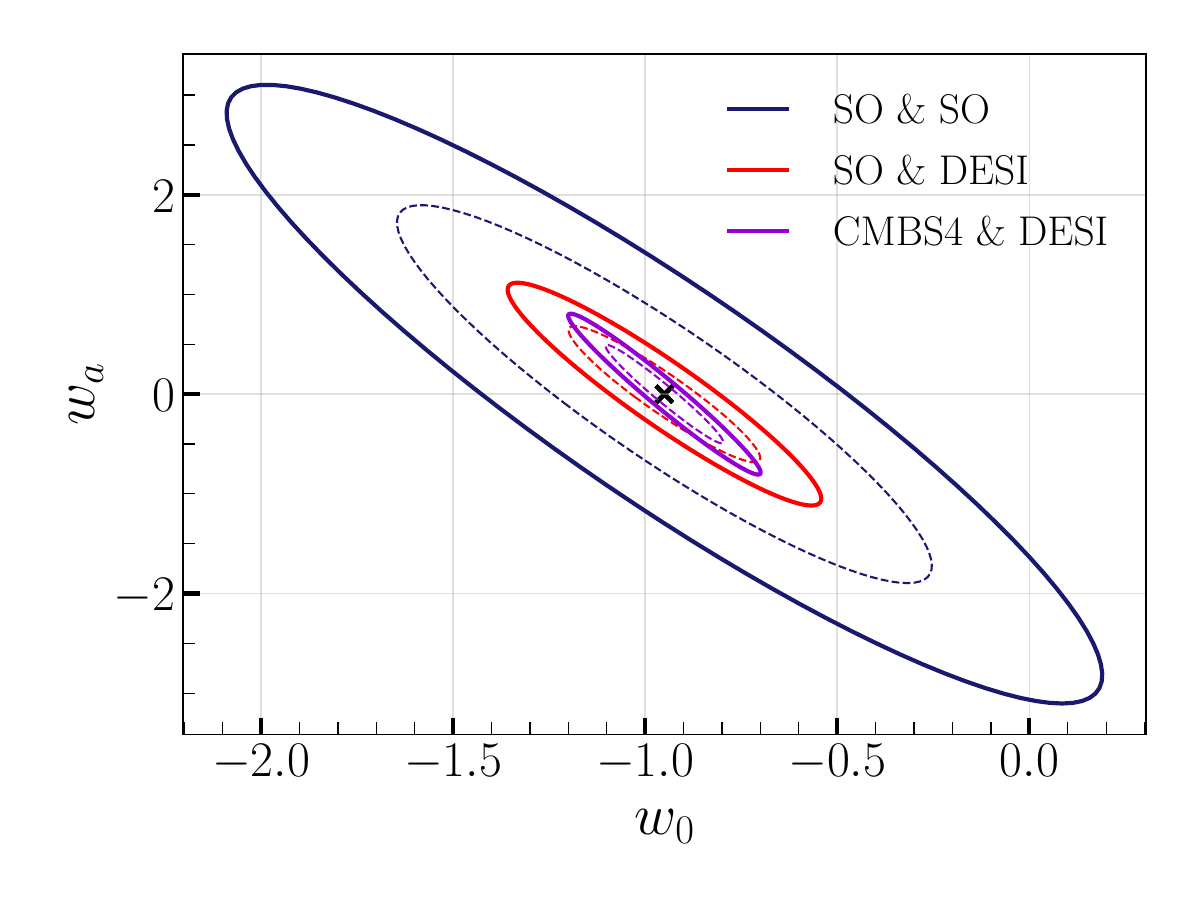}
			\caption{Planck (CMB)+CMB lensing prior}
			\label{fig:wo_wa_planck_for_SO}
		\end{subfigure}%
		\begin{subfigure}[t]{0.51\textwidth}
			\centering
			\hspace{-0.7cm}
			\includegraphics[width=\linewidth]{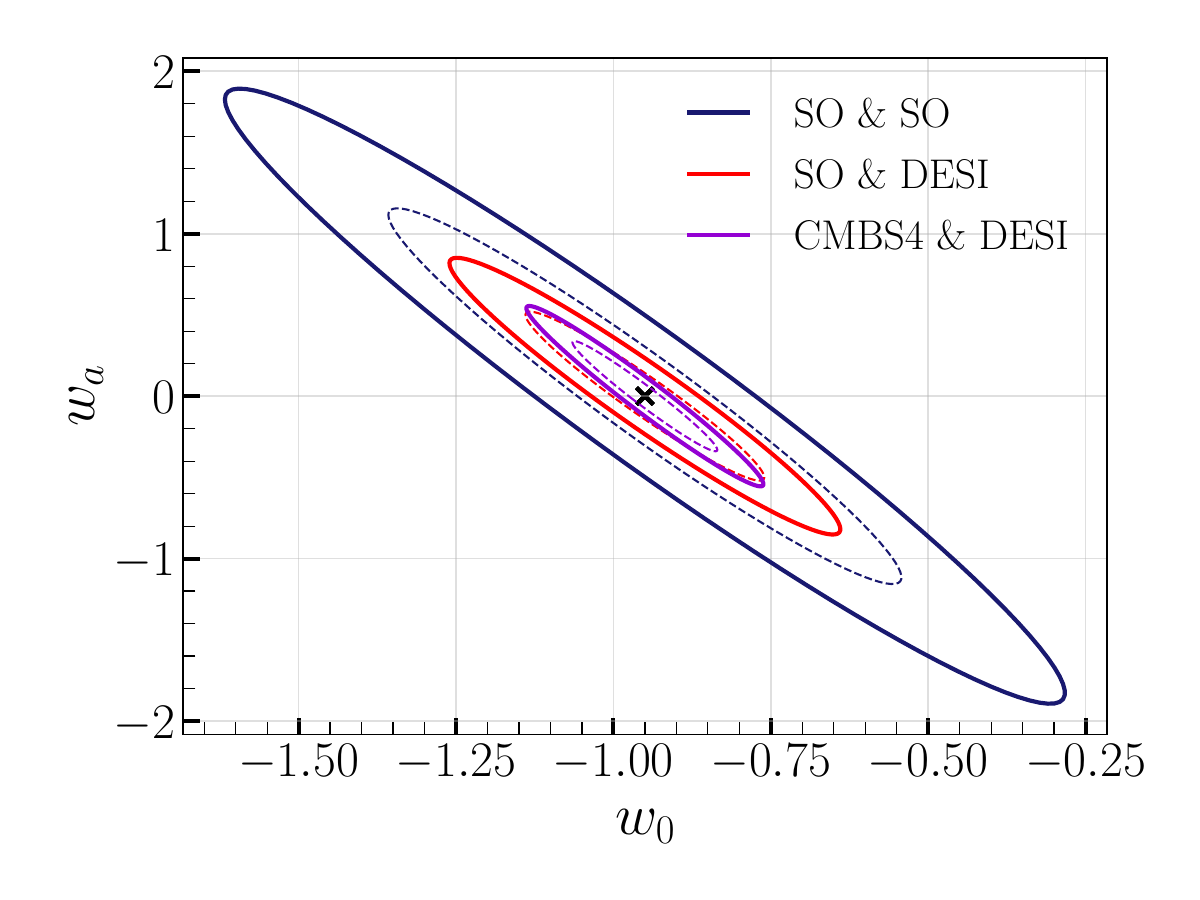}
			\caption{Planck (CMB)+CMB lensing+DESI BAO prior}
			\label{fig:wo_wa_planck_BAO_for_SO}
		\end{subfigure}
		
		\caption{{\color{black}The dashed and solid lines represent the $1\sigma$ and $2\sigma$ confidence ellipses respectively. The upper panel provides the constraints in the $w_0-\gamma$ plane from two different prior choices: on the left with only Planck (CMB)+CMB lensing prior and on the right with Planck (CMB)+CMB lensing+DESI BAO prior. The bottom panel gives the constraints for the $w_0-w_a$ plane. In blue, we show the results for just the pairwise estimator using clusters from SO. In red, we show the constraints from the cross-pairwise estimator with clusters from SO and galaxies from DESI. For comparison, we also show (in purple) the constraints from using the cross-pairwise estimator with CMB-S4 and DESI.}}
		\label{fig:Conf_ellipse_for_SO}
	\end{figure}
	\begin{table}
		\centering
		\begin{tabular}{|c|c|c|c|c|c|}
			\hline
			\multirow{2}{*}{Parameters} & \multirow{2}{*}{\begin{tabular}[c]{@{}c@{}}SO \& \\ SO\end{tabular}} & \multirow{2}{*}{\begin{tabular}[c]{@{}c@{}}CMB-S4 \&\\  CMB-S4\end{tabular}} & \multirow{2}{*}{\begin{tabular}[c]{@{}c@{}}SO \& \\ DESI\end{tabular}} & \multirow{2}{*}{\begin{tabular}[c]{@{}c@{}}CMB-S4 \&\\  DESI\end{tabular}} & \multirow{2}{*}{\begin{tabular}[c]{@{}c@{}}CMB-S4 \&\\  LSST+\end{tabular}} \\
			&                                                                      &                                                                              &                                                                        &                                                                            &                                                                             \\ \hline
			$\Omega_bh^2$               & 0.00014                                                              & 0.00014                                                                      & 0.00013                                                                & 0.00011                                                                    & 0.000080                                                                    \\ \hline
			$\Omega_m$                  & 0.028                                                                & 0.023                                                                        & 0.013                                                                  & 0.0071                                                                     & 0.0042                                                                      \\ \hline
			$\Omega_k$                  & 0.0020                                                               & 0.0019                                                                       & 0.0017                                                                 & 0.0014                                                                     & 0.0012                                                                      \\ \hline
			$H_0$                       & 2.7                                                                  & 2.3                                                                          & 1.4                                                                    & 0.81                                                                       & 0.47                                                                        \\ \hline
			$\ln(10^{10}A_s)$           & 0.014                                                                & 0.014                                                                        & 0.014                                                                  & 0.0073                                                                     & 0.0052                                                                      \\ \hline
			$n_s$                       & 0.0042                                                               & 0.0042                                                                       & 0.004                                                                  & 0.0035                                                                     & 0.0027                                                                      \\ \hline
			$w_0$                       & 0.27                                                                 & 0.24                                                                         & 0.13                                                                   & 0.075                                                                      & 0.052                                                                       \\ \hline
			$w_a$                       & 0.76                                                                 & 0.71                                                                         & 0.34                                                                   & 0.22                                                                       & 0.16                                                                        \\ \hline
			$\tau_{\mathrm{reion}}$     & 0.0071                                                               & 0.0071                                                                       & 0.0069                                                                 & 0.0044                                                                     & 0.0037                                                                      \\ \hline
			$\gamma$                    & 0.065                                                                & 0.043                                                                        & 0.015                                                                  & 0.0070                                                                     & 0.0042                                                                      \\ \hline
		\end{tabular}
		\caption{ Here we provide the marginalised $1\sigma$ errors on individual cosmological parameters as obtained from the Fisher martix of the pairwise and cross-pairwise kSZ estimators with Planck (CMB)+CMB lensing+DESI BAO prior choice. }
		\label{tab:cosmo_constraint}
	\end{table}where $T$ can be either $T_{\mathrm{pairwise}}$ or $T_{\mathrm{cl-gal}}$. The midpoint of our comoving distance (redshift) bins are $\chi$, the objects in each pair have comoving separations falling in bins centered at $r_a$, and $N_\chi$ and $N_r$ are the number of redshift and spatial separation bins, respectively. \\\\
	\textcolor{black} { The authors  in \cite{2012_Hand} have used the 5000 most luminous BOSS  (Baryon Oscillation Spectroscopic Survey) DR9 galaxies in the ACT ( Atacama Cosmology Telescope) sky region as the proxies for clusters for the  calculation of the pairwise correlation. In our analysis, galaxies are not proxies for clusters but are independent tracers of the large-scale structure. We create a mock catalog of clusters using the mass-redshift selection function for the CMB-S4 wide and the SO, given by \cite{2019_cmbs4_science_case,2022ApJ_raghunathan,2019JCAP_Simons_Obs,2022_Raghunathan} and the halo mass function given in \cite{2008_Tinker}. The CMB-S4 survey will have a sensitivity of $2 \mathrm{\mu K\,arcmin}$ sensitivity while the SO-Goal experiment will have a sensitivity of  $6 \mathrm{\mu K\,arcmin}$. In the case of cluster count, CMB-S4 will detect around $10^5$ cluster while SO-Goal has a prediction of detecting around 26500 clusters using their SZ surveys respectively. Both these surveys have a similar angular resolution of $1 \, \mathrm{arcmin}$}.   For the cross-pairwise case, we have considered galaxy samples from DESI and LSST ``gold sample" (assuming spectroscopic data from future follow-ups for LSST). For DESI we use the  bright galaxy survey (BGS), emission line galaxies (ELG), luminous red galaxies (LRG), and quasars (QSO). We note that the ELG sample only goes upto $z=1.6$. We put a high redshift cut-off of $z=3.0$ for all the samples. {\color{black} The number of clusters expected to be detected using CMB-S4 and and galaxies that DESI and LSST will detect in different redshift bin is given in table \ref{tab:cl_gal_no}. DESI and CMB-S4 have a sky overlap of $25\%$  of the sky while LSST and CMB-S4 have a $45\%$ of the sky common to both surveys, so, we have to reduce the number of clusters used in the calculation cross-pairwise covariance matrix accordingly}.  {\color{black} Similarly, SO has a sky coverage of $40\%$ with complete overlap with DESI. While pairing, we only consider cluster-galaxy pairs which are separated in space by more than 20 Mpc, so that the galaxies do not  belong to the cluster we are pairing them with. } \\\\
	There is not enough information in the kSZ effect to constrain all the 10 cosmological parameters. Therefore, we need to combine the kSZ measurements with other complementary experiments. We have considered our results in combination with a Planck (CMB)+ CMB lensing prior and with a Planck (CMB)+ CMB lensing+DESI BAO prior. To calculate the prior we use the the code Cobaya \cite{2020_Cobaya} to run the Markov Chain Monte Carlo (MCMC) pipeline to obtain the posterior distribution of the nine parameters excluding the growth rate exponent $\gamma$. {\color{black} We note that the priors are obtained assuming General Relativity to be the correct theory of gravity, in which case the growth rate is a derived parameter.} This is a good approximation since the parameter $\gamma$ encapsulates late time deviations from $\Lambda$CDM and general relativity to which CMB is not very sensitive. A deviation of $\gamma$ from the $\Lambda$CDM value, when including the kSZ effect, would thus signal new physics beyond the $\Lambda$CDM cosmology.  
	\begin{figure}
		\hspace{-1.cm}
		\begin{subfigure}{1.0\textwidth}
			\vfill
			\centering
			\includegraphics[width=1.02\linewidth]{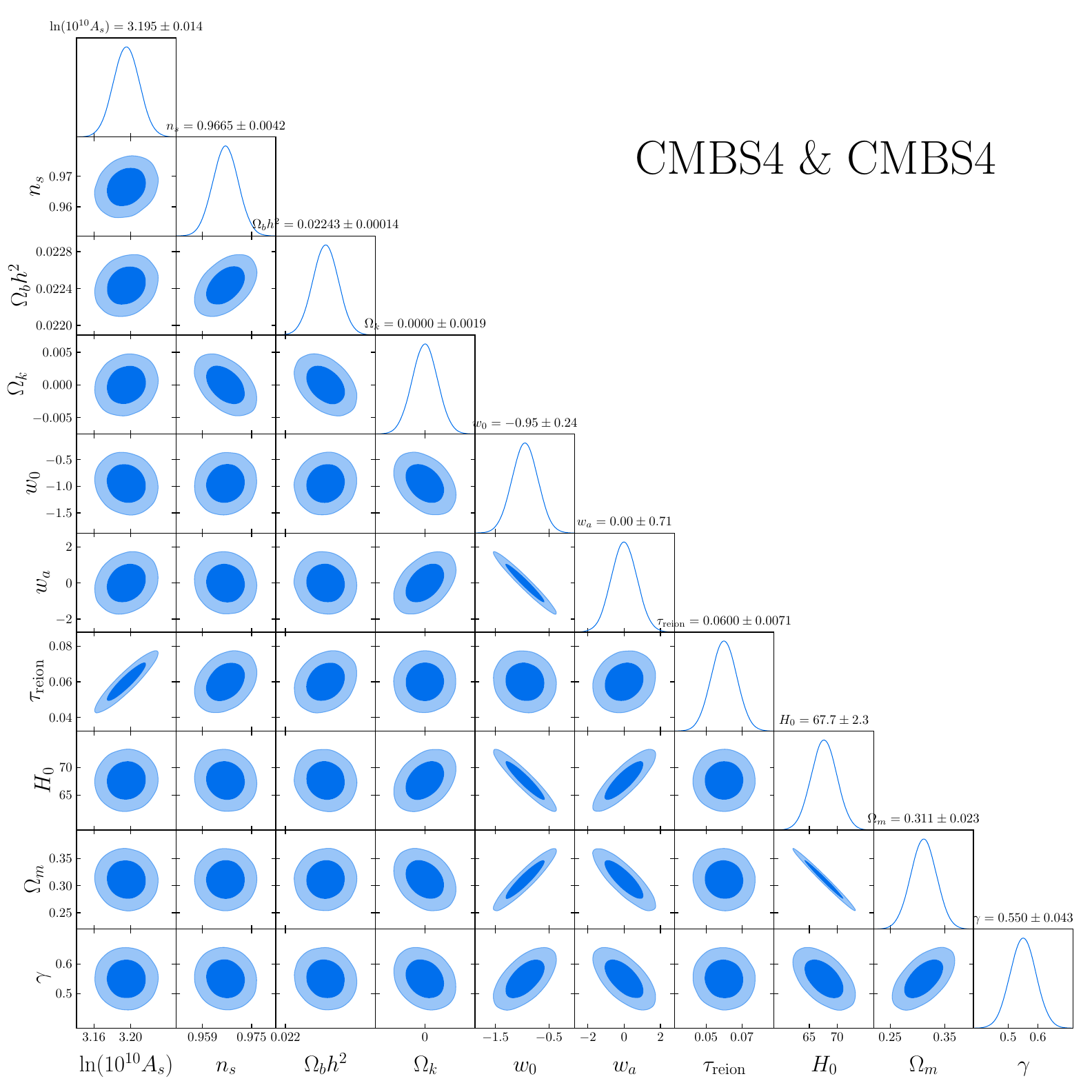}
		\end{subfigure}
		\caption{ The $1\sigma$ and $2\sigma$ constraints on the 10 cosmological parameters for the pairwise kSZ estimator for Planck (CMB)+CMB lensing+DESI BAO prior choice. }
		\label{fig:corner_CMB-S4_CMB-S4}
	\end{figure}\begin{figure}
		\hspace{-1.cm}
		\begin{subfigure}{1.0\textwidth}
			\vfill
			\centering
			\includegraphics[width=1.02\linewidth]{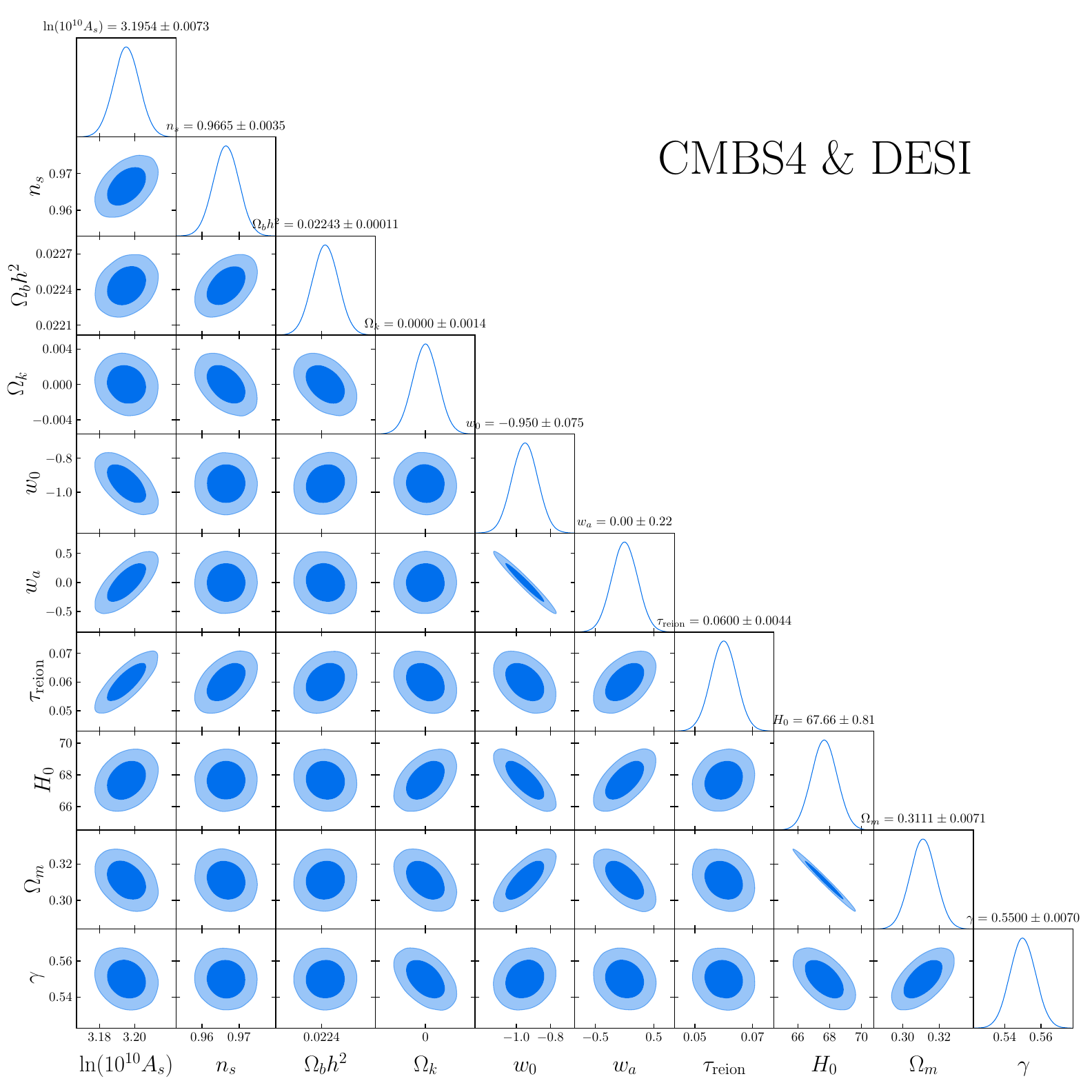}
		\end{subfigure}
		\caption{Same as \ref{fig:corner_CMB-S4_CMB-S4}, but for the cross-pairwise kSZ effect using the DESI galaxies.}
		\label{fig:corner_CMB-S4_desi}
	\end{figure}\begin{figure}
		\hspace{-1.cm}
		\begin{subfigure}{1.0\textwidth}
			\vfill
			\centering
			\includegraphics[width=1.02\linewidth]{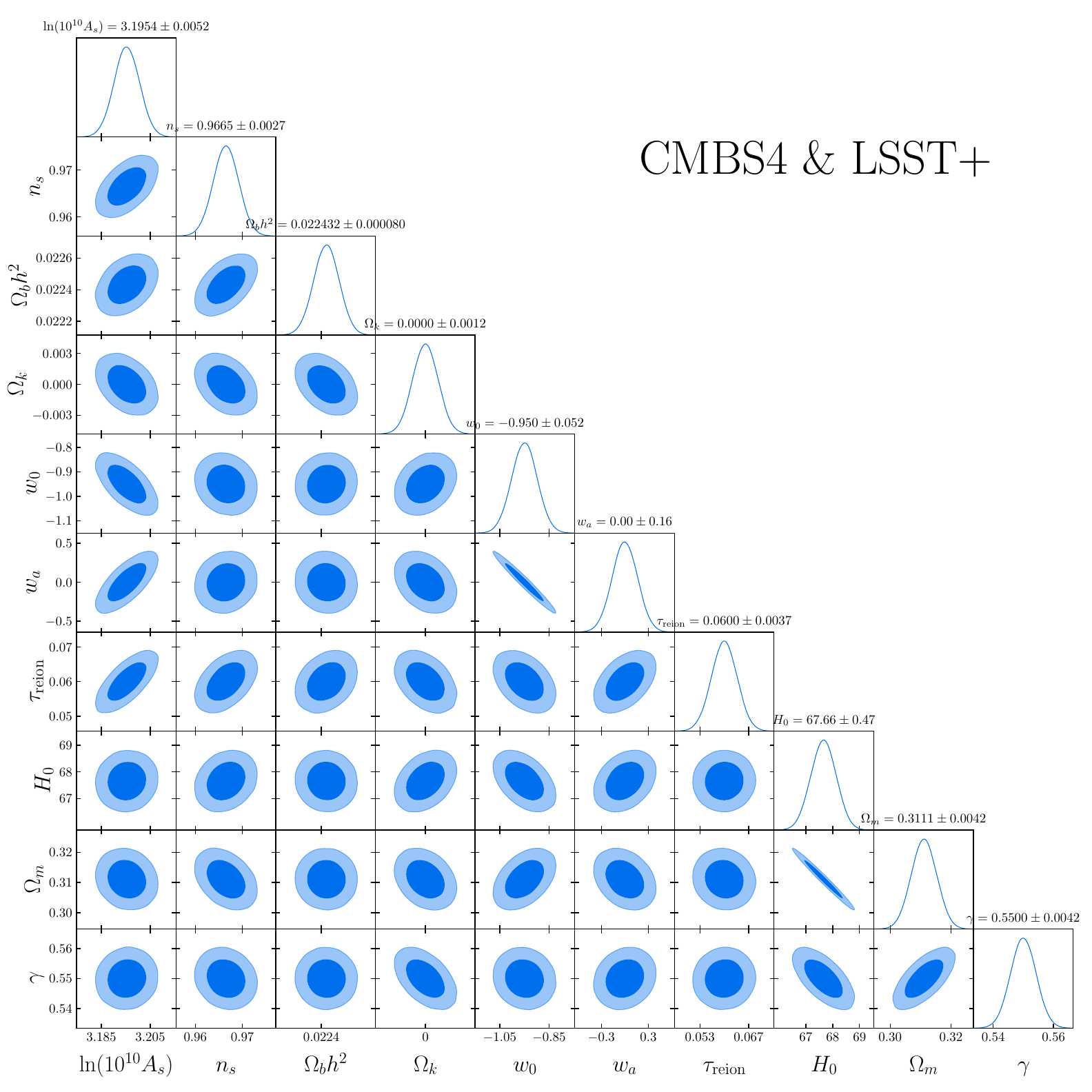}
		\end{subfigure}
		\caption{Same as \ref{fig:corner_CMB-S4_CMB-S4}, but for the cross-pairwise kSZ effect using the LSST ``gold sample'' galaxies. }
		\label{fig:corner_CMB-S4_lsst}
	\end{figure}
	\begin{figure}
		\hspace{-1.cm}
		\begin{subfigure}{1.0\textwidth}
			\vfill
			\centering
			\includegraphics[width=1.02\linewidth]{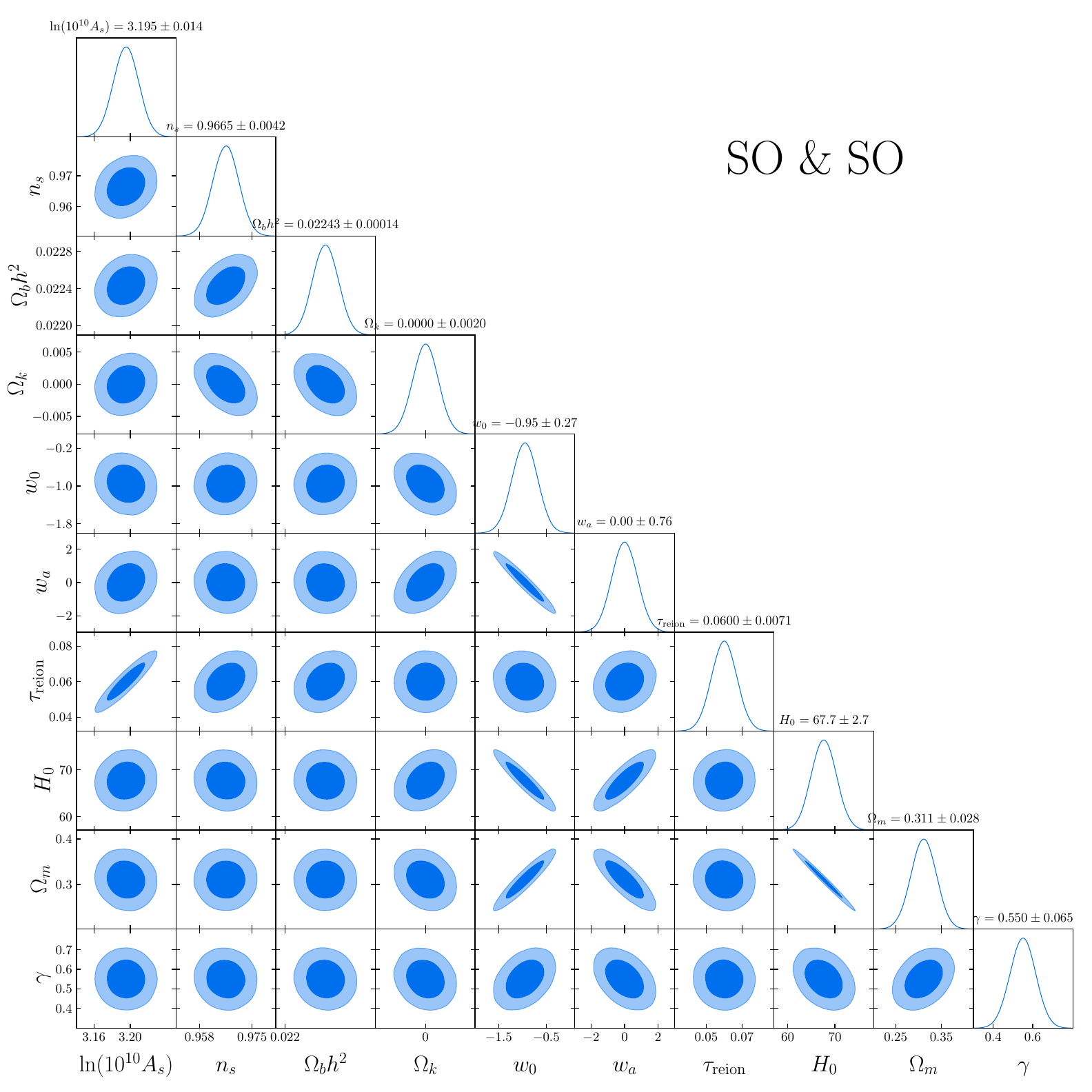}
		\end{subfigure}
		\caption{Same as \ref{fig:corner_CMB-S4_CMB-S4}, but for the pairwise kSZ effect using the SO clusters.}
		\label{fig:corner_so_so}
	\end{figure}
	\begin{figure}
		\hspace{-1.cm}
		\begin{subfigure}{1.0\textwidth}
			\vfill
			\centering
			\includegraphics[width=1.02\linewidth]{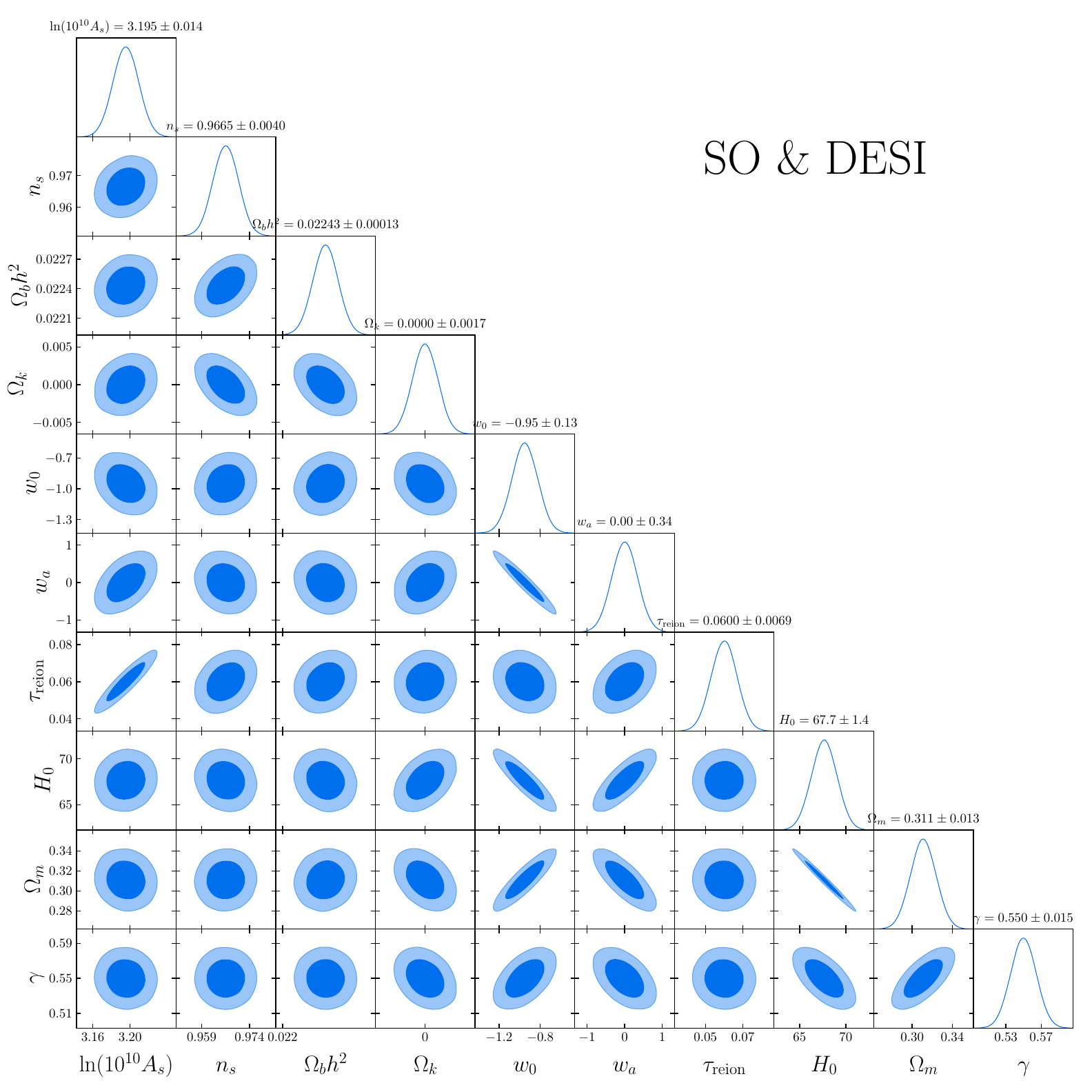}
		\end{subfigure}
		\caption{Same as \ref{fig:corner_CMB-S4_CMB-S4}, but for the cross-pairwise kSZ effect using the SO clusters and DESI galaxies.}
		\label{fig:corner_so_desi}
	\end{figure}
	
	We use the following likelihood function for obtaining the MC chains:  Planck 2018 lowl - TT and EE, Planck NPIPE highl CamSpec-TTTEEE \cite{2020_Planck_likelihoods}, and Planck pr4 lensing \cite{2020_Planck_lensing}. To include the BAO information, we use the year-1 data of the DESI survey \cite{2024_DESI_BAO_Gal,2024_DESI_BAO_Gal,2024_DESI_BAO_Lya}. From the Monte Carlo chains, we obtain the required covariance matrix by marginalising over the nuisance parameters.\\\\ 
	The Fisher matrix is evaluated at the fiducial values of the parameters. In this work, we are interested in the measurement of the dark energy parameters $w_0,w_a,$ and the growth rate exponent $\gamma$.  In order to present the result in the $w_0-w_a$ and $w_a-\gamma$ plane, we marginalise \cite{2009_Albrecht,2013_fisher_Khatri} over the rest of the 8 parameters for each case. The marginalised Fisher matrix can be represented by an ellipse given by $\Delta p^T F\Delta p=\chi^2(\nu)$, where $\Delta p=p-p_{\mathrm{fiducial}}$. For two parameters, $\chi^2(1) = 2.3$ and $\chi^2(2) = 6.17$ respectively which represent the $68.3\%$ ($1\sigma$) and  $95.4\%$ ($2\sigma$) confidence intervals. In figure \ref{fig:wo_gamma_planck} and figure \ref{fig:wo_gamma_planck_BAO} we show the results in the $w_0-\gamma$ plane with Planck (CMB)+CMB lensing and Planck (CMB)+CMB lensing+DESI BAO priors respectively. The same is shown for $w_0-w_a$ plane in figure \ref{fig:wo_wa_planck} and figure \ref{fig:wo_wa_planck_BAO}.  {\color{black}Similarly, in figure \ref{fig:Conf_ellipse_for_SO}, we show the results where we use clusters that will be observed by the SO LAT survey. }We can see that the cross-pairwise estimator provides much tighter constraints on the cosmological parameters. As expected, cross-pairing clusters with galaxies from DESI (which will provide us with spectroscopic information about galaxies) provides a significant improvement over the previously studied pairwise kSZ effect.  Using the LSST ``gold sample'' galaxies produces even better constraints, assuming there will be spectroscopic follow-up available, since we need accurate distances to galaxies to calculate the optimised weights as well as the sign of the weights. We also need distances to assign each cluster galaxy pair to appropriate separation bin. We note that photometric redshifts may be sufficient since the accuracy requirement for each of these tasks is not very stringent \cite{2013_fabian_photo_z}. We leave the study of using photometric redshifts for galaxies in cross-pairwise estimator for future work. To understand the constraints on the full 10-dimensional parameter space, we also show the corner plots for the pairwise estimator in figure \ref{fig:corner_CMB-S4_CMB-S4} \textcolor{black}{(referred as "CMB-S4 \& CMB-S4" to signify that both the objects in a pair are clusters which will be detected by CMB-S4 wide survey.)} and for the cross-pairwise estimator in figure \ref{fig:corner_CMB-S4_desi} and figure \ref{fig:corner_CMB-S4_lsst}  \textcolor{black}{(referred as "CMB-S4 \& DESI" and "CMB-S4 \& LSST+"  to signify that the pairing between a cluster from either the SO-Gold survey or the CMB-S4 wide survey and  galaxies which will be detected by DESI and  LSST+ surveys respectively.) In figure \ref{fig:corner_so_so} and figure \ref{fig:corner_so_desi}, we show the pairwise and cross-pairwise results using clusters from SO and galaxies from DESI. (reffered as  "SO \& SO" and "SO \& DESI" respectively.)} with the Planck (CMB)+CMB lensing+ DESI BAO prior. The $1-\sigma$ errors on each parameter for the pairwise and cross-pairwise estimator are listed above the 1-d marginalised distribution of each parameters, which we have also tabulated in table \ref{tab:cosmo_constraint}.\\\\
	It is clear from these plots that cross-pairing clusters with galaxies can yield better measurement of the cosmological parameters than the pairwise kSZ signal. In the most pessimistic but unrealistic case, where we have considered a fully correlated Poisson noise between the cluster and galaxy sample, the cross-pairwise method is still more effective in constraining cosmological parameters as shown in Appendix \ref{App:Noise_case}. The $1-\sigma$ error forecasts for all cases are compared in table \ref{tab:cosmo_constraint}. 
	{\color{black}Comparing our results for cross-pairwise estimator with the pairwise estimator, we see a improvement in error bars on dark energy parameters by a factor of $\sim2$ when pairing DESI galaxies with SO clusters and by a factor of $\sim3(4)$ when pairing DESI(LSST+) galaxies  with CMB-S4 clusters. Similarly,  we see a factor of  $\sim4$ and $\sim 6(10)$ for $\gamma$, for the respective cases.}
	\section{Conclusions\label{conclusion}}
	We present a new estimator for the kSZ signal, the cross-pairwise estimator which is based on the cross-correlation of the peculiar velocity of galaxy clusters with neighbouring galaxies, a concept that we had first developed for the polarised kSZ effect. In the existing pairwise kSZ method, a cluster is paired with another cluster and then pairs of such clusters which are separated by a given distance are stacked to obtain the average pairwise signal. Our new estimator is based on the idea that, instead of using cluster-cluster pair, we can pair a cluster with any other tracers with which the peculiar velocity of the cluster is correlated, and galaxies being superior in number are a much better tracer of the underlying gravitational field and therefore, the peculiar motion of the cluster. \\\\
	We have shown that using the cross-pairwise estimator has two main advantages. Firstly, even though we lose some signal as shown in figure \ref{fig:pairwisecrosspairwisesignal}, as the galaxies have negligible kSZ effect and smaller bias, the reduction in measurement and instrument errors is more significant due to the greater number of cluster-galaxy pairs compared to cluster-cluster pairs. Secondly, the Poisson (shot) noise contribution is absent in the case of the cross-pairwise kSZ estimator, since the clusters and galaxies are two independent tracers of the large scale structure. This decreases the statistical noise significantly and the combination of these two effects increases the signal-to-noise ratio compared to the pairwise kSZ effect. Of course, we are using extra data, i.e. a galaxy survey, for the cross-pairwise estimator.\\\\
	Our Fisher forecasts for the optimised pairwise estimator show that a significant improvement should be expected over the un-optimised estimator  used in \cite{2015ApJ_DarkEnergy}.  We  note that an exact comparison is not possible, because we use a different cosmological model and priors as well as different mass cutoffs for the cluster sample. { \color{black} We can improve the measurement of $w_0$, $w_a$, and $\gamma$ by a factor of 2.07, 2.23, and 4.33 when we use the cross-pairwise estimator with  SO and DESI compared to using pairwise estimator with SO. The result with cross-pairwise estimator with SO and DESI (stage 3 surveys) is even better that the pairwise estimator using CMB-S4 clusters (a stage 4 survey).  Similarly, we get an improvement in the measurement of $w_0$, $w_a$, and $\gamma$ by a factor of 3.2, 3.2, and 6.1 respectively, when using cross-pairwise estimator for DESI galaxies with CMB-S4 clusters and a factor of 4.8, 4.4, and 10 for LSST+ galaxies  with DESI compared to the optimised pairwise estimator using CMB-S4.} Thus, combining CMB surveys with the galaxy surveys shows great promise in future to probe the cosmic velocity fields and discover new physics beyond the $\Lambda$CDM model.
	\acknowledgments
	This work is supported by the Department of Atomic Energy, Government of India, under Project Identification Number RTI 4002. We acknowledge the use of computational facilities of the Department of Theoretical Physics at Tata Institute of Fundamental Research, Mumbai. We want to specially thank Somnath Bharadwaj for his valuable comments about the connection between the BBGKY hierarchy and the ensemble averages. AKG is also thankful to Aseem Paranjape, Subhabrata Majumdar, Shadab Alam, Ravi Sheth, and Anoma Ganguly for useful discussions.
	\appendix
	\section{Derivation of pairwise kSZ effect\label{App:Review_kSZ}}
	The expectation value $\Big\langle \mathbf{v_{1}}({\mathbf{x_1}})-\mathbf{v_{2}}({\mathbf{x_2}})\Big\rangle_2$ can be written as,
	\begin{align}
		\label{ensemble_ksz}
		\Big\langle \mathbf{v_{1}}({\mathbf{x_1}})-\mathbf{v_{2}}({\mathbf{x_2}})\Big\rangle_2 =\frac{\Big\langle \left(\mathbf{v_{1}}-\mathbf{v_{2}}\right)(1+\delta^h_1)(1+\delta^h_2)\Big\rangle}{\Big\langle (1+\delta^h_1)(1+\delta^h_2)\Big\rangle}.
	\end{align}
	Let us first evaluate the numerator.
	\begin{align}
		\label{ensemble_expn_terms}
		\Big\langle \left(\mathbf{v_{1}}-\mathbf{v_{2}}\right)(1+\delta^h_1)(1+\delta^h_2)\Big\rangle&=\Big\langle\mathbf{v_{1}}\delta^h_2\Big\rangle-\Big\langle\mathbf{v_{2}}\delta^h_1\Big\rangle+\mathrm{higher\,order\,terms}\nonumber\\
		&=b_1\Big\langle\mathbf{v_{1}}^{(1)}\,\delta_2^{(1)}\Big\rangle-b_1\Big\langle\mathbf{v_{2}}^{(1)}\,\delta_1^{(1)}\Big\rangle+\mathrm{higher\,order\,terms}.
	\end{align}
	Moving to Fourier space, we can write,
	\begin{align}
		&\delta^{(1)}(\mathbf{x},\chi)=D(\chi)\int\frac{d^3\mathbf{k}}{(2\pi)^3}\exp\left(i \mathbf{k}\cdot\mathbf{x}\right)\delta^{(1)}(\mathbf{k}),
	\end{align}
	\begin{align}
		&\mathbf{v}^{(1)}(\mathbf{x},\chi)=D(\chi)[afH](\chi)\int\frac{d^3\mathbf{k}}{(2\pi)^3}\exp\left(i \mathbf{k}\cdot\mathbf{x}\right) \frac{i\mathbf{k}}{k^2}\delta^{(1)}(\mathbf{k}),
	\end{align}
	If we make a substitution $\mathbf{k_1}\rightarrow-\mathbf{k_1}$, then $\Big\langle\mathbf{v_{2}}^{(1)}\,\delta_1^{(1)}\Big\rangle$ becomes equal to the negative of $\Big\langle\mathbf{v_{1}}^{(1)}\,\delta_2^{(1)}\Big\rangle$. Therefore, we need to evaluate only $\Big\langle\mathbf{v_{1}}^{(1)}\,\delta_2^{(1)}\Big\rangle$.
	\begin{align}
		\Big\langle\mathbf{v_{1}}^{(1)}\,\delta_2^{(1)}\Big\rangle&=D^2(afH)\int\int\frac{d^3\mathbf{k_1}d^3\mathbf{k_2}}{(2\pi)^6}\left(\frac{i\mathbf{k_1}}{k_1^2}\right)\exp\left(i\mathbf{k_1}\cdot\mathbf{x_1}\right)\exp\left(i\mathbf{k_2}\cdot\mathbf{x_2}\right)\Big\langle\delta_1^{(1)}(\mathbf{k_1})\delta_1^{(1)}(\mathbf{k_2})\Big\rangle\nonumber\\
		&=D^2(afH)\int\frac{d^3\mathbf{k_1}}{(2\pi)^3}\left(\frac{i\mathbf{k_1}}{k_1^2}\right)\exp\left(i\mathbf{k_1}\cdot(\mathbf{x_1}-\mathbf{x_2})\right)P(k_1)\nonumber\\
		&=D^2(afH)\int\frac{ dk_1}{(2\pi)^3}d\Omega_{\mathbf{\hat{k}_1}}\,i\mathbf{k_1}\exp\left(i\mathbf{k_1}\cdot\mathbf{r}\right)P(k_1)\nonumber\\
		&=D^2(afH)\int\frac{ dk_1}{(2\pi)^3}\,d\Omega_{\mathbf{\hat{k}_1}}\,\frac{\partial}{\partial\mathbf{r}}\exp\left(i\mathbf{k_1}\cdot\mathbf{r}\right)P(k_1),
	\end{align}
	where $\mathbf{r}=\mathbf{x_1}-\mathbf{x_2}$. Expanding $\exp\left(i\mathbf{k_1}\cdot\mathbf{r}\right)$ in terms of the spherical harmonics and the Bessel function we get,
	\begin{align}
		\exp(i\mathbf{k_1}\cdot\mathbf{r})=4\pi\sum_{L_1,M_1}i^{L_1}Y^{*}_{L_1M_1}(\mathbf{\hat{k}_1})Y_{L_1M_1}(\mathbf{\hat{r}})j_{L_1}(k_1r).
	\end{align}
	Therefore,
	\begin{align}
		\int d\Omega_{\mathbf{\hat{k}_1}}\frac{\partial}{\partial\mathbf{r}}\exp(i\mathbf{k_1}\cdot\mathbf{r})&=4\pi\sum_{L_1,M_1}i^{L_1}\int d\Omega_{\mathbf{\hat{k}_1}}Y^{*}_{L_1M_1}(\mathbf{\hat{k}_1})\frac{\partial}{\partial\mathbf{r}}\left(Y_{L_1M_1}(\mathbf{\hat{r}})j_{L_1}(k_1r)\right)\nonumber\\
		&=4\pi\sum_{L_1,M_1}i^{L_1}\sqrt{4\pi}\,\delta_{L_1,0}\,\delta_{M_1,0}\frac{\partial}{\partial\mathbf{r}}\left(Y_{L_1M_1}(\mathbf{\hat{r}})j_{L_1}(k_1r)\right)\nonumber\\
		&=(4\pi)\frac{\partial}{\partial\mathbf{r}}j_{0}(k_1x)=-(4\pi)k_1\,j_{1}(k_1r)\mathbf{\hat{r}}.
	\end{align}
	Therefore,
	\begin{align}
		\label{corr_vd}
		\Big\langle\mathbf{v_{1}}^{(1)}\,\delta_2^{(1)}\Big\rangle&=-D^2(afH) \frac{1}{2\pi^2}\int dk_1 k_1\,j_{1}(k_1r)P(k_1)\;\mathbf{\hat{r}}.
	\end{align}
	Let us now look at the volume average of the matter density two-point correlation function $\xi(r)=\big\langle\delta_1^{(1)}\delta_2^{(1)}\big\rangle$.
	\begin{align}
		\xi(r)=\frac{D^2}{2\pi^2}\int dk_1 k^2_1\,j_{0}(k_1x)P(k_1).
	\end{align}
	Therefore, the volume average of $\xi(r)$ is,
	\begin{align}
		\label{xi_bar2}
		\bar{\xi}(r)&=\frac{3}{r^3}\int_{0}^{r}\xi(y)y^2dy\nonumber\\
		&=D^2\frac{1}{2\pi^2}\int dk_1 k^2_1\,P(k_1)\;\frac{3}{r^3}\int_{0}^{r}\,j_{0}(k_1y)y^2dy\nonumber\\
		&=D^2\frac{1}{2\pi^2}\int dk_1 k^2_1\,P(k_1)\;\frac{3}{r^3}\left(\frac{r^2}{k_1}j_{1}(k_1r)\right)\nonumber\\
		&=D^2\frac{1}{2\pi^2}\int dk_1 k_1\,P(k_1)\;\frac{3}{r}\left(j_{1}(k_1r)\right).
	\end{align}
	Comparing Eq.(\ref{corr_vd}) and Eq.(\ref{xi_bar2}) we get,
	\begin{align}
		\Big\langle\mathbf{v_{1}}^{(1)}\,\delta_2^{(1)}\Big\rangle=-\frac{1}{3}	(afH)\bar{\xi}(r)\;\mathbf{r}.
	\end{align}
	Redefining the separation vector as, $\mathbf{r}=\mathbf{x_2}-\mathbf{x_1}$, we can absorb the minus sign. This choice of separation vector is necessary to give consistent result with our sign convention. Thus we get,
	\begin{align}
		\Big\langle \left(\mathbf{v_{1}}-\mathbf{v_{2}}\right)(1+\delta^h_1)(1+\delta^h_2)\Big\rangle=\frac{2}{3}[afH](\chi)\,b_1(\chi,m)	\bar{\xi}(r)\;\mathbf{r}.
	\end{align}
	Similarly we can solve the denominator in Eq.(\ref{ensemble_ksz}).
	\begin{align}
		\label{deno_expansion_2}
		\Big\langle (1+\delta^h_1)(1+\delta^h_2)\Big\rangle&=1+D^2b^2_1\int\int\frac{d^3\mathbf{k_1}d^3\mathbf{k_2}}{(2\pi)^6}\exp\left(i\mathbf{k_1}\cdot\mathbf{x_1}\right)\exp\left(i\mathbf{k_2}\cdot\mathbf{x_2}\right)\Big\langle\delta_1^{(1)}(\mathbf{k_1})\delta_1^{(1)}(\mathbf{k_2})\Big\rangle\nonumber\\
		&=1+\frac{D^2b^2_1}{2\pi^2}\int dk_1k_1^2j_0(k_1r)P(k_1)=1+b^2_1\,\xi(r).
	\end{align}
	Therefore, the expectation value $\Big\langle \mathbf{v_{1}}({\mathbf{x_1}})-\mathbf{v_{2}}({\mathbf{x_2}})\Big\rangle_2$ becomes,
	\begin{align}
		\Big\langle \mathbf{v_{1}}({\mathbf{x_1}})-\mathbf{v_{2}}({\mathbf{x_2}})\Big\rangle_2=\frac{2}{3}[afH](\chi)\frac{\,b_1(\chi,m).	\bar{\xi}(r)}{1+b^2_1\,\xi(r)}\;\mathbf{r}.
	\end{align}
	Finally, we can write the unpolarised pairwise kSZ signal as,
	\begin{align}
		\label{ksz_final}
		T_{\mathrm{pairwise}}(r)=-\frac{2}{3}\sum_{i}\;w_i\;\tau_{\mathrm{eff}}(m_i,\chi_i)[afH](\chi_i)\tau_{\mathrm{eff}}(m_i,\chi_i)\frac{\,b_1(\chi_i,m_i)	\bar{\xi}(r)}{1+b^2_1(m_i,\chi_i)\,\xi(r)}\;\mathbf{r}\cdot\mathbf{\hat{n}_{i}}.
	\end{align}
	We can similarly repeat the entire calculation for the cross-pairwise effect. In the case of cross-pairwise kSZ effect, the only surviving term in Eq.(\ref{ensemble_expn_terms}) is $\Big\langle\mathbf{v_{1}}\delta^h_2\Big\rangle$. This is because the first term is multiplied with the Thomson optical depth of the cluster, while the second term is multiplied with the Thomson optical depth of the galaxy. Since the Optical depth of clusters is almost two orders of magnitude larger than that of galaxies, we will get negligible contribution from the second term. Hence, we consider the second term to be equal to zero. This reduces the sum by a factor of 2. We also need to consider the bias factor correctly. The bias factor in the numerator of Eq.(\ref{ensemble_ksz}) will now be the galaxy bias factors. The final expression of the ensemble average of the cross-pairwise estimator is,
	\begin{align}
		\label{cross_pair_ksz_app}
		T_{\mathrm{cl-gal}}(r)=-\frac{1}{3}\sum_{i}w_i[afH](\chi_i)\tau_{\mathrm{eff}}(m_i,\chi_i)\frac{r\,b^{\mathrm{gal}}_1\;	\bar{\xi}(r)}{1+b^{\mathrm{gal}}_1\,b^{\mathrm{cl}}_1(m_i,\chi_i)\,\xi(r)}\;\cos\theta_i,
	\end{align}
	\section{Calculation of the statistical noise \label{pairwise_cosmic_variance}}
	For pairwise kSZ between two cluster pairs in separation and redshift bins $[r,\chi]$ and $[r',\chi']$, the statistical noise contribution to the covariance matrix can be written as:	
	\begin{align}
		\label{cov_start_appendix}
		C_{T_{\mathrm{pairwise}}}(r,r')=&\Big\langle\int \frac{d^3\mathbf{x}d^3\mathbf{x}'}{V_s^2}\int \frac{ d\Omega_{r} d\Omega_{r'}}{(4\pi)^2}\hat{T}_{\mathrm{pairwise}}\;\hat{T'}_{\mathrm{pairwise}}\Big\rangle_4\nonumber\\
		&\hspace{6cm}-T_{\mathrm{pairwise}}(r)T'_{\mathrm{pairwise}}(r'),
	\end{align}
	where
	\begin{align}
		\hat{T}_{\mathrm{pairwise}}&=\sum_{i}\;w_i\;\tau_{\mathrm{eff}}(m_i,\chi_i) \left(\mathbf{v_{i1}}({\mathbf{x}})-\mathbf{v_{i2}}({\mathbf{x+r}})\right)\cdot\mathbf{\hat{n}_{i}}.
	\end{align}
	Note that instead of using  $\mathbf{ x_1}$ and $\mathbf{ x_2}$, we are using $\mathbf{ x}$ and $\mathbf{x+r}$ as our cluster position, just for calculation convenience. As we have shown, $w_i=\mathbf{\hat{r}_i}\cdot\mathbf{\hat{n}_{i}}$ and
	\begin{align}
		\label{4_point_defn}
		\int \frac{d^3\mathbf{x}d^3\mathbf{x}'}{V_s^2}\int \frac{ d\Omega_{r} d\Omega_{r'}}{(4\pi)^2}\Big\langle\hat{T}_{\mathrm{pairwise}}\hat{T}_{\mathrm{pairwise}}\Big\rangle_4&=\sum_{i,j}\;w_iw_j\,\tau_{\mathrm{eff}}(m_i,\chi_i)\tau_{\mathrm{eff}}(m_j,\chi_j)\nonumber \\
		&\hspace{1.5cm}\int \frac{d^3\mathbf{x}d^3\mathbf{x}'}{V_s^2}\int \frac{ d\Omega_{r} d\Omega_{r'}}{(4\pi)^2}\;\left(\frac{G_1}{G_0}\right),
	\end{align}
	where
	\begin{align}
		\label{num_avg}
		{G_1}=&\int \left(\mathbf{v_{i1}}({\mathbf{x}})-\mathbf{v_{i2}}({\mathbf{x+r}})\right)\cdot \mathbf{\hat{n}_{i}}\;\left(\mathbf{v_{j3}}({\mathbf{x'}})-\mathbf{v_{j4}}({\mathbf{x'+r'}})\right)\cdot \mathbf{\hat{n}'_{j}}\nonumber\\
		&\hspace{0.0cm}f_4\left(\mathbf{x},(\mathbf{x+r}),\mathbf{x'},(\mathbf{x'+r'}),\mathbf{v_{i1}},\mathbf{v_{i2}},\mathbf{v_{j3}},\mathbf{v_{j4}}\big|\chi_i,\chi_j,m_i,m_j\right)d^3\mathbf{v_{i1}}d^3\mathbf{v_{i2}}d^3\mathbf{v_{j3}}d^3\mathbf{v_{j4}},
	\end{align}
	and
	\begin{align}
		\label{deno_avg}
		{G_0}=\int f_4\left(\mathbf{x},(\mathbf{x+r}),\mathbf{x'},(\mathbf{x'+r'}),\mathbf{v_{i1}},\mathbf{v_{i2}},\mathbf{v_{j3}},\mathbf{v_{j4}}\big|\chi_i,\chi_j,m_i,m_j\right)d^3\mathbf{v_{i1}}d^3\mathbf{v_{i2}}d^3\mathbf{v_{j3}}d^3\mathbf{v_{j4}}.
	\end{align}
	Henceforth, we will suppress the subscript `i' and `j' in the velocity and density fields  and also in the LoS direction for brevity. The four particle phase space distribution of halos can be related to the local joint probability distribution of halo number density and velocity fields
	as,
	\begin{align}
		\label{f4_defn}
		&f_4\left(\mathbf{x},(\mathbf{x+r}),\mathbf{x'},(\mathbf{x'+r'}),\mathbf{v_{1}},\mathbf{v_{2}},\mathbf{v_{3}},\mathbf{v_{4}}\big|\chi_i,\chi_j,m_i,m_j\right)=\nonumber\\
		&\int n_1(\mathbf{x})n_2(\mathbf{x+r})	n_3(\mathbf{x'})	n_4(\mathbf{x'+r'})		\rho\left(n_1,n_2,n_3,n_4,\mathbf{v_{1}},\mathbf{v_{2}},\mathbf{v_{3}},\mathbf{v_{4}}\big|\chi_i,\chi_j,m_i,m_j\right)\nonumber\\
		&\hspace{12cm}dn_1\,dn_2\,dn_3\,dn_4.
	\end{align}
	The local number density $n(x|m, \chi)$ can be expressed as,	
	\begin{align}
		n(\mathbf{x}|m, \chi) = \bar{n}(m, \chi)(1 + \delta^h(\mathbf{x}, m, \chi)),
	\end{align}
	Therefore, using  Eq.(\ref{f4_defn}) in Eq.(\ref{deno_avg}), we get,
	\begin{align}
		\label{deno_final_avg}
		G_0&=\int f_4\left(\mathbf{x},(\mathbf{x+r}),\mathbf{x'},(\mathbf{x'+r'}),\mathbf{v_{1}},\mathbf{v_{2}},\mathbf{v_{3}},\mathbf{v_{4}}\big|\chi_i,\chi_j,m_i,m_j\right)d^3\mathbf{v_{1}}d^3\mathbf{v_{2}}d^3\mathbf{v_{3}}d^3\mathbf{v_{4}}\nonumber\\
		&=\bar{n}^2(\chi_i,m_i)\bar{n}^2(\chi_j,m_j)\Big\langle (1 + \delta_1^h(\mathbf{x}, m_i, \chi_i))(1 + \delta_2^h(\mathbf{x+r}, m_i, \chi_i))\nonumber\\
		&\hspace{5cm}(1 + \delta_3^h(\mathbf{x'}, m_j, \chi_j))(1 + \delta_4^h(\mathbf{x'+r'}, m_j, \chi_j))\Big\rangle .
	\end{align}
	Using  Eq.(\ref{f4_defn}) in Eq.(\ref{num_avg}), we get,
	\begin{align}
		\label{num_final_avg}
		&G_1=\bar{n}^2(\chi_i,m_i)\bar{n}^2(\chi_j,m_j)\Big\langle\left(\mathbf{v_{1}}({\mathbf{x}})-\mathbf{v_{2}}({\mathbf{x+r}})\right)\cdot \mathbf{\hat{n}}\;\left(\mathbf{v_{3}}({\mathbf{x'}})-\mathbf{v_{4}}({\mathbf{x'+r'}})\right)\cdot \mathbf{\hat{n}'}\nonumber\\
		&(1 + \delta_1^h(\mathbf{x}, m_i, \chi_i))(1 + \delta_2^h(\mathbf{x+r}, m_i, \chi_i))(1 + \delta_3^h(\mathbf{x'}, m_j, \chi_j))(1 + \delta_4^h(\mathbf{x'+r'}, m_j, \chi_j))\Big\rangle.
	\end{align}
	Since the factors of number density cancel out from the numerator and denominator, we will ignore them hereafter. To linear order,  the local halo overdensity is related to the local dark matter overdensity as, 
	\begin{align}
		\label{halo_dm_reln}
		\delta^h(\mathbf{x}, m, \chi)=b_1(m,\chi)\delta^{(1)}(\mathbf{x},\chi)\equiv b_1(m,\chi)\delta(\mathbf{x},\chi).
	\end{align}
	We start by simplifying the denominator of Eq.(\ref{4_point_defn}), i.e. $G_0$. Expanding Eq.(\ref{deno_final_avg}), and using Eq.(\ref{halo_dm_reln}) we get,
	\begin{align}
		&G_0=1+b^2_1(m_i,\chi_i)\Big\langle\delta_1(\mathbf{x}, \chi_i) \delta_2(\mathbf{x+r}, \chi_i)\Big\rangle+b^2_1(m_j,\chi_j)\Big\langle \delta_3(\mathbf{x'}, \chi_j)\delta_4(\mathbf{x'+r'}, \chi_j)\Big\rangle+\nonumber\\
		&b^2_1(m_i,\chi_i)b^2_1(m_j,\chi_j)\Big\langle\delta_1(\mathbf{x}, \chi_i) \delta_2(\mathbf{x+r},\chi_i) \delta_3(\mathbf{x'}, \chi_j)\delta_4(\mathbf{x'+r'},\chi_j)\Big\rangle+\mathrm{less\; significant \;terms}.
	\end{align}
	Evaluating the two point correlation function we get,
	\begin{align}
		\Big\langle\delta_1(\mathbf{x}, \chi_i) \delta_2(\mathbf{x+r}, \chi_i)\Big\rangle=\xi(r,\chi_i).
	\end{align}
	Therefore,
	\begin{align}
		&G_0=1+b^2_1(m_i,\chi_i)\xi(r,\chi_i)+b^2_1(m_j,\chi_j)\xi(r',\chi_j)+\nonumber\\
		&\hspace{2cm}b^2_1(m_i,\chi_i)b^2_1(m_j,\chi_j)\xi(r,\chi_i)\xi(r',\chi_j)+\mathrm{less\; significant \;terms}
	\end{align}
	Here we assume that since $r,r'\ll|x-x'|$,therefore, $\xi(|x-x'|,\chi)\ll\xi(r,\chi),\xi(r',\chi)$. Thus, the only contributing terms will be those involving correlation function of $r$ and $r'$. Therefore,
	\begin{align}
		&G_0=\left(1+b^2_1(m_i,\chi_i)\xi(r,\chi_i)\right)\left(1+b^2_1(m_j,\chi_j)\xi(r',\chi_j)\right)+\mathrm{less\; significant \;terms}.
	\end{align}
	Similarly we need to simplify the numerator in Eq.(\ref{4_point_defn}), i.e. $G_1$. Expanding Eq.(\ref{num_final_avg}), we get,
	\begin{align}
		G_1=&\Big\langle\Big(\mathbf{v_{1}}({\mathbf{x}})\cdot\mathbf{\hat{n}}\;\mathbf{v_{3}}({\mathbf{x'}})\cdot\mathbf{\hat{n}'}-\mathbf{v_{1}}({\mathbf{x}})\cdot\mathbf{\hat{n}}\;\mathbf{v_{4}}({\mathbf{x'+r'}})\cdot\mathbf{\hat{n}'}\nonumber\\
		&-\mathbf{v_{2}}({\mathbf{x+r}})\cdot\mathbf{\hat{n}}\;\mathbf{v_{3}}({\mathbf{x'}})\cdot\mathbf{\hat{n}'}+\mathbf{v_{2}}({\mathbf{x+r}})\cdot\mathbf{\hat{n}}\;\mathbf{v_{4}}({\mathbf{x'+r'}})\cdot\mathbf{\hat{n}'}\Big)\nonumber\\
		&\Big(1 + \delta_1^h(\mathbf{x}, m_i, \chi_i)\delta_2^h(\mathbf{x+r}, m_i, \chi_i)+\delta_1^h(\mathbf{x}, m_i, \chi_i)\delta_3^h(\mathbf{x'}, m_j, \chi_j)+\nonumber\\
		&\delta_1^h(\mathbf{x}, m_i, \chi_i)\delta_4^h(\mathbf{x'+r'}, m_j, \chi_j)+\delta_2^h(\mathbf{x+r}, m_i, \chi_i)\delta_3^h(\mathbf{x'}, m_j, \chi_j)+\nonumber\\
		&\delta_2^h(\mathbf{x+r}, m_i, \chi_i)\delta_4^h(\mathbf{x'+r'}, m_j, \chi_j)+\delta_3^h(\mathbf{x'}, m_j, \chi_j)\delta_4^h(\mathbf{x'+r'}, m_j, \chi_j)\Big)\Big\rangle\nonumber\\
		&+\mathrm{higher\; order \;terms}.
	\end{align} 
	There are 4 two-point function and 24 four-point functions to be evaluated and simplified. We are going to group these terms into 5 different group. One group will have all the two-point function and the rest 4 groups will have different set of four-point function. Thus we write $G_1$ as,
	\begin{align}
		G_1=\left( \mathcal{X}+\mathcal{A}+\mathcal{B}+\mathcal{C}+\mathcal{D}\right),
	\end{align}
	where
	\begin{align}
		\label{term_X}
		\mathcal{X}=&\Big\langle\mathbf{v_{1}}({\mathbf{x}})\cdot\mathbf{\hat{n}}\;\mathbf{v_{3}}({\mathbf{x'}})\cdot\mathbf{\hat{n}'}-\mathbf{v_{1}}({\mathbf{x}})\cdot\mathbf{\hat{n}}\;\mathbf{v_{4}}({\mathbf{x'+r'}})\cdot\mathbf{\hat{n}'}-\nonumber\\
		&\mathbf{v_{2}}({\mathbf{x+r}})\cdot\mathbf{\hat{n}}\;\mathbf{v_{3}}({\mathbf{x'}})\cdot\mathbf{\hat{n}'}+\mathbf{v_{2}}({\mathbf{x+r}})\cdot\mathbf{\hat{n}}\;\mathbf{v_{4}}({\mathbf{x'+r'}})\cdot\mathbf{\hat{n}'}\Big\rangle,
	\end{align}
	\begin{align}
		\label{term_A}
		&\mathcal{A}= \Big\langle\mathbf{v_{1}}({\mathbf{x}})\cdot\mathbf{\hat{n}}\;\mathbf{v_{3}}({\mathbf{x'}})\cdot\mathbf{\hat{n}'}\;\big(\delta_1^h(\mathbf{x}, m_i, \chi_i)\delta_2^h(\mathbf{x+r}, m_i, \chi_i)+\delta_1^h(\mathbf{x}, m_i, \chi_i)\delta_3^h(\mathbf{x'}, m_j, \chi_j)+\nonumber\\
		&\hspace{1cm}\delta_1^h(\mathbf{x}, m_i, \chi_i)\delta_4^h(\mathbf{x'+r'}, m_j, \chi_j)+\delta_2^h(\mathbf{x+r}, m_i, \chi_i)\delta_3^h(\mathbf{x'}, m_j, \chi_j)+\nonumber\\
		&\hspace{1cm}\delta_2^h(\mathbf{x+r}, m_i, \chi_i)\delta_4^h(\mathbf{x'+r'}, m_j, \chi_j)+\delta_3^h(\mathbf{x'}, m_j, \chi_j)\delta_4^h(\mathbf{x'+r'}, m_j, \chi_j)\big)\Big\rangle,
	\end{align}
	\begin{align}
		\label{term_B}
		&\mathcal{B}= -\Big\langle\mathbf{v_{1}}({\mathbf{x}})\cdot\mathbf{\hat{n}}\;\mathbf{v_{4}}({\mathbf{x'+r'}})\cdot\mathbf{\hat{n}'}\;\big(\delta_1^h(\mathbf{x}, m_i, \chi_i)\delta_2^h(\mathbf{x+r}, m_i, \chi_i)+\delta_1^h(\mathbf{x}, m_i, \chi_i)\delta_3^h(\mathbf{x'}, m_j, \chi_j)+\nonumber\\
		&\hspace{1cm}\delta_1^h(\mathbf{x}, m_i, \chi_i)\delta_4^h(\mathbf{x'+r'}, m_j, \chi_j)+\delta_2^h(\mathbf{x+r}, m_i, \chi_i)\delta_3^h(\mathbf{x'}, m_j, \chi_j)+\nonumber\\
		&\hspace{1cm}\delta_2^h(\mathbf{x+r}, m_i, \chi_i)\delta_4^h(\mathbf{x'+r'}, m_j, \chi_j)+\delta_3^h(\mathbf{x'}, m_j, \chi_j)\delta_4^h(\mathbf{x'+r'}, m_j, \chi_j)\big)\Big\rangle,
	\end{align}
	\begin{align}
		\label{term_C}
		&\mathcal{C}= -\Big\langle\mathbf{v_{2}}({\mathbf{x+r}})\cdot\mathbf{\hat{n}}\;\mathbf{v_{3}}({\mathbf{x'}})\cdot\mathbf{\hat{n}'}\;\big(\delta_1^h(\mathbf{x}, m_i, \chi_i)\delta_2^h(\mathbf{x+r}, m_i, \chi_i)+\delta_1^h(\mathbf{x}, m_i, \chi_i)\delta_3^h(\mathbf{x'}, m_j, \chi_j)+\nonumber\\
		&\hspace{1cm}\delta_1^h(\mathbf{x}, m_i, \chi_i)\delta_4^h(\mathbf{x'+r'}, m_j, \chi_j)+\delta_2^h(\mathbf{x+r}, m_i, \chi_i)\delta_3^h(\mathbf{x'}, m_j, \chi_j)+\nonumber\\
		&\hspace{1cm}\delta_2^h(\mathbf{x+r}, m_i, \chi_i)\delta_4^h(\mathbf{x'+r'}, m_j, \chi_j)+\delta_3^h(\mathbf{x'}, m_j, \chi_j)\delta_4^h(\mathbf{x'+r'}, m_j, \chi_j)\big)\Big\rangle,
	\end{align}
	and
	\begin{align}
		\label{term_D}
		&\mathcal{D}=\Big\langle\mathbf{v_{2}}({\mathbf{x+r}})\cdot\mathbf{\hat{n}}\;\mathbf{v_{4}}({\mathbf{x'+r'}})\cdot\mathbf{\hat{n}'}\;\big(\delta_1^h(\mathbf{x}, m_i, \chi_i)\delta_2^h(\mathbf{x+r}, m_i, \chi_i)+\delta_1^h(\mathbf{x}, m_i, \chi_i)\delta_3^h(\mathbf{x'}, m_j, \chi_j)+\nonumber\\
		&\hspace{1cm}\delta_1^h(\mathbf{x}, m_i, \chi_i)\delta_4^h(\mathbf{x'+r'}, m_j, \chi_j)+\delta_2^h(\mathbf{x+r}, m_i, \chi_i)\delta_3^h(\mathbf{x'}, m_j, \chi_j)+\nonumber\\
		&\hspace{1cm}\delta_2^h(\mathbf{x+r}, m_i, \chi_i)\delta_4^h(\mathbf{x'+r'}, m_j, \chi_j)+\delta_3^h(\mathbf{x'}, m_j, \chi_j)\delta_4^h(\mathbf{x'+r'}, m_j, \chi_j)\big)\Big\rangle.
	\end{align}
	We shall evaluate each term separately. We first transfer to Fourier space. the density fields and the velocity fields at first order in perturbation theory are given by,
	\begin{align}
		\label{delta1}
		&\delta^{(1)}(\mathbf{x},\chi)=D(\chi)\int\frac{d^3\mathbf{k}}{(2\pi)^3}\exp\left(i \mathbf{k}\cdot\mathbf{x}\right)\delta^{(1)}(\mathbf{k}),
	\end{align}
	\begin{align}
		\label{vel1}
		&\mathbf{v}^{(1)}(\mathbf{x},\chi)=D(\chi)\,[afH](\chi)\int\frac{d^3\mathbf{k}}{(2\pi)^3}\exp\left(i \mathbf{k}\cdot\mathbf{x}\right) \frac{i\mathbf{k}}{k^2}\delta^{(1)}(\mathbf{k}).
	\end{align}
	\subsection*{Simplification of term $\mathcal{A}$}
	The expression for $\mathcal{A}$ is given in Eq.(\ref{term_A}),
	\begin{align}
		\label{A_main_fourier}	
		\mathcal{A}&=i^2\,D(\chi_i)\,[afH](\chi_i)D(\chi_j)\,[afH](\chi_j)\int \frac{d^3\mathbf{k}d^3\mathbf{k'}d^3\mathbf{k_1}d^3\mathbf{k_1'}}{(2\pi)^{12}}\,e^{i \mathbf{k}\cdot\mathbf{x}}\,e^{i \mathbf{k'}\cdot\mathbf{x'}} \frac{\mathbf{k}\cdot\mathbf{\hat{n}}\,\mathbf{k'}\cdot\mathbf{\hat{n}'}}{k^2{k'}^2}\nonumber\\
		&\Big\langle \delta(\mathbf{k})\delta(\mathbf{k'})\delta(\mathbf{k_1})\delta(\mathbf{k_1'})  \Big\rangle \Big[ b^2_1(m_i,\chi_i)D^2(\chi_i)e^{i \mathbf{k_1}\cdot\mathbf{x}}\,e^{i \mathbf{k_1'}\cdot(\mathbf{x+r})}+ b_1(m_i,\chi_i)D(\chi_i)b_1(m_j,\chi_j)D(\chi_j)\nonumber\\
		&\Big(e^{i \mathbf{k_1}\cdot\mathbf{x}}\,e^{i \mathbf{k_1'}\cdot\mathbf{x'}}+e^{i \mathbf{k_1}\cdot\mathbf{x}}\,e^{i \mathbf{k_1'}\cdot(\mathbf{x'+r'})}+e^{i \mathbf{k_1}\cdot(\mathbf{x+r})}\,e^{i \mathbf{k_1'}\cdot\mathbf{x'}}+
		e^{i \mathbf{k_1}\cdot(\mathbf{x+r})}\,e^{i \mathbf{k_1'}\cdot(\mathbf{x'+r'})}
		\Big)\nonumber\\
		&\hspace{3cm} +b^2_1(m_j,\chi_j)D^2(\chi_j)e^{i \mathbf{k_1}\cdot\mathbf{x'}}\,e^{i \mathbf{k_1'}\cdot(\mathbf{x'+r'})}
		\Big].
	\end{align}
	Considering Gaussian field, the ensemble average can be written as,
	\begin{align}
		\label{4p_isserlis}
		&\Big\langle \delta(\mathbf{k})\delta(\mathbf{k'})\delta(\mathbf{k_1})\delta(\mathbf{k_1'})  \Big\rangle=(2\pi)^6 P(k)P(k_1) \delta_D(\mathbf{k}+\mathbf{k'})\delta_D(\mathbf{k_1}+\mathbf{k_1'})+\nonumber\\
		&(2\pi)^6 P(k)P(k') \delta_D(\mathbf{k}+\mathbf{k_1})\delta_D(\mathbf{k'}+\mathbf{k_1'})+(2\pi)^6 P(k)P(k') \delta_D(\mathbf{k}+\mathbf{k_1'})\delta_D(\mathbf{k'}+\mathbf{k_1}).
	\end{align}
	Therefore we can further split the expression in $\mathcal{A}$ into three different terms, each with one terms from the ensemble average. Thus we have,
	\begin{align}
		\mathcal{A}=\mathcal{A}_1+\mathcal{A}_2+\mathcal{A}_3
	\end{align} 
	We first simplify $\mathcal{A}_1$.
	\begin{align}
		\mathcal{A}_1&=i^2\,D(\chi_i)\,[afH](\chi_i)D(\chi_j)\,[afH](\chi_j)\int \frac{d^3\mathbf{k}d^3\mathbf{k'}d^3\mathbf{k_1}d^3\mathbf{k_1'}}{(2\pi)^{12}}\,e^{i \mathbf{k}\cdot\mathbf{x}}\,e^{i \mathbf{k'}\cdot\mathbf{x'}} \frac{\mathbf{k}\cdot\mathbf{\hat{n}}\,\mathbf{k'}\cdot\mathbf{\hat{n}'}}{k^2{k'}^2} (2\pi)^6 \nonumber\\
		& P(k)P(k_1)\delta_D(\mathbf{k}+\mathbf{k'})\delta_D(\mathbf{k_1}+\mathbf{k_1'})\bigg[ b^2_1(m_i,\chi_i)D^2(\chi_i)e^{i \mathbf{k_1}\cdot\mathbf{x}}\,e^{i \mathbf{k_1'}\cdot(\mathbf{x+r})}+ b_1(m_i,\chi_i)D(\chi_i)\nonumber\\
		&b_1(m_j,\chi_j)D(\chi_j)\Big(e^{i \mathbf{k_1}\cdot\mathbf{x}}\,e^{i \mathbf{k_1'}\cdot\mathbf{x'}}+e^{i \mathbf{k_1}\cdot\mathbf{x}}\,e^{i \mathbf{k_1'}\cdot(\mathbf{x'+r'})}+e^{i \mathbf{k_1}\cdot(\mathbf{x+r})}\,e^{i \mathbf{k_1'}\cdot\mathbf{x'}}+
		e^{i \mathbf{k_1}\cdot(\mathbf{x+r})}\,e^{i \mathbf{k_1'}\cdot(\mathbf{x'+r'})}
		\Big)\nonumber\\
		&\hspace{3cm} +b^2_1(m_j,\chi_j)D^2(\chi_j)e^{i \mathbf{k_1}\cdot\mathbf{x'}}\,e^{i \mathbf{k_1'}\cdot(\mathbf{x'+r'})}
		\bigg],
	\end{align}
	\begin{align}
		\mathcal{A}_1&=i^2\,D(\chi_i)\,[afH](\chi_i)D(\chi_j)\,[afH](\chi_j)\int \frac{d^3\mathbf{k}d^3\mathbf{k_1}}{(2\pi)^{6}}\,e^{i \mathbf{k}\cdot\mathbf{x}}\,e^{-i \mathbf{k}\cdot\mathbf{x'}} \frac{-\mathbf{k}\cdot\mathbf{\hat{n}}\,\mathbf{k}\cdot\mathbf{\hat{n}'}}{k^2{k}^2} P(k)P(k_1)\nonumber\\
		& \Big[ b^2_1(m_i,\chi_i)D^2(\chi_i)e^{i \mathbf{k_1}\cdot\mathbf{x}}\,e^{-i \mathbf{k_1}\cdot(\mathbf{x+r})}+ b_1(m_i,\chi_i)D(\chi_i)b_1(m_j,\chi_j)D(\chi_j)\nonumber\\
		&\Big(e^{i \mathbf{k_1}\cdot\mathbf{x}}\,e^{- \mathbf{k_1}\cdot\mathbf{x'}}+e^{i \mathbf{k_1}\cdot\mathbf{x}}\,e^{-i \mathbf{k_1}\cdot(\mathbf{x'+r'})}+e^{i \mathbf{k_1}\cdot(\mathbf{x+r})}\,e^{-i \mathbf{k_1}\cdot\mathbf{x'}}+
		e^{i \mathbf{k_1}\cdot(\mathbf{x+r})}\,e^{-i \mathbf{k_1}\cdot(\mathbf{x'+r'})}
		\Big)\nonumber\\
		&\hspace{3cm} +b^2_1(m_j,\chi_j)D^2(\chi_j)e^{i \mathbf{k_1}\cdot\mathbf{x'}}\,e^{i \mathbf{k_1'}\cdot(\mathbf{x'+r'})}
		\Big].
	\end{align}
	We observe the first and the last term in the expression of $\mathcal{A}_1$ only contributes to the $k=0$ mode, so we can ignore it as $P(k=0)=0$. Therefore we get,
	\begin{align}
		\mathcal{A}_1&=i^2\,D(\chi_i)\,[afH](\chi_i)D(\chi_j)\,[afH](\chi_j)\int \frac{d^3\mathbf{k}d^3\mathbf{k_1}}{(2\pi)^{6}}\, \frac{-\mathbf{k}\cdot\mathbf{\hat{n}}\,\mathbf{k}\cdot\mathbf{\hat{n}'}}{k^2{k}^2} P(k)P(k_1)\nonumber\\
		& e^{i(\mathbf{k+k_1})\cdot\mathbf{x}}\,e^{-i (\mathbf{k+k_1})\cdot\mathbf{x'}}\Big[  b_1(m_i,\chi_i)D(\chi_i)b_1(m_j,\chi_j)D(\chi_j)\nonumber\\
		&\Big(1+e^{-i \mathbf{k_1}\cdot\mathbf{r'}}+e^{i \mathbf{k_1}\cdot\mathbf{r}}+
		e^{i \mathbf{k_1}\cdot\mathbf{r}}\,e^{-i \mathbf{k_1}\cdot\mathbf{r'}}
		\Big)
		\Big].
	\end{align}
	To proceed further, we include the integral over $\mathbf{x}$ and $\mathbf{x'}$ which is present in Eq.(\ref{cov_start_appendix}). Therefore we have,
	\begin{align}
		&\int \frac{d^3\mathbf{x}d^3\mathbf{x}'}{V_s^2}\,\mathcal{A}_1 = i^2\,D(\chi_i)\,[afH](\chi_i)D(\chi_j)\,[afH](\chi_j)\int \frac{d^3\mathbf{k}d^3\mathbf{k_1}}{(2\pi)^{6}}\, \frac{-\mathbf{k}\cdot\mathbf{\hat{n}}\,\mathbf{k}\cdot\mathbf{\hat{n}'}}{k^2{k}^2} P(k)P(k_1)\nonumber\\
		&\int \frac{d^3\mathbf{x}d^3\mathbf{x}'}{V_s^2} e^{i(\mathbf{k+k_1})\cdot\mathbf{x}}\,e^{-i (\mathbf{k+k_1})\cdot\mathbf{x'}}\Big[  b_1(m_i,\chi_i)D(\chi_i)b_1(m_j,\chi_j)D(\chi_j)\nonumber\\
		&\Big(1+e^{-i \mathbf{k_1}\cdot\mathbf{r'}}+e^{i \mathbf{k_1}\cdot\mathbf{r}}+
		e^{i \mathbf{k_1}\cdot\mathbf{r}}\,e^{-i \mathbf{k_1}\cdot\mathbf{r'}}
		\Big)
		\Big],
	\end{align}
	\begin{align}
		&\int \frac{d^3\mathbf{x}d^3\mathbf{x}'}{V_s^2}\,\mathcal{A}_1 = i^2\,D(\chi_i)\,[afH](\chi_i)D(\chi_j)\,[afH](\chi_j)\int \frac{d^3\mathbf{k}}{(2\pi)^{3}}\, \frac{-\mathbf{k}\cdot\mathbf{\hat{n}}\,\mathbf{k}\cdot\mathbf{\hat{n}'}}{k^2{k}^2} P(k)P(k)\nonumber\\
		&\frac{1}{V_s} \Big[  b_1(m_i,\chi_i)D(\chi_i)b_1(m_j,\chi_j)D(\chi_j)\Big(1+e^{i \mathbf{k}\cdot\mathbf{r'}}+e^{-i \mathbf{k}\cdot\mathbf{r}}+
		e^{-i \mathbf{k}\cdot\mathbf{r}}\,e^{i \mathbf{k}\cdot\mathbf{r'}}
		\Big)
		\Big],\\\nonumber\\
		&\int \frac{d^3\mathbf{x}d^3\mathbf{x}'}{V_s^2}\,\mathcal{A}_1=i^2\,D^2(\chi_i)\,[afH](\chi_i)D^2(\chi_j)\,[afH](\chi_j)b_1(m_i,\chi_i)b_1(m_j,\chi_j)\nonumber\\
		&\int \frac{d^3\mathbf{k}}{(2\pi)^{3}}\, \frac{-\mathbf{k}\cdot\mathbf{\hat{n}}\,\mathbf{k}\cdot\mathbf{\hat{n}'}}{k^2{k}^2} P(k)P(k) \frac{1}{V_s} \Big[ 1+e^{i \mathbf{k}\cdot\mathbf{r'}}+e^{-i \mathbf{k}\cdot\mathbf{r}}+
		e^{-i \mathbf{k}\cdot\mathbf{r}}\,e^{i \mathbf{k}\cdot\mathbf{r'}}
		\Big].
	\end{align}
	Similarly, we can simplify $\mathcal{A}_2$ and $\mathcal{A}_3$.
	\begin{align}
		&\int\frac{d^3\mathbf{x}d^3\mathbf{x}'}{V_s^2}\,\mathcal{A}_2=i^2\,D(\chi_i)\,[afH](\chi_i)D(\chi_j)\,[afH](\chi_j)\int \frac{d^3\mathbf{k}d^3\mathbf{k'}}{(2\pi)^{6}}\,\frac{d^3\mathbf{x}d^3\mathbf{x}'}{V_s^2} \frac{\mathbf{k}\cdot\mathbf{\hat{n}}\,\mathbf{k'}\cdot\mathbf{\hat{n}'}}{k^2{k'}^2} P(k)P(k')\nonumber\\
		&e^{i \mathbf{k}\cdot\mathbf{x}}\,e^{i \mathbf{k'}\cdot\mathbf{x'}} 
		\Big[ b^2_1(m_i,\chi_i)D^2(\chi_i)e^{-i \mathbf{k}\cdot\mathbf{x}}\,e^{-i \mathbf{k'}\cdot(\mathbf{x+r})}+ b_1(m_i,\chi_i)D(\chi_i)b_1(m_j,\chi_j)D(\chi_j)\nonumber\\
		&\Big(e^{-i \mathbf{k}\cdot\mathbf{x}}\,e^{-i \mathbf{k'}\cdot\mathbf{x'}}+e^{-i \mathbf{k}\cdot\mathbf{x}}\,e^{-i \mathbf{k'}\cdot(\mathbf{x'+r'})}+e^{-i \mathbf{k}\cdot(\mathbf{x+r})}\,e^{-i \mathbf{k'}\cdot\mathbf{x'}}+
		e^{-i \mathbf{k}\cdot(\mathbf{x+r})}\,e^{-i \mathbf{k'}\cdot(\mathbf{x'+r'})}
		\Big)\nonumber\\
		&\hspace{3cm} +b^2_1(m_j,\chi_j)D^2(\chi_j)e^{-i \mathbf{k}\cdot\mathbf{x'}}\,e^{-i \mathbf{k'}\cdot(\mathbf{x'+r'})}
		\Big].
	\end{align}
	The non-zero terms, i.e. which doesn't only contribute to $k=0$ mode are,
	\begin{align}
		&\int\frac{d^3\mathbf{x}d^3\mathbf{x}'}{V_s^2}\,\mathcal{A}_2=i^2\,D(\chi_i)\,[afH](\chi_i)D(\chi_j)\,[afH](\chi_j)\int \frac{d^3\mathbf{k}d^3\mathbf{k'}}{(2\pi)^{6}}\,\frac{d^3\mathbf{x}d^3\mathbf{x}'}{V_s^2} \frac{\mathbf{k}\cdot\mathbf{\hat{n}}\,\mathbf{k'}\cdot\mathbf{\hat{n}'}}{k^2{k'}^2} P(k)P(k')\nonumber\\
		&
		\Big[  b_1(m_i,\chi_i)D(\chi_i)b_1(m_j,\chi_j)D(\chi_j)\Big(1+e^{-i \mathbf{k'}\cdot\mathbf{r'}}+e^{-i \mathbf{k}\cdot\mathbf{r}}+
		e^{-i \mathbf{k}\cdot\mathbf{r}}\,e^{-i \mathbf{k'}\cdot\mathbf{r'}}
		\Big)
		\Big].
	\end{align}
	Note that the exponential are independent of $\mathbf{x}$ and $\mathbf{x'}$. Integrating over the survey volume we get,
	\begin{align}
		&\int\frac{d^3\mathbf{x}d^3\mathbf{x}'}{V_s^2}\,\mathcal{A}_2=i^2\,D^2(\chi_i)\,[afH](\chi_i)D^2(\chi_j)\,[afH](\chi_j)b_1(m_i,\chi_i)b_1(m_j,\chi_j)\int \frac{d^3\mathbf{k}d^3\mathbf{k'}}{(2\pi)^{6}}\nonumber\\
		&\hspace{2cm} \frac{\mathbf{k}\cdot\mathbf{\hat{n}}\,\mathbf{k'}\cdot\mathbf{\hat{n}'}}{k^2{k'}^2} P(k)P(k')
		\Big[  \Big(1+e^{-i \mathbf{k'}\cdot\mathbf{r'}}+e^{-i \mathbf{k}\cdot\mathbf{r}}+
		e^{-i \mathbf{k}\cdot\mathbf{r}}\,e^{-i \mathbf{k'}\cdot\mathbf{r'}}
		\Big)
		\Big].
	\end{align}
	Similarly, we simplify $\mathcal{A}_3$.
	\begin{align}
		&\int\frac{d^3\mathbf{x}d^3\mathbf{x}'}{V_s^2}\,\mathcal{A}_3=i^2\,D(\chi_i)\,[afH](\chi_i)D(\chi_j)\,[afH](\chi_j)\int \frac{d^3\mathbf{k}d^3\mathbf{k'}}{(2\pi)^{6}}\,\frac{d^3\mathbf{x}d^3\mathbf{x}'}{V_s^2} \frac{\mathbf{k}\cdot\mathbf{\hat{n}}\,\mathbf{k'}\cdot\mathbf{\hat{n}'}}{k^2{k'}^2} P(k)P(k')\nonumber\\
		&e^{i \mathbf{k}\cdot\mathbf{x}}\,e^{i \mathbf{k'}\cdot\mathbf{x'}} 
		\Big[ b^2_1(m_i,\chi_i)D^2(\chi_i)e^{-i \mathbf{k'}\cdot\mathbf{x}}\,e^{-i \mathbf{k}\cdot(\mathbf{x+r})}+ b_1(m_i,\chi_i)D(\chi_i)b_1(m_j,\chi_j)D(\chi_j)\nonumber\\
		&\Big(e^{-i \mathbf{k'}\cdot\mathbf{x}}\,e^{-i \mathbf{k}\cdot\mathbf{x'}}+e^{-i \mathbf{k'}\cdot\mathbf{x}}\,e^{-i \mathbf{k}\cdot(\mathbf{x'+r'})}+e^{-i \mathbf{k'}\cdot(\mathbf{x+r})}\,e^{-i \mathbf{k}\cdot\mathbf{x'}}+
		e^{-i \mathbf{k'}\cdot(\mathbf{x+r})}\,e^{-i \mathbf{k}\cdot(\mathbf{x'+r'})}
		\Big)\nonumber\\
		&\hspace{3cm} +b^2_1(m_j,\chi_j)D^2(\chi_j)e^{-i \mathbf{k'}\cdot\mathbf{x'}}\,e^{-i \mathbf{k}\cdot(\mathbf{x'+r'})}
		\Big].
	\end{align}
	The non-zero terms are,
	\begin{align}
		&\int\frac{d^3\mathbf{x}d^3\mathbf{x}'}{V_s^2}\,\mathcal{A}_3=i^2\,D(\chi_i)\,[afH](\chi_i)D(\chi_j)\,[afH](\chi_j)\int \frac{d^3\mathbf{k}d^3\mathbf{k'}}{(2\pi)^{6}}\,\frac{\mathbf{k}\cdot\mathbf{\hat{n}}\,\mathbf{k'}\cdot\mathbf{\hat{n}'}}{k^2{k'}^2} P(k)P(k')\nonumber\\
		&
		\Big[ \int \frac{d^3\mathbf{x}d^3\mathbf{x}'}{V_s^2}  e^{i(\mathbf{k}-\mathbf{k'})\cdot\mathbf{x}}e^{-i(\mathbf{k}-\mathbf{k'})\cdot\mathbf{x'}}b_1(m_i,\chi_i)D(\chi_i)b_1(m_j,\chi_j)D(\chi_j)\nonumber\\
		&
		\Big(1+e^{-i \mathbf{k'}\cdot\mathbf{r'}}+e^{-i \mathbf{k}\cdot\mathbf{r}}+
		e^{-i \mathbf{k}\cdot\mathbf{r}}\,e^{-i \mathbf{k'}\cdot\mathbf{r'}}
		\Big)
		\Big],
	\end{align}
	\begin{align}
		&\int\frac{d^3\mathbf{x}d^3\mathbf{x}'}{V_s^2}\,\mathcal{A}_3=i^2\,D^2(\chi_i)\,[afH](\chi_i)D^2(\chi_j)\,[afH](\chi_j)b_1(m_i,\chi_i)b_1(m_j,\chi_j)\nonumber\\
		& \int \frac{d^3\mathbf{k}}{(2\pi)^{3}}\,\frac{\mathbf{k}\cdot\mathbf{\hat{n}}\,\mathbf{k}\cdot\mathbf{\hat{n}'}}{k^2{k}^2} P(k)P(k)\frac{1}{V_s}  
		\Big[
		\Big(1+e^{-i \mathbf{k}\cdot\mathbf{r'}}+e^{-i \mathbf{k}\cdot\mathbf{r}}+
		e^{-i \mathbf{k}\cdot\mathbf{r}}\,e^{-i \mathbf{k}\cdot\mathbf{r'}}
		\Big)
		\Big].
	\end{align}
	In the same way we can simplify the expressions in $\mathcal{B},\mathcal{C} $ and $\mathcal{D}$. 
	\subsection*{Simplification of term $\mathcal{B}$}
	Transforming the expression in $\mathcal{B}$, i.e. Eq.(\ref{term_B}) in the Fourier space we have,
	\begin{align}
		\label{B_main_fourier}
		\mathcal{B}&=-i^2\,D(\chi_i)\,[afH](\chi_i)D(\chi_j)\,[afH](\chi_j)\int \frac{d^3\mathbf{k}d^3\mathbf{k'}d^3\mathbf{k_1}d^3\mathbf{k_1'}}{(2\pi)^{12}}\,e^{i \mathbf{k}\cdot\mathbf{x}}\,e^{i \mathbf{k'}\cdot(\mathbf{x'+r'})} \frac{\mathbf{k}\cdot\mathbf{\hat{n}}\,\mathbf{k'}\cdot\mathbf{\hat{n}'}}{k^2{k'}^2}\nonumber\\
		&\Big\langle \delta(\mathbf{k})\delta(\mathbf{k'})\delta(\mathbf{k_1})\delta(\mathbf{k_1'})  \Big\rangle \Big[ b^2_1(m_i,\chi_i)D^2(\chi_i)e^{i \mathbf{k_1}\cdot\mathbf{x}}\,e^{i \mathbf{k_1'}\cdot(\mathbf{x+r})}+ b_1(m_i,\chi_i)D(\chi_i)b_1(m_j,\chi_j)D(\chi_j)\nonumber\\
		&\Big(e^{i \mathbf{k_1}\cdot\mathbf{x}}\,e^{i \mathbf{k_1'}\cdot\mathbf{x'}}+e^{i \mathbf{k_1}\cdot\mathbf{x}}\,e^{i \mathbf{k_1'}\cdot(\mathbf{x'+r'})}+e^{i \mathbf{k_1}\cdot(\mathbf{x+r})}\,e^{i \mathbf{k_1'}\cdot\mathbf{x'}}+
		e^{i \mathbf{k_1}\cdot(\mathbf{x+r})}\,e^{i \mathbf{k_1'}\cdot(\mathbf{x'+r'})}
		\Big)\nonumber\\
		&\hspace{3cm} +b^2_1(m_j,\chi_j)D^2(\chi_j)e^{i \mathbf{k_1}\cdot\mathbf{x'}}\,e^{i \mathbf{k_1'}\cdot(\mathbf{x'+r'})}
		\Big].
	\end{align}
	The only difference from the expression of $\mathcal{A}$ in Eq.(\ref{A_main_fourier}) is a factor of $e^{i \mathbf{k'}\cdot\mathbf{r'}}$.
	Using Eq.(\ref{4p_isserlis}), we can again break $\mathcal{B}$ into 3 different parts, $\mathcal{B}=\mathcal{B}_1+\mathcal{B}_2+\mathcal{B}_3$. Therefore we get,
	\begin{align}
		&\int \frac{d^3\mathbf{x}d^3\mathbf{x}'}{V_s^2}\,\mathcal{B}_1=-i^2\,D^2(\chi_i)\,[afH](\chi_i)D^2(\chi_j)\,[afH](\chi_j)b_1(m_i,\chi_i)b_1(m_j,\chi_j)\nonumber\\
		&\int \frac{d^3\mathbf{k}}{(2\pi)^{3}}\, \frac{-\mathbf{k}\cdot\mathbf{\hat{n}}\,\mathbf{k}\cdot\mathbf{\hat{n}'}}{k^2{k}^2} P(k)\,P(k) \,\frac{1}{V_s}\,
		e^{-i \mathbf{k}\cdot\mathbf{r'}} \Big[ 1+e^{i \mathbf{k}\cdot\mathbf{r'}}+e^{-i \mathbf{k}\cdot\mathbf{r}}+
		e^{-i \mathbf{k}\cdot\mathbf{r}}\,e^{i \mathbf{k}\cdot\mathbf{r'}}
		\Big],	
	\end{align}
	\begin{align}
		&\int\frac{d^3\mathbf{x}d^3\mathbf{x}'}{V_s^2}\,\mathcal{B}_2=-i^2\,D^2(\chi_i)\,[afH](\chi_i)D^2(\chi_j)\,[afH](\chi_j)b_1(m_i,\chi_i)b_1(m_j,\chi_j)\int \frac{d^3\mathbf{k}d^3\mathbf{k'}}{(2\pi)^{6}}\nonumber\\
		&\hspace{2cm} \frac{\mathbf{k}\cdot\mathbf{\hat{n}}\,\mathbf{k'}\cdot\mathbf{\hat{n}'}}{k^2{k'}^2} P(k)\,P(k')\,
		e^{i \mathbf{k'}\cdot\mathbf{r'}}	\Big[  \Big(1+e^{-i \mathbf{k'}\cdot\mathbf{r'}}+e^{-i \mathbf{k}\cdot\mathbf{r}}+
		e^{-i \mathbf{k}\cdot\mathbf{r}}\,e^{-i \mathbf{k'}\cdot\mathbf{r'}}
		\Big)
		\Big],
	\end{align}
	and 
	\begin{align}
		&\int\frac{d^3\mathbf{x}d^3\mathbf{x}'}{V_s^2}\,\mathcal{B}_3=-i^2\,D^2(\chi_i)\,[afH](\chi_i)D^2(\chi_j)\,[afH](\chi_j)b_1(m_i,\chi_i)b_1(m_j,\chi_j)\nonumber\\
		& \int \frac{d^3\mathbf{k}}{(2\pi)^{3}}\,\frac{\mathbf{k}\cdot\mathbf{\hat{n}}\,\mathbf{k}\cdot\mathbf{\hat{n}'}}{k^2{k}^2} P(k)\,P(k)\,\frac{1}{V_s}  \,
		e^{i \mathbf{k}\cdot\mathbf{r'}}	\Big[ \Big(1+e^{-i \mathbf{k}\cdot\mathbf{r'}}+e^{-i \mathbf{k}\cdot\mathbf{r}}+
		e^{-i \mathbf{k}\cdot\mathbf{r}}\,e^{-i \mathbf{k}\cdot\mathbf{r'}}
		\Big)
		\Big].
	\end{align}
	\subsection*{Simplification of term $\mathcal{C}$}
	From Eq.(\ref{term_C}), we have,
	\begin{align}
		\label{C_main_fourier}
		\mathcal{C}&=-i^2\,D(\chi_i)\,[afH](\chi_i)D(\chi_j)\,[afH](\chi_j)\int \frac{d^3\mathbf{k}d^3\mathbf{k'}d^3\mathbf{k_1}d^3\mathbf{k_1'}}{(2\pi)^{12}}\,e^{i \mathbf{k}\cdot(\mathbf{x+r})}\,e^{i \mathbf{k'}\cdot\mathbf{x'}} \frac{\mathbf{k}\cdot\mathbf{\hat{n}}\,\mathbf{k'}\cdot\mathbf{\hat{n}'}}{k^2{k'}^2}\nonumber\\
		&\Big\langle \delta(\mathbf{k})\delta(\mathbf{k'})\delta(\mathbf{k_1})\delta(\mathbf{k_1'})  \Big\rangle \Big[ b^2_1(m_i,\chi_i)D^2(\chi_i)e^{i \mathbf{k_1}\cdot\mathbf{x}}\,e^{i \mathbf{k_1'}\cdot(\mathbf{x+r})}+ b_1(m_i,\chi_i)D(\chi_i)b_1(m_j,\chi_j)D(\chi_j)\nonumber\\
		&\Big(e^{i \mathbf{k_1}\cdot\mathbf{x}}\,e^{i \mathbf{k_1'}\cdot\mathbf{x'}}+e^{i \mathbf{k_1}\cdot\mathbf{x}}\,e^{i \mathbf{k_1'}\cdot(\mathbf{x'+r'})}+e^{i \mathbf{k_1}\cdot(\mathbf{x+r})}\,e^{i \mathbf{k_1'}\cdot\mathbf{x'}}+
		e^{i \mathbf{k_1}\cdot(\mathbf{x+r})}\,e^{i \mathbf{k_1'}\cdot(\mathbf{x'+r'})}
		\Big)\nonumber\\
		&\hspace{3cm} +b^2_1(m_j,\chi_j)D^2(\chi_j)e^{i \mathbf{k_1}\cdot\mathbf{x'}}\,e^{i \mathbf{k_1'}\cdot(\mathbf{x'+r'})}
		\Big].
	\end{align}
	Here the difference is a factor of $e^{i \mathbf{k}\cdot\mathbf{r}}$ in the beginning.
	As before we split $\mathcal{C}$ into three parts. Thus we have, 
	\begin{align}
		&\int \frac{d^3\mathbf{x}d^3\mathbf{x}'}{V_s^2}\,\mathcal{C}_1=-i^2\,D^2(\chi_i)\,[afH](\chi_i)D^2(\chi_j)\,[afH](\chi_j)b_1(m_i,\chi_i)b_1(m_j,\chi_j)\nonumber\\
		&\int \frac{d^3\mathbf{k}}{(2\pi)^{3}}\, \frac{-\mathbf{k}\cdot\mathbf{\hat{n}}\,\mathbf{k}\cdot\mathbf{\hat{n}'}}{k^2{k}^2} P(k)\,P(k) \,\frac{1}{V_s}\,
		e^{i \mathbf{k}\cdot\mathbf{r}} \Big[ 1+e^{i \mathbf{k}\cdot\mathbf{r'}}+e^{-i \mathbf{k}\cdot\mathbf{r}}+
		e^{-i \mathbf{k}\cdot\mathbf{r}}\,e^{i \mathbf{k}\cdot\mathbf{r'}}
		\Big],	
	\end{align}
	\begin{align}
		&\int\frac{d^3\mathbf{x}d^3\mathbf{x}'}{V_s^2}\,\mathcal{C}_2=-i^2\,D^2(\chi_i)\,[afH](\chi_i)D^2(\chi_j)\,[afH](\chi_j)b_1(m_i,\chi_i)b_1(m_j,\chi_j)\int \frac{d^3\mathbf{k}d^3\mathbf{k'}}{(2\pi)^{6}}\nonumber\\
		&\hspace{2cm} \frac{\mathbf{k}\cdot\mathbf{\hat{n}}\,\mathbf{k'}\cdot\mathbf{\hat{n}'}}{k^2{k'}^2} P(k)\,P(k')\,
		e^{i \mathbf{k}\cdot\mathbf{r}}	\Big[  \Big(1+e^{-i \mathbf{k'}\cdot\mathbf{r'}}+e^{-i \mathbf{k}\cdot\mathbf{r}}+
		e^{-i \mathbf{k}\cdot\mathbf{r}}\,e^{-i \mathbf{k'}\cdot\mathbf{r'}}
		\Big)
		\Big],
	\end{align}
	and 
	\begin{align}
		&\int\frac{d^3\mathbf{x}d^3\mathbf{x}'}{V_s^2}\,\mathcal{C}_3=-i^2\,D^2(\chi_i)\,[afH](\chi_i)D^2(\chi_j)\,[afH](\chi_j)b_1(m_i,\chi_i)b_1(m_j,\chi_j)\nonumber\\
		& \int \frac{d^3\mathbf{k}}{(2\pi)^{3}}\,\frac{\mathbf{k}\cdot\mathbf{\hat{n}}\,\mathbf{k}\cdot\mathbf{\hat{n}'}}{k^2{k}^2} P(k)\,P(k)\,\frac{1}{V_s}  \,
		e^{i \mathbf{k}\cdot\mathbf{r}}	\Big[ \Big(1+e^{-i \mathbf{k}\cdot\mathbf{r'}}+e^{-i \mathbf{k}\cdot\mathbf{r}}+
		e^{-i \mathbf{k}\cdot\mathbf{r}}\,e^{-i \mathbf{k}\cdot\mathbf{r'}}
		\Big)
		\Big],
	\end{align}
	\subsection*{Simplification of term $\mathcal{D}$}
	Similarly, from Eq.(\ref{term_D}) we have,
	\begin{align}
		\label{D_main_fourier}
		\mathcal{D}&=i^2\,D(\chi_i)\,[afH](\chi_i)D(\chi_j)\,[afH](\chi_j)\int \frac{d^3\mathbf{k}d^3\mathbf{k'}d^3\mathbf{k_1}d^3\mathbf{k_1'}}{(2\pi)^{12}}\,e^{i \mathbf{k}\cdot(\mathbf{x+r})}\,e^{i \mathbf{k'}\cdot(\mathbf{x'+r'})} \frac{\mathbf{k}\cdot\mathbf{\hat{n}}\,\mathbf{k'}\cdot\mathbf{\hat{n}'}}{k^2{k'}^2}\nonumber\\
		&\Big\langle \delta(\mathbf{k})\delta(\mathbf{k'})\delta(\mathbf{k_1})\delta(\mathbf{k_1'})  \Big\rangle \Big[ b^2_1(m_i,\chi_i)D^2(\chi_i)e^{i \mathbf{k_1}\cdot\mathbf{x}}\,e^{i \mathbf{k_1'}\cdot(\mathbf{x+r})}+ b_1(m_i,\chi_i)D(\chi_i)b_1(m_j,\chi_j)D(\chi_j)\nonumber\\
		&\Big(e^{i \mathbf{k_1}\cdot\mathbf{x}}\,e^{i \mathbf{k_1'}\cdot\mathbf{x'}}+e^{i \mathbf{k_1}\cdot\mathbf{x}}\,e^{i \mathbf{k_1'}\cdot(\mathbf{x'+r'})}+e^{i \mathbf{k_1}\cdot(\mathbf{x+r})}\,e^{i \mathbf{k_1'}\cdot\mathbf{x'}}+
		e^{i \mathbf{k_1}\cdot(\mathbf{x+r})}\,e^{i \mathbf{k_1'}\cdot(\mathbf{x'+r'})}
		\Big)\nonumber\\
		&\hspace{3cm} +b^2_1(m_j,\chi_j)D^2(\chi_j)e^{i \mathbf{k_1}\cdot\mathbf{x'}}\,e^{i \mathbf{k_1'}\cdot(\mathbf{x'+r'})}
		\Big].
	\end{align}
	Here the difference is a factor of $e^{i \mathbf{k}\cdot\mathbf{r}}e^{i \mathbf{k'}\cdot\mathbf{r'}}$ in the beginning compared to Eq.(\ref{A_main_fourier}).
	As before we split $\mathcal{D}$ into three parts. Thus we have, 
	\begin{align}
		&\int \frac{d^3\mathbf{x}d^3\mathbf{x}'}{V_s^2}\,\mathcal{D}_1=i^2\,D^2(\chi_i)\,[afH](\chi_i)D^2(\chi_j)\,[afH](\chi_j)b_1(m_i,\chi_i)b_1(m_j,\chi_j)\nonumber\\
		&\int \frac{d^3\mathbf{k}}{(2\pi)^{3}}\, \frac{-\mathbf{k}\cdot\mathbf{\hat{n}}\,\mathbf{k}\cdot\mathbf{\hat{n}'}}{k^2{k}^2} P(k)\,P(k) \,\frac{1}{V_s}\,
		e^{i \mathbf{k}\cdot\mathbf{r}}\,	e^{-i \mathbf{k}\cdot\mathbf{r'}} \Big[ 1+e^{i \mathbf{k}\cdot\mathbf{r'}}+e^{-i \mathbf{k}\cdot\mathbf{r}}+
		e^{-i \mathbf{k}\cdot\mathbf{r}}\,e^{i \mathbf{k}\cdot\mathbf{r'}}
		\Big]	,
	\end{align}
	\begin{align}
		&\int\frac{d^3\mathbf{x}d^3\mathbf{x}'}{V_s^2}\,\mathcal{D}_2=i^2\,D^2(\chi_i)\,[afH](\chi_i)D^2(\chi_j)\,[afH](\chi_j)b_1(m_i,\chi_i)b_1(m_j,\chi_j)\int \frac{d^3\mathbf{k}d^3\mathbf{k'}}{(2\pi)^{6}}\nonumber\\
		&\hspace{2cm} \frac{\mathbf{k}\cdot\mathbf{\hat{n}}\,\mathbf{k'}\cdot\mathbf{\hat{n}'}}{k^2{k'}^2} P(k)\,P(k')\,
		e^{i \mathbf{k}\cdot\mathbf{r}}	\,e^{i \mathbf{k'}\cdot\mathbf{r'}}\Big[  \Big(1+e^{-i \mathbf{k'}\cdot\mathbf{r'}}+e^{-i \mathbf{k}\cdot\mathbf{r}}+
		e^{-i \mathbf{k}\cdot\mathbf{r}}\,e^{-i \mathbf{k'}\cdot\mathbf{r'}}
		\Big)
		\Big],
	\end{align}
	and
	\begin{align}
		&\int\frac{d^3\mathbf{x}d^3\mathbf{x}'}{V_s^2}\,\mathcal{D}_3=i^2\,D^2(\chi_i)\,[afH](\chi_i)D^2(\chi_j)\,[afH](\chi_j)b_1(m_i,\chi_i)b_1(m_j,\chi_j)\nonumber\\
		& \int \frac{d^3\mathbf{k}}{(2\pi)^{3}}\,\frac{\mathbf{k}\cdot\mathbf{\hat{n}}\,\mathbf{k}\cdot\mathbf{\hat{n}'}}{k^2{k}^2} P(k)\,P(k)\,\frac{1}{V_s}  \,
		e^{i \mathbf{k}\cdot\mathbf{r}}	\,	e^{i \mathbf{k}\cdot\mathbf{r'}}	\Big[ \Big(1+e^{-i \mathbf{k}\cdot\mathbf{r'}}+e^{-i \mathbf{k}\cdot\mathbf{r}}+
		e^{-i \mathbf{k}\cdot\mathbf{r}}\,e^{-i \mathbf{k}\cdot\mathbf{r'}}
		\Big)
		\Big].
	\end{align}
	\subsection*{Addition of simplified terms, $\mathcal{A}_1,\mathcal{B}_1,\mathcal{C}_1$ and $\mathcal{D}_1$}
	We can now add the terms in a systematic way. We first add $\mathcal{A}_1,\mathcal{B}_1,\mathcal{C}_1$ and $\mathcal{D}_1$. Thus we get,
	\begin{align}
		&\int\frac{d^3\mathbf{x}d^3\mathbf{x}'}{V_s^2}\left(\mathcal{A}_1+\mathcal{B}_1+\mathcal{C}_1+\mathcal{D}_1 \right)= i^2\,D^2(\chi_i)\,[afH](\chi_i)D^2(\chi_j)\,[afH](\chi_j)b_1(m_i,\chi_i)b_1(m_j,\chi_j)\nonumber\\
		&\int \frac{d^3\mathbf{k}}{(2\pi)^{3}}\, \frac{-\mathbf{k}\cdot\mathbf{\hat{n}}\,\mathbf{k}\cdot\mathbf{\hat{n}'}}{k^2{k}^2} P(k)\,P(k) \,\frac{1}{V_s}\,
		\Big[ 1+e^{i \mathbf{k}\cdot\mathbf{r'}}+e^{-i \mathbf{k}\cdot\mathbf{r}}+
		e^{-i \mathbf{k}\cdot\mathbf{r}}\,e^{i \mathbf{k}\cdot\mathbf{r'}}
		\Big]	\nonumber\\
		&\hspace{8cm}\Big[1-e^{-i \mathbf{k}\cdot\mathbf{r'}}-e^{i \mathbf{k}\cdot\mathbf{r}}+e^{i \mathbf{k}\cdot\mathbf{r}}\,	e^{-i \mathbf{k}\cdot\mathbf{r'}}\Big],
	\end{align}
	\begin{align}
		&\int\frac{d^3\mathbf{x}d^3\mathbf{x}'}{V_s^2}\left(\mathcal{A}_1+B_1+\mathcal{C}_1+\mathcal{D}_1 \right)= i^2\,D^2(\chi_i)\,[afH](\chi_i)D^2(\chi_j)\,[afH](\chi_j)b_1(m_i,\chi_i)b_1(m_j,\chi_j)\nonumber\\
		&\int \frac{d^3\mathbf{k}}{(2\pi)^{3}}\, \frac{\mathbf{k}\cdot\mathbf{\hat{n}}\,\mathbf{k}\cdot\mathbf{\hat{n}'}}{k^2{k}^2} P(k)\,P(k) \,\frac{1}{V_s}\,
		\Big[ e^{i \mathbf{k}\cdot\mathbf{r}}\,e^{i \mathbf{k}\cdot\mathbf{r'}}+e^{-i \mathbf{k}\cdot\mathbf{r}}\,e^{-i \mathbf{k}\cdot\mathbf{r'}}-e^{i \mathbf{k}\cdot\mathbf{r}}e^{-i \mathbf{k}\cdot\mathbf{r'}}-e^{-i \mathbf{k}\cdot\mathbf{r}}e^{i \mathbf{k}\cdot\mathbf{r'}}
		\Big].	
	\end{align}
	In the $2^{nd}$ and the $4^{th}$ term we make a transformation $\mathbf{k}\longrightarrow-\mathbf{k}$ without changing the integral result. This way we can rewrite the expression as,
	\begin{align}
		\label{1st_term_sum}
		&\int\frac{d^3\mathbf{x}d^3\mathbf{x}'}{V_s^2}\left(\mathcal{A}_1+\mathcal{B}_1+\mathcal{C}_1+\mathcal{D}_1 \right)=2\, i^2\,D^2(\chi_i)\,[afH](\chi_i)D^2(\chi_j)\,[afH](\chi_j)b_1(m_i,\chi_i)b_1(m_j,\chi_j)\nonumber\\
		&\int \frac{d^3\mathbf{k}}{(2\pi)^{3}}\, \frac{\mathbf{k}\cdot\mathbf{\hat{n}}\,\mathbf{k}\cdot\mathbf{\hat{n}'}}{k^2{k}^2} P(k)\,P(k) \,\frac{1}{V_s}\,
		\Big[ e^{i \mathbf{k}\cdot(\mathbf{r}+\mathbf{r'})}-e^{i \mathbf{k}\cdot(\mathbf{r}-\mathbf{r'})}
		\Big].	
	\end{align}
	\subsection*{Addition of simplified terms, $\mathcal{A}_3,\mathcal{B}_3,\mathcal{C}_3$ and $\mathcal{D}_3$}
	Similarly, we can add  $\mathcal{A}_3,\mathcal{B}_3,\mathcal{C}_3$ and $\mathcal{D}_3$
	\begin{align}
		&\int\frac{d^3\mathbf{x}d^3\mathbf{x}'}{V_s^2}\left(\mathcal{A}_3+\mathcal{B}_3+\mathcal{C}_3+\mathcal{D}_3 \right)= i^2\,D^2(\chi_i)\,[afH](\chi_i)D^2(\chi_j)\,[afH](\chi_j)b_1(m_i,\chi_i)b_1(m_j,\chi_j)\nonumber\\
		&\int \frac{d^3\mathbf{k}}{(2\pi)^{3}}\, \frac{\mathbf{k}\cdot\mathbf{\hat{n}}\,\mathbf{k}\cdot\mathbf{\hat{n}'}}{k^2{k}^2} P(k)\,P(k) \,\frac{1}{V_s}\,
		\Big[ 1+e^{-i \mathbf{k}\cdot\mathbf{r'}}+e^{-i \mathbf{k}\cdot\mathbf{r}}+
		e^{-i \mathbf{k}\cdot\mathbf{r}}\,e^{-i \mathbf{k}\cdot\mathbf{r'}}
		\Big]	\nonumber\\
		&\hspace{8cm}\Big[1-e^{i \mathbf{k}\cdot\mathbf{r'}}-e^{i \mathbf{k}\cdot\mathbf{r}}+e^{i \mathbf{k}\cdot\mathbf{r}}\,	e^{i \mathbf{k}\cdot\mathbf{r'}}\Big].
	\end{align}
	This term also simplifies to the same expression as in Eq.(\ref{1st_term_sum}).
	\begin{align}
		\label{3rd_term_sum}
		&\int\frac{d^3\mathbf{x}d^3\mathbf{x}'}{V_s^2}\left(\mathcal{A}_3+\mathcal{B}_3+\mathcal{C}_3+\mathcal{D}_3 \right)=2\, i^2\,D^2(\chi_i)\,[afH](\chi_i)D^2(\chi_j)\,[afH](\chi_j)b_1(m_i,\chi_i)b_1(m_j,\chi_j)\nonumber\\
		&\int \frac{d^3\mathbf{k}}{(2\pi)^{3}}\, \frac{\mathbf{k}\cdot\mathbf{\hat{n}}\,\mathbf{k}\cdot\mathbf{\hat{n}'}}{k^2{k}^2} P(k)\,P(k) \,\frac{1}{V_s}\,
		\Big[ e^{i \mathbf{k}\cdot(\mathbf{r}+\mathbf{r'})}-e^{i \mathbf{k}\cdot(\mathbf{r}-\mathbf{r'})}
		\Big].	
	\end{align}
	The expression in Eq.(\ref{3rd_term_sum}) is exactly the same as the expression in Eq.(\ref{1st_term_sum}).
	\subsection*{Addition of simplified terms, $\mathcal{A}_2,\mathcal{B}_2,\mathcal{C}_2$ and $\mathcal{D}_2$}
	The summation of $\mathcal{A}_2,\mathcal{B}_2,\mathcal{C}_2$ and $\mathcal{D}_2$ yields a different result from the previous two sums.
	\begin{align}
		&\int\frac{d^3\mathbf{x}d^3\mathbf{x}'}{V_s^2}\left(\mathcal{A}_2+\mathcal{B}_2+\mathcal{C}_2+\mathcal{D}_2 \right)= i^2\,D^2(\chi_i)\,[afH](\chi_i)D^2(\chi_j)\,[afH](\chi_j)b_1(m_i,\chi_i)b_1(m_j,\chi_j)\nonumber\\
		&\int \frac{d^3\mathbf{k}}{(2\pi)^{3}}\,\frac{d^3\mathbf{k'}}{(2\pi)^{3}}\, \frac{\mathbf{k}\cdot\mathbf{\hat{n}}\,\mathbf{k'}\cdot\mathbf{\hat{n}'}}{k^2{k'}^2} P(k)\,P(k')\,
		\Big[ 1+e^{-i \mathbf{k'}\cdot\mathbf{r'}}+e^{-i \mathbf{k}\cdot\mathbf{r}}+
		e^{-i \mathbf{k'}\cdot\mathbf{r'}}\,e^{-i \mathbf{k}\cdot\mathbf{r}}
		\Big]	\nonumber\\
		&\hspace{8cm}\Big[1-e^{i \mathbf{k'}\cdot\mathbf{r'}}-e^{i \mathbf{k}\cdot\mathbf{r}}+e^{i \mathbf{k}\cdot\mathbf{r}}\,	e^{i \mathbf{k'}\cdot\mathbf{r'}}\Big],
	\end{align}
	\begin{align}
		&\int\frac{d^3\mathbf{x}d^3\mathbf{x}'}{V_s^2}\left(\mathcal{A}_2+\mathcal{B}_2+\mathcal{C}_2+\mathcal{D}_2 \right)= i^2\,D^2(\chi_i)\,[afH](\chi_i)D^2(\chi_j)\,[afH](\chi_j)b_1(m_i,\chi_i)b_1(m_j,\chi_j)\nonumber\\
		&\int \frac{d^3\mathbf{k}}{(2\pi)^{3}}\,\frac{d^3\mathbf{k'}}{(2\pi)^{3}}\, \frac{\mathbf{k}\cdot\mathbf{\hat{n}}\,\mathbf{k'}\cdot\mathbf{\hat{n}'}}{k^2{k'}^2} P(k)\,P(k')\,
		\Big[ e^{i \mathbf{k}\cdot\mathbf{r}}\,e^{i \mathbf{k'}\cdot\mathbf{r'}}+e^{-i \mathbf{k'}\cdot\mathbf{r'}}\,e^{-i \mathbf{k}\cdot\mathbf{r}}-e^{i \mathbf{k}\cdot\mathbf{r}}e^{-i \mathbf{k'}\cdot\mathbf{r'}}-e^{-i \mathbf{k}\cdot\mathbf{r}}e^{i \mathbf{k'}\cdot\mathbf{r'}}
		\Big]	,
	\end{align}
	\begin{align}
		\label{2nd_term_sum}
		&\int\frac{d^3\mathbf{x}d^3\mathbf{x}'}{V_s^2}\left(\mathcal{A}_2+\mathcal{B}_2+\mathcal{C}_2+\mathcal{D}_2 \right)= i^2\,D^2(\chi_i)\,[afH](\chi_i)D^2(\chi_j)\,[afH](\chi_j)b_1(m_i,\chi_i)b_1(m_j,\chi_j)\nonumber\\
		&\int \frac{d^3\mathbf{k}}{(2\pi)^{3}}\,\frac{d^3\mathbf{k'}}{(2\pi)^{3}}\, \frac{\mathbf{k}\cdot\mathbf{\hat{n}}\,\mathbf{k'}\cdot\mathbf{\hat{n}'}}{k^2{k'}^2} P(k)\,P(k')\,
		\Big[ \Big(e^{i \mathbf{k}\cdot\mathbf{r}}-e^{-i \mathbf{k}\cdot\mathbf{r}}\Big)\,\Big(e^{i \mathbf{k'}\cdot\mathbf{r'}}-e^{-i \mathbf{k'}\cdot\mathbf{r'}}\Big)
		\Big]	.
	\end{align}
	We can simplify the above expression further, by performing angular part of the integral in k-space.
	\begin{align}
		\label{term2_k_ang_int}
		\int d{\Omega_{\mathbf{k}}} &\;i\mathbf{k}\cdot\mathbf{\hat{n}}  \Big(e^{i \mathbf{k}\cdot\mathbf{r}}-e^{-i \mathbf{k}\cdot\mathbf{r}}\Big)	\int d{\Omega_{k'}}\; i\mathbf{k'}\cdot\mathbf{\hat{n}'} \Big(e^{i \mathbf{\mathbf{k'}}\cdot\mathbf{r'}}-e^{-i \mathbf{k'}\cdot\mathbf{r'}}\Big)\nonumber\\
		&=	\int d{\Omega_{\mathbf{k}}} \;\mathbf{\hat{n}}\cdot\frac{\partial}{\partial \mathbf{r}} \Big(e^{i \mathbf{k}\cdot\mathbf{r}}+e^{-i \mathbf{k}\cdot\mathbf{r}}\Big)	\int d{\Omega_{\mathbf{k'}}}\; \mathbf{\hat{n}'}\cdot\frac{\partial}{\partial \mathbf{r'}}  \Big(e^{i \mathbf{k'}\cdot\mathbf{r'}}+e^{-i \mathbf{k'}\cdot\mathbf{r'}}\Big)\nonumber\\\nonumber\\
		&=\Big[\mathbf{\hat{n}}\cdot\frac{\partial}{\partial \mathbf{r}} \,\int d{\Omega_{\mathbf{k}}} \;\sum_{L,M}(4\pi)(i^{L}+(-i)^{L})Y^{*}_{LM}(\mathbf{\hat{k}}) Y_{LM}(\mathbf{\hat{r}})j_L(kr)\Big]\nonumber\\
		&\hspace{1cm}\Big[\mathbf{\hat{n}'}\cdot\frac{\partial}{\partial \mathbf{r'}}\,\int d{\Omega_{\mathbf{k'}}}\;\sum_{L',M'}(4\pi)(i^{L'}+(-i)^{L'})Y^{*}_{L'M'}(\mathbf{\hat{k}'}) Y_{L'M'}(\mathbf{\hat{r}'})j_{L'}(k'r') \Big]\nonumber\\
		&= 4\,(4\pi)^2 \,\mathbf{\hat{n}}\cdot\frac{\partial}{\partial \mathbf{r}} j_0(kr) \,\mathbf{\hat{n}'}\cdot\frac{\partial}{\partial \mathbf{r'}} j_0(k'r')\nonumber\\\nonumber\\
		&= 4\,(4\pi)^2 k\,k'\,  j_1(kr) j_1(k'r')\mathbf{\hat{r}}\cdot\mathbf{\hat{n}}\;\mathbf{\hat{r}'}\cdot\mathbf{\hat{n}'}.
	\end{align}
	Therefore from Eq.(\ref{2nd_term_sum}), we get,
	\begin{align}
		&\int\frac{d^3\mathbf{x}d^3\mathbf{x}'}{V_s^2}\left(\mathcal{A}_2+\mathcal{B}_2+\mathcal{C}_2+\mathcal{D}_2 \right)=4\,(4\pi)^2\, i^2\,D^2(\chi_i)\,[afH](\chi_i)D^2(\chi_j)\,[afH](\chi_j)\nonumber\\
		&b_1(m_i,\chi_i)b_1(m_j,\chi_j)\,k \, k' \, P(k) \, P(k')\,  j_1(kr) \, j_1(k'r')\mathbf{\hat{r}}\cdot\mathbf{\hat{n}}\;\mathbf{\hat{r}'}\cdot\mathbf{\hat{n}'}.
	\end{align}
	In Appendix \ref{App:Review_kSZ}, we have shown that the matter density two-point correlation function $\xi(r)$ is,
	\begin{align}
		\xi(r)=\frac{D^2}{2\pi^2}\int dk k^2\,j_{0}(kr)P(k).
	\end{align}
	Therefore, the volume average of $\xi(x)$ is,
	\begin{align}
		\label{xi_bar}
		\bar{\xi}(r)=D^2\frac{1}{2\pi^2}\int dk k\;\frac{3}{r}\left(j_{1}(kr)\right).
	\end{align}
	Using Eq.(\ref{term2_k_ang_int}) and Eq.(\ref{xi_bar}) in Eq.(\ref{2nd_term_sum}) we get,
	\begin{align}
		\label{term2_simple}
		\int\frac{d^3\mathbf{x}d^3\mathbf{x}'}{V_s^2}\left(\mathcal{A}_2+\mathcal{B}_2+\mathcal{C}_2+\mathcal{D}_2 \right)=&\frac{2}{3}\,[afH](\chi_i)b_1(m_i,\chi_i)\bar{\xi}(r)\nonumber\\
		&\hspace{1cm}\frac{2}{3}\,[afH](\chi_j)b_1(m_j,\chi_j)\bar{\xi}(r')\mathbf{r}\cdot\mathbf{\hat{n}}\;\mathbf{r'}\cdot\mathbf{\hat{n}'}.
	\end{align}
	Therefore, the contribution to Eq.(\ref{4_point_defn}) is,
	\begin{align}
		\label{contri_term2}
		\int \frac{ d\Omega_{r} d\Omega_{r'}}{(4\pi)^2} \sum_{i,j}\;w_iw_j&\,\tau_{\mathrm{eff}}(m_i,\chi_i)\tau_{\mathrm{eff}}(m_j,\chi_j)\frac{2}{3}\,[afH](\chi_i)\frac{b_1(m_i,\chi_i)\bar{\xi}(r)}{\left(1+b^2_1(m_i,\chi_i)\xi(r,\chi_i)\right)}\nonumber\\
		&\frac{2}{3}\,[afH](\chi_j)\frac{b_1(m_j,\chi_j)\bar{\xi}(r')}{\left(1+b^2_1(m_j,\chi_j)\xi(r',\chi_j)\right)}\mathbf{r}\cdot\mathbf{\hat{n}_{i}}\;\mathbf{r'}\cdot\mathbf{\hat{n}'_{j}}.
	\end{align}
	If we perform the summation over i and j, then the expression will be independent of $\mathbf{\hat{r}}$ and $\mathbf{\hat{r}'}$. Hence, the angular integral $\int \frac{ d\Omega_{r} d\Omega_{r'}}{(4\pi)^2}$ will be trivial and equal to unity. 
	Therefore, we can rewrite the above expression as,
	\begin{align}
		\label{contri_term2a}
		\sum_{i,j}\;w_iw_j\,\tau_{\mathrm{eff}}(m_i,\chi_i)&\tau_{\mathrm{eff}}(m_j,\chi_j)\frac{2}{3}\,[afH](\chi_i)\frac{b_1(m_i,\chi_i)\bar{\xi}(r)}{\left(1+b^2_1(m_i,\chi_i)\xi(r,\chi_i)\right)}\nonumber\\
		&\frac{2}{3}\,[afH](\chi_j)\frac{b_1(m_j,\chi_j)\bar{\xi}(r')}{\left(1+b^2_1(m_j,\chi_j)\xi(r',\chi_j)\right)}\mathbf{r}\cdot\mathbf{\hat{n}_{i}}\;\mathbf{r'}\cdot\mathbf{\hat{n}'_{j}}.
	\end{align}
	\\
	\subsection*{Further simplification of terms, $\mathcal{A}_1,\mathcal{B}_1,\mathcal{C}_1$, $\mathcal{D}_1$, $\mathcal{A}_3,\mathcal{B}_3,\mathcal{C}_3$ and $\mathcal{D}_3$}
	Next, we further simplify Eq.(\ref{1st_term_sum}) and Eq.(\ref{3rd_term_sum}).
	\begin{align}
		\int\frac{d^3\mathbf{x}d^3\mathbf{x}'}{V_s^2}&\left(\mathcal{A}_1+\mathcal{B}_1+\mathcal{C}_1+\mathcal{D}_1 \right)+\int\frac{d^3\mathbf{x}d^3\mathbf{x}'}{V_s^2}\left(\mathcal{A}_3+\mathcal{B}_3+\mathcal{C}_3+\mathcal{D}_3 \right)=\nonumber\\
		&4\, i^2\,D^2(\chi_i)\,[afH](\chi_i)D^2(\chi_j)\,[afH](\chi_j)b_1(m_i,\chi_i)b_1(m_j,\chi_j)\nonumber\\
		&\int \frac{d^3\mathbf{k}}{(2\pi)^{3}}\, \frac{\mathbf{k}\cdot\mathbf{\hat{n}}\,\mathbf{k}\cdot\mathbf{\hat{n}'}}{k^2{k}^2} P(k)\,P(k) \,\frac{1}{V_s}\,
		\Big[ e^{i \mathbf{k}\cdot(\mathbf{r}+\mathbf{r'})}-e^{i \mathbf{k}\cdot(\mathbf{r}-\mathbf{r'})}
		\Big],
	\end{align}
	\begin{align}
		\int\frac{d^3\mathbf{x}d^3\mathbf{x}'}{V_s^2}&\left(\mathcal{A}_1+\mathcal{B}_1+\mathcal{C}_1+\mathcal{D}_1 \right)+\int\frac{d^3\mathbf{x}d^3\mathbf{x}'}{V_s^2}\left(\mathcal{A}_3+\mathcal{B}_3+\mathcal{C}_3+\mathcal{D}_3 \right)=\nonumber\\
		&4\,D^2(\chi_i)\,[afH](\chi_i)D^2(\chi_j)\,[afH](\chi_j)b_1(m_i,\chi_i)b_1(m_j,\chi_j)\nonumber\\
		&\mathbf{\hat{n}}\cdot\frac{\partial}{\partial \mathbf{r}}\;\mathbf{\hat{n}'}\cdot\frac{\partial}{\partial \mathbf{r'}}\int \frac{d^3\mathbf{k}}{(2\pi)^{3}}\, \frac{1}{k^4} P(k)\,P(k) \,\frac{1}{V_s}\,
		\Big[ e^{i \mathbf{k}\cdot(\mathbf{r}+\mathbf{r'})}+e^{i \mathbf{k}\cdot(\mathbf{r}-\mathbf{r'})}
		\Big]	.
	\end{align}
	The angular integral in k-space is,
	\begin{align}
		\int d{\Omega_{\mathbf{k}}}\,\Big(e^{i \mathbf{k}\cdot(\mathbf{r}+\mathbf{r'})}+e^{i \mathbf{k}\cdot(\mathbf{r}-\mathbf{r'})}\Big)=(4\pi)^2\,\sum_{L,M}Y^{*}_{L,M}(\mathbf{\hat{r}})Y_{L,M}(\mathbf{\hat{r}'})\;j_{L}(kr)j_{L}(kr')\left(1+(-1)^L\right).
	\end{align}
	Therefore,
	\begin{align}
		&\int\frac{d^3\mathbf{x}d^3\mathbf{x}'}{V_s^2}\left(\mathcal{A}_1+\mathcal{B}_1+\mathcal{C}_1+\mathcal{D}_1 \right)+\int\frac{d^3\mathbf{x}d^3\mathbf{x}'}{V_s^2}\left(\mathcal{A}_3+\mathcal{B}_3+\mathcal{C}_3+\mathcal{D}_3 \right)=\nonumber\\
		&4\,D^2(\chi_i)\,[afH](\chi_i)D^2(\chi_j)\,[afH](\chi_j)b_1(m_i,\chi_i)b_1(m_j,\chi_j)\int \frac{dk}{(2\pi)^{3}}\, \frac{1}{k^2} P^2(k)\,\frac{1}{V_s}\,\nonumber\\
		&
		(4\pi)^2\,\sum_{L,M}\left(1+(-1)^L\right)\left\{\mathbf{\hat{n}}\cdot\frac{\partial}{\partial \mathbf{r}}\,\left(Y^{*}_{L,M}(\mathbf{\hat{r}})j_{L}(kr)\right)\right\}\left\{\mathbf{\hat{n}'}\cdot\frac{\partial}{\partial \mathbf{r'}}\left(Y_{L,M}(\mathbf{\hat{r}'})\;j_{L}(kr')\right)\right\}.
	\end{align}
	Performing the partial derivatives we get,
	\begin{align}
		&\left\{\mathbf{\hat{n}}\cdot\frac{\partial}{\partial \mathbf{r}}\,\left(Y^{*}_{L,M}(\mathbf{\hat{r}})j_{L}(kr)\right)\right\}\left\{\mathbf{\hat{n}'}\cdot\frac{\partial}{\partial \mathbf{r'}}\left(Y_{L,M}(\mathbf{\hat{r}'})\;j_{L}(kr')\right)\right\}=k^2\nonumber\\
		&\left\{ \sqrt{\frac{L+1}{2L+1} }\;j_{L+1}(kr)\;\mathbf{\hat{n}}\cdot\mathbf{Y}^{\,^*L+1}_{\,LM}(\mathbf{\hat{r}}) +\sqrt{\frac{L}{2L+1}}j_{L-1}(kr)\;\mathbf{\hat{n}}\cdot\mathbf{Y}^{\,^*L-1}_{\,LM}(\mathbf{\hat{r}})\right\}\nonumber\\
		&\left\{  \sqrt{\frac{L+1}{2L+1} }\;j_{L+1}(kr')\;\mathbf{\hat{n}'}\cdot\mathbf{Y}^{\,L+1}_{\,LM}(\mathbf{\hat{r}'}) +\sqrt{\frac{L}{2L+1}}j_{L-1}(kr')\;\mathbf{\hat{n}'}\cdot\mathbf{Y}^{\,L-1}_{\,LM}(\mathbf{\hat{r}'})  \right\},
	\end{align}
	where $\mathbf{Y}^{\,L}_{\,JM}(\mathbf{\hat{r}})$ are the vector spherical harmonics. They can be defined in different way, but for us the following definition is required,
	\begin{align}
		\mathbf{Y}^{\,L}_{\,JM}(\mathbf{\hat{r}})=\sum_{\mu}C^{JM}_{LM-\mu\,1\mu}Y_{LM-\mu}(\mathbf{\hat{r}})\mathbf{e_\mu},
	\end{align}
	where, $\mathbf{e_\mu} \{\mu=-1,0,1\}$ are the spherical covariant basis vectors and the Clebsch-Gordan coefficients are related to Wigner 3J symbols as.
	.\begin{align}
		C^{j_3m_3}_{j_1m_i\,j_2m_j}= (-1)^{(j_1-j_2+m_3)}\;\sqrt{2j_3+1}
		\left(\begin{array}{ccc}
			j_1& j_2 & j_3\\ 
			m_i& m_j & -m_3
		\end{array}\right).
	\end{align}
	To proceed further, we perform the angular integral $\int \frac{ d\Omega_{r} d\Omega_{r'}}{(4\pi)^2}$. In the weighted estimator we are weighting by  $\mathbf{\hat{r}}\cdot\mathbf{\hat{n}}$. Therefore, the angular integral becomes,
	\begin{align}
		\label{ang_int_omr}
		\int \frac{ d\Omega_{r} d\Omega_{r'}}{(4\pi)^2}&\;(\mathbf{\hat{r}}\cdot\mathbf{\hat{n}})\;(\mathbf{\hat{r}'}\cdot\mathbf{\hat{n}'})\,\bigg[
		\frac{L+1}{2L+1} \;j_{L+1}(kr)\;j_{L+1}(kr')\,\mathbf{\hat{n}}\cdot\mathbf{Y}^{\,^*L+1}_{\,LM}(\mathbf{\hat{r}})\;\mathbf{\hat{n}'}\cdot\mathbf{Y}^{\,L+1}_{\,LM}(\mathbf{\hat{r}'})\nonumber\\
		&+\frac{\sqrt{(L+1)L}}{(2L+1)}\;j_{L+1}(kr)\;j_{L-1}(kr')\;\mathbf{\hat{n}}\cdot\mathbf{Y}^{\,^*L+1}_{\,LM}(\mathbf{\hat{r}})\;\mathbf{\hat{n}'}\cdot\mathbf{Y}^{\,L-1}_{\,LM}(\mathbf{\hat{r}'})\nonumber\\
		&+\frac{\sqrt{(L+1)L}}{(2L+1)}\;j_{L-1}(kr)\;j_{L+1}(kr')\;\mathbf{\hat{n}}\cdot\mathbf{Y}^{\,^*L-1}_{\,LM}(\mathbf{\hat{r}})\;\mathbf{\hat{n}'}\cdot\mathbf{Y}^{\,L+1}_{\,LM}(\mathbf{\hat{r}'})\nonumber\\
		&  +\frac{L}{2L+1} \;j_{L-1}(kr)\;j_{L-1}(kr')\;\mathbf{\hat{n}}\cdot\mathbf{Y}^{\,^*L-1}_{\,LM}(\mathbf{\hat{r}})\;\mathbf{\hat{n}'}\cdot\mathbf{Y}^{\,L-1}_{\,LM}(\mathbf{\hat{r}'})
		\bigg].
	\end{align} 
	We can do the angular integrals separately over $\mathbf{\hat{r}}$ and $\mathbf{\hat{r}'}$ variables. Therefore,
	\begin{align}
		\int \frac{ d\Omega_{r}}{4\pi} \;(\mathbf{\hat{r}}\cdot\mathbf{\hat{n}})\,\mathbf{\hat{n}}\cdot\mathbf{Y}^{\,^*L+1}_{\,LM}(\mathbf{\hat{r}})&=\frac{4\pi}{3}\,\sum_{M_1}
		Y^{*}_{1M_1}(\mathbf{\hat{n}})\int \frac{ d\Omega_{r}}{4\pi}\,Y_{1M_1}(\mathbf{\hat{r}})\;\mathbf{\hat{n}}\cdot\mathbf{Y}^{\,^*L+1}_{\,LM}(\mathbf{\hat{r}})\nonumber\\
		&=\frac{4\pi}{3}\,\sum_{M_1}
		Y^{*}_{1M_1}(\mathbf{\hat{n}})\;\sum_{\mu}\;\mathbf{\hat{n}}\cdot\mathbf{e^{*}_\mu}\,C^{\,^*LM}_{L+1M-\mu\,1\mu}
		\int \frac{ d\Omega_{r}}{4\pi}\,Y_{1M_1}(\mathbf{\hat{r}})Y^{*}_{L+1M-\mu}(\mathbf{\hat{r}})\nonumber\\
		&=\frac{1}{3}\,\sum_{M_1}
		Y^{*}_{1M_1}(\mathbf{\hat{n}})\;\sum_{\mu}\;\mathbf{\hat{n}}\cdot\mathbf{e^{*}_\mu}\,C^{\,^*LM}_{L+1M-\mu\,1\mu}\;\delta_{L+1,1}\;\delta_{M_1,M-\mu}.
	\end{align}
	Because of the Kronecker Delta functions, we have $L=0$ and $M_1=M-\mu$. Since $L=0$, then only possibility for $M$ is also $M=0$. Therefore we get,
	\begin{align}
		\int \frac{ d\Omega_{r}}{4\pi} \;(\mathbf{\hat{r}}\cdot\mathbf{\hat{n}})\;\mathbf{\hat{n}}\cdot\mathbf{Y}^{\,^*L+1}_{\,LM}(\mathbf{\hat{r}})&=\frac{1}{3}\;\sum_{\mu}\;\mathbf{\hat{n}}\cdot\mathbf{e^{*}_\mu}\;
		Y^{*}_{1-\mu}(\mathbf{\hat{n}})\;C^{\,^*00}_{1-\mu\,1\mu}\nonumber\\
		&=\frac{1}{3}\;\mathbf{\hat{n}}\cdot\mathbf{Y}^{\,^*1}_{\,00}(\mathbf{\hat{n}})=-\frac{1}{3}\;\frac{1}{\sqrt{4\pi}}\;\mathbf{\hat{n}}\cdot\mathbf{\hat{n}}\;\delta_{L+1,1}.
	\end{align}
	Similarly, we have,
	\begin{align}
		\int \frac{ d\Omega_{r'}}{4\pi} \;(\mathbf{\hat{r}'}\cdot\mathbf{\hat{n}'})\;\mathbf{\hat{n}'}\cdot\mathbf{Y}^{\,L+1}_{\,LM}(\mathbf{\hat{r}'})=-\frac{1}{3}\;\frac{1}{\sqrt{4\pi}}\;\mathbf{\hat{n}'}\cdot\mathbf{\hat{n}'}\;\delta_{L+1,1}
	\end{align}
	We can immediately observe that the cross terms vanish, as the two Kronnecker Delta functions $\;\delta_{L+1,1}$ and $\;\delta_{L-1,1}$ cannot be simultaneously satisfied. Therefore we only need to calculate last last term in Eq.(\ref{ang_int_omr}).
	\begin{align}
		\int \frac{ d\Omega_{r}}{4\pi} \;(\mathbf{\hat{r}}\cdot\mathbf{\hat{n}})\;\mathbf{\hat{n}}\cdot\mathbf{Y}^{\,^*L-1}_{\,LM}(\mathbf{\hat{r}})&=\frac{4\pi}{3}\,\sum_{M_1}
		Y^{*}_{1M_1}(\mathbf{\hat{n}})\int \frac{ d\Omega_{r}}{4\pi}\,Y_{1M_1}(\mathbf{\hat{r}})\;\mathbf{\hat{n}}\cdot\mathbf{Y}^{\,^*L-1}_{\,LM}(\mathbf{\hat{r}})\nonumber\\
		&=\frac{4\pi}{3}\,\sum_{M_1}
		Y^{*}_{1M_1}(\mathbf{\hat{n}})\;\sum_{\mu}\;\mathbf{\hat{n}}\cdot\mathbf{e^{*}_\mu}\,C^{\,^*LM}_{L-1M-\mu\,1\mu}
		\int \frac{ d\Omega_{r}}{4\pi}\,Y_{1M_1}(\mathbf{\hat{r}})Y^{*}_{L-1M-\mu}(\mathbf{\hat{r}})\nonumber\\
		&=\frac{1}{3}\,\sum_{M_1}
		Y^{*}_{1M_1}(\mathbf{\hat{n}})\;\sum_{\mu}\;\mathbf{\hat{n}}\cdot\mathbf{e^{*}_\mu}\,C^{\,^*LM}_{L-1M-\mu\,1\mu}\;\delta_{L-1,1}\;\delta_{M_1,M-\mu}\nonumber\\
		&=\frac{1}{3}
		\;\sum_{\mu}\;\mathbf{\hat{n}}\cdot\mathbf{e^{*}_\mu}\,C^{\,^*2M}_{1M-\mu\,1\mu}Y^{*}_{1M-\mu}(\mathbf{\hat{n}})\nonumber\\
		&=\frac{1}{3}\;\mathbf{\hat{n}}\cdot\mathbf{Y}^{\,^*1}_{\,2M}(\mathbf{\hat{n}})\;\delta_{L-1,1}.
	\end{align}
	Similarly, we have,
	\begin{align}
		\int \frac{ d\Omega_{r'}}{4\pi} \;(\mathbf{\hat{r}'}\cdot\mathbf{\hat{n}'})\;\mathbf{\hat{n}'}\cdot\mathbf{Y}^{\,L-1}_{\,LM}(\mathbf{\hat{r}'})=\frac{1}{3}\;\mathbf{\hat{n}'}\cdot\mathbf{Y}^{\,1}_{\,2M}(\mathbf{\hat{n}'})\;\delta_{L-1,1}.
	\end{align}
	Therefore, combining the results of the partial derivatives and then the angular integrations we get,
	\begin{align}
		&\int \frac{ d\Omega_{r} d\Omega_{r'}}{(4\pi)^2}\;(\mathbf{\hat{r}}\cdot\mathbf{\hat{n}})\;(\mathbf{\hat{r}'}\cdot\mathbf{\hat{n}'})\;\Bigg[	\int\frac{d^3\mathbf{x}d^3\mathbf{x}'}{V_s^2}\left(\mathcal{A}_1+\mathcal{B}_1+\mathcal{C}_1+\mathcal{D}_1 \right)+\int\frac{d^3\mathbf{x}d^3\mathbf{x}'}{V_s^2}\left(\mathcal{A}_3+\mathcal{B}_3+\mathcal{C}_3+\mathcal{D}_3 \right)\Bigg]\nonumber\\
		&=8\,(4\pi)^2\,D^2(\chi_i)\,[afH](\chi_i)D^2(\chi_j)\,[afH](\chi_j)b_1(m_i,\chi_i)b_1(m_j,\chi_j)\nonumber\\
		&\hspace{3cm}\,\frac{1}{V_s}\;
		\left(\frac{1}{3}\right)^2	\frac{dk}{(2\pi)^{3}}P^2(k)\;j_{1}(kr)\;j_{1}(kr')\bigg[\frac{1}{4\pi}\;(\mathbf{\hat{n}}\cdot\mathbf{\hat{n}})\,(\mathbf{\hat{n}'}\cdot\mathbf{\hat{n}'})\nonumber\\
		&\hspace{5cm}+\frac{2}{5}\;\sum_M\;\;\left(\mathbf{\hat{n}}\cdot\mathbf{Y}^{\,^*1}_{\,2M}(\mathbf{\hat{n}})\right)\;\left(\mathbf{\hat{n}'}\cdot\mathbf{Y}^{\,1}_{\,2M}(\mathbf{\hat{n}'})\right)\bigg]\nonumber
	\end{align}
	\begin{align}
		&=8\,(4\pi)^2\,D^2(\chi_i)\,[afH](\chi_i)D^2(\chi_j)\,[afH](\chi_j)b_1(m_i,\chi_i)b_1(m_j,\chi_j)
		\nonumber\\
		&\hspace{3cm}\,\frac{1}{V_s}\;	\left(\frac{1}{3}\right)^2	\frac{dk}{(2\pi)^{3}}P^2(k)\;j_{1}(kr)\;j_{1}(kr')\bigg[\frac{1}{4\pi}
		+\left(\frac{2}{5}\right)^2\;\sum_M\;Y^{*}_{2M}(\mathbf{\hat{n}})\;Y_{\,2M}(\mathbf{\hat{n}'})\bigg]\nonumber
	\end{align}
	\begin{align}
		=8\,(4\pi)^2&\,D^2(\chi_i)\,[afH](\chi_i)D^2(\chi_j)\,[afH](\chi_j)b_1(m_i,\chi_i)b_1(m_j,\chi_j)\nonumber\\
		&\hspace{2cm}\,\frac{1}{V_s}\;	\left(\frac{1}{3}\right)^2	\frac{dk}{(2\pi)^{3}}P^2(k)\;j_{1}(kr)\;j_{1}(kr')\bigg[\frac{1}{4\pi}
		+\left(\frac{2}{5}\right)^2\frac{5}{4\pi}P_2(\mathbf{\hat{n}}\cdot\mathbf{\hat{n}'})\bigg]\nonumber
	\end{align}
	\begin{align}
		=8\,(4\pi)^2&\,D^2(\chi_i)\,[afH](\chi_i)D^2(\chi_j)\,[afH](\chi_j)b_1(m_i,\chi_i)b_1(m_j,\chi_j)\nonumber\\
		&\hspace{2.0cm}\,\frac{1}{V_s}\;	\left(\frac{1}{3}\right)^2 \frac{1}{4\pi} \frac{3}{5}	\frac{dk}{(2\pi)^{3}}P^2(k)\;j_{1}(kr)\;j_{1}(kr')\bigg[
		1+2(\mathbf{\hat{n}}\cdot\mathbf{\hat{n}'})^2\bigg].
	\end{align}
	Therefore, the contribution to Eq.(\ref{4_point_defn}) is,
	\begin{align}
		\label{contri_term1and3}
		&\frac{1}{V_s}\; \frac{4}{15\pi^2}\sum_{i,j}\;m_i\;m_j\,\tau_{\mathrm{eff}}(m_i,\chi_i)\tau_{\mathrm{eff}}(m_j,\chi_j)\,\bigg[
		1+2(\mathbf{\hat{n}_{i}}\cdot\mathbf{\hat{n}'_{j}})^2\bigg]\, [afH](\chi_i)\nonumber\\
		&\,[afH](\chi_j)\int dk\,\left(b_1(m_i,\chi_i)P(k,\chi_i)\right)\left(b_1(m_j,\chi_j)P(k,\chi_j)\right)\;j_{1}(kr)\;j_{1}(kr')\nonumber\\
		&\left[\left(1+b^2_1(m_i,\chi_i)\xi(x,\chi_i)\right)\left(1+b^2_1(m_j,\chi_j)\xi(x',\chi_j)\right)\right]^{-1}.
	\end{align}
	\subsection*{Simplification of term $\mathcal{X}$}
	Next, we simply the term given in Eq.(\ref{term_X}).
	\begin{align}
		\mathcal{X}=\Big\langle\mathbf{v_{1}}({\mathbf{x}})\mathbf{v_{3}}({\mathbf{x'}})-\mathbf{v_{1}}({\mathbf{x}})\mathbf{v_{4}}({\mathbf{x'+r'}})-\mathbf{v_{2}}({\mathbf{x+r}})\mathbf{v_{3}}({\mathbf{x'}})+\mathbf{v_{2}}({\mathbf{x+r}})\mathbf{v_{4}}({\mathbf{x'+r'}})\Big\rangle,
	\end{align}
	\begin{align}
		&=i^2\,D(\chi_i)\,[afH](\chi_i)D(\chi_j)\,[afH](\chi_j)\int \frac{d^3\mathbf{k}d^3\mathbf{k'}}{(2\pi)^{6}}\,e^{i \mathbf{k}\cdot\mathbf{x}}\,e^{i \mathbf{k'}\cdot\mathbf{x'}} \frac{\mathbf{k}\mathbf{k'}}{k^2{k'}^2}\Big\langle \delta(\mathbf{k})\delta(\mathbf{k'}) \Big\rangle\nonumber\\
		&\hspace{1cm}	\Big[ 1-e^{i \mathbf{k'}\cdot\mathbf{r'}}-e^{i \mathbf{k}\cdot\mathbf{r}}+e^{i \mathbf{k}\cdot\mathbf{r}}e^{i \mathbf{k'}\cdot\mathbf{r'}}\Big],	
	\end{align}
	\begin{align}
		&=i^2\,D(\chi_i)\,[afH](\chi_i)D(\chi_j)\,[afH](\chi_j)\int \frac{d^3\mathbf{k}}{(2\pi)^{3}}\,e^{i \mathbf{k}\cdot(\mathbf{x}-\mathbf{x'})}\frac{\mathbf{k}\mathbf{k}}{k^2{k}^2}P(k)\nonumber\\
		&\hspace{1cm}	\Big[ 1-e^{i \mathbf{k}\cdot\mathbf{r'}}-e^{i \mathbf{k}\cdot\mathbf{r}}+e^{i \mathbf{k}\cdot\mathbf{r}}e^{i \mathbf{k}\cdot\mathbf{r'}}\Big].	
	\end{align}
	But, doing the volume integral over $\mathbf{x}$ and $\mathbf{x'}$ we get,
	\begin{align}
		\int\frac{d^3\mathbf{x}d^3\mathbf{x}'}{V_s^2}e^{i \mathbf{k}\cdot(\mathbf{x}-\mathbf{x'})}=(2\pi)^3\frac{1}{V_s}\delta_D(\mathbf{k}).
	\end{align}
	This shows that it only contributes to $\mathbf{k}=0$ mode which is 0 for the expression in the term $\mathcal{X}$. The term $\mathcal{X}$ does not contribute to the covariance matrix. Thus the expressions given in Eq.(\ref{contri_term2a}) and in Eq.(\ref{contri_term1and3}) are the only non-zero contributions to Eq.(\ref{4_point_defn}), which is the first part of Eq.(\ref{cov_start_appendix}). \\\\
	Now the last part in Eq.(\ref{cov_start_appendix}) have already been solved in Appendix \ref{App:Review_kSZ},
	\begin{align}
		\label{Tp_Tp}
		T_{\mathrm{pairwise}}(r)T'_{\mathrm{pairwise}}(r')&=\sum_{i,j}\;w_iw_j\,\tau_{\mathrm{eff}}(m_i,\chi_i)\tau_{\mathrm{eff}}(m_j,\chi_j)\frac{2}{3}\,[afH](\chi_i)\frac{b_1(m_i,\chi_i)\bar{\xi}(r)}{\left(1+b^2_1(m_i,\chi_i)\xi(r,\chi_i)\right)}\nonumber\\
		&\frac{2}{3}\,[afH](\chi_j)\frac{b_1(m_j,\chi_j)\bar{\xi}(r')}{\left(1+b^2_1(m_j,\chi_j)\xi(r',\chi_j)\right)}\mathbf{r}\cdot\mathbf{\hat{n}_{i}}\;\mathbf{r'}\cdot\mathbf{\hat{n}'_{j}}.
	\end{align}
	The expression given in Eq.(\ref{contri_term2a}) and the term given by Eq.(\ref{Tp_Tp}) are exactly the same with different signs as we can see from Eq.(\ref{cov_start_appendix}). Therefore, they cancel each other exactly leaving the expression in Eq.(\ref{contri_term1and3}) as the only contribution to Eq.(\ref{cov_start_appendix}). Therefore, the statistical noise contribution to the covariance matrix is given by,
	\begin{align}
		C_{T_{\mathrm{pairwise}}}(r,r')=&\frac{1}{V_s}\; \frac{4}{15\pi^2}\sum_{i,j}\;m_i\;m_j\,\tau_{\mathrm{eff}}(m_i,\chi_i)\tau_{\mathrm{eff}}(m_j,\chi_j)\,\bigg[
		1+2(\mathbf{\hat{n}_{i}}\cdot\mathbf{\hat{n}'_{j}})^2\bigg]\, [afH](\chi_i)\nonumber\\
		&\,[afH](\chi_j)\int dk\,\left(b_1(m_i,\chi_i)P(k,\chi_i)\right)\left(b_1(m_j,\chi_j)P(k,\chi_j)\right)\;j_{1}(kr)\;j_{1}(kr')\nonumber\\
		&\left[\left(1+b^2_1(m_i,\chi_i)\xi(x,\chi_i)\right)\left(1+b^2_1(m_j,\chi_j)\xi(x',\chi_j)\right)\right]^{-1}.
	\end{align}
	To include the Poisson (shot) noise contribution we have to just rewrite,
	\begin{align}
		b_1(m_i,\chi_i) P(k,\chi_i)\rightarrow\left(b_1(m_i,\chi_i)P(k,\chi_i)+\frac{1}{n_\mathrm{cl}}\right), 
	\end{align}
	where $n_\mathrm{cl}$ is the galaxy cluster number density. Therefore, the final expression for the statistical uncertainty in the covariance matrix is,
	\begin{align}
		\label{C_pair_appendix}
		&C_{T_{\mathrm{pairwise}}}(r,r')=\frac{1}{V_s}\; \frac{4}{15\pi^2}\sum_{i,j}\;m_i\;m_j\,\tau_{\mathrm{eff}}(m_i,\chi_i)\tau_{\mathrm{eff}}(m_j,\chi_j)\,\bigg[
		1+2(\mathbf{\hat{n}_{i}}\cdot\mathbf{\hat{n}'_{j}})^2\bigg]\, [afH](\chi_i)\nonumber\\
		&\,[afH](\chi_j)\int dk\,\left(b_1(m_i,\chi_i)P(k,\chi_i)+\frac{1}{n_{\mathrm{cl}}}\right)\left(b_1(m_j,\chi_j)P(k,\chi_j)+\frac{1}{n_{\mathrm{cl}}}\right)\;j_{1}(kr)\;j_{1}(kr')\nonumber\\
		&\left[\left(1+b^2_1(m_i,\chi_i)\xi(x,\chi_i)\right)\left(1+b^2_1(m_j,\chi_j)\xi(x',\chi_j)\right)\right]^{-1}.
	\end{align}
	{\color{black}The expression given in Eq.(\ref{C_pair_appendix}) agrees with  the expression for the covariance matrix given in \cite{2015ApJ_DarkEnergy} after we take into account the extra weight factors in our optimised estimator. In particular, we have an extra geometric  prefactor of $ 1/15 \big[1+2(\mathbf{\hat{n}_{i}}\cdot\mathbf{\hat{n}'_{j}})^2\big]$.} Similarly, we repeat the calculation for the optimised cross-pairwise estimator. Similar to Eq.(\ref{cov_start_appendix}), the statistical part of the covariance matrix for the cross-pairwise case can be written as,
	\begin{align}
		\label{crosscov_start_appendix}
		C_{T_{\mathrm{pairwise}}}(r,r')=&\Big\langle\int \frac{d^3\mathbf{x}d^3\mathbf{x}'}{V_s^2}\int \frac{ d\Omega_{r} d\Omega_{r'}}{(4\pi)^2}\hat{T}_{\mathrm{cl-gal}}\;\hat{T'}_{\mathrm{cl-gal}}\Big\rangle_4\nonumber\\
		&\hspace{6cm}-T_{\mathrm{cl-gal}}(r)T'_{\mathrm{cl-gal}}(r').
	\end{align}
	In this case, only the equivalent term in $\mathcal{A}$, i.e. Eq.(\ref{term_A})  survives.  This reduces the covariance matrix by a factor of 4. We just need to simplify the term $\mathcal{A}_1$, $\mathcal{A}_2$, and $\mathcal{A}_3$. After simplifying, the term $\mathcal{A}_2$ will completely cancel the term we get from $T_{\mathrm{cl-gal}}(r)T'_{\mathrm{cl-gal}}(r')$ in Eq.(\ref{crosscov_start_appendix}). Then the terms $\mathcal{A}_1$ and $\mathcal{A}_3$ adds up to give the final expression,
	\begin{align}
		\label{cov_crosspair}
		&C_{T_{\mathrm{cl-gal}}}(r,r')=\frac{1}{V_s}\; \frac{1}{15\pi^2}\sum_{i,j}\;m_i\;m_j\,\tau_{\mathrm{eff}}(m_i,\chi_i)\tau_{\mathrm{eff}}(m_j,\chi_j)\,\bigg[
		1+2(\mathbf{\hat{n}_{i}}\cdot\mathbf{\hat{n}'_{j}})^2\bigg]\, [afH](\chi_i)\nonumber\\
		&\hspace{1cm}\,[afH](\chi_j)\int dk\,\left(b^{\mathrm{gal}}_1(\chi_i)P(k,\chi_i)\right)\left(b^{\mathrm{gal}}_1(\chi_j)P(k,\chi_j)\right)\;j_{1}(kr)\;j_{1}(kr')\nonumber\\
		&\hspace{1cm}\left[\left(1+b^{\mathrm{gal}}_1(\chi_i)b^{\mathrm{cl}}_1(m_i,\chi_i)\xi(x,\chi_i)\right)\left(1+b^{\mathrm{gal}}_1(\chi_j)b^{\mathrm{cl}}_1(m_j,\chi_j)\xi(x',\chi_j)\right)\right]^{-1}.
	\end{align}
	Furthermore, in the case of cross-pairwise effect, the Poisson term is absent.
	\begin{figure}
		\hspace{-0.5cm}
		\centering
		\begin{subfigure}[t]{0.51\textwidth}
			\centering
			\includegraphics[width=\linewidth]{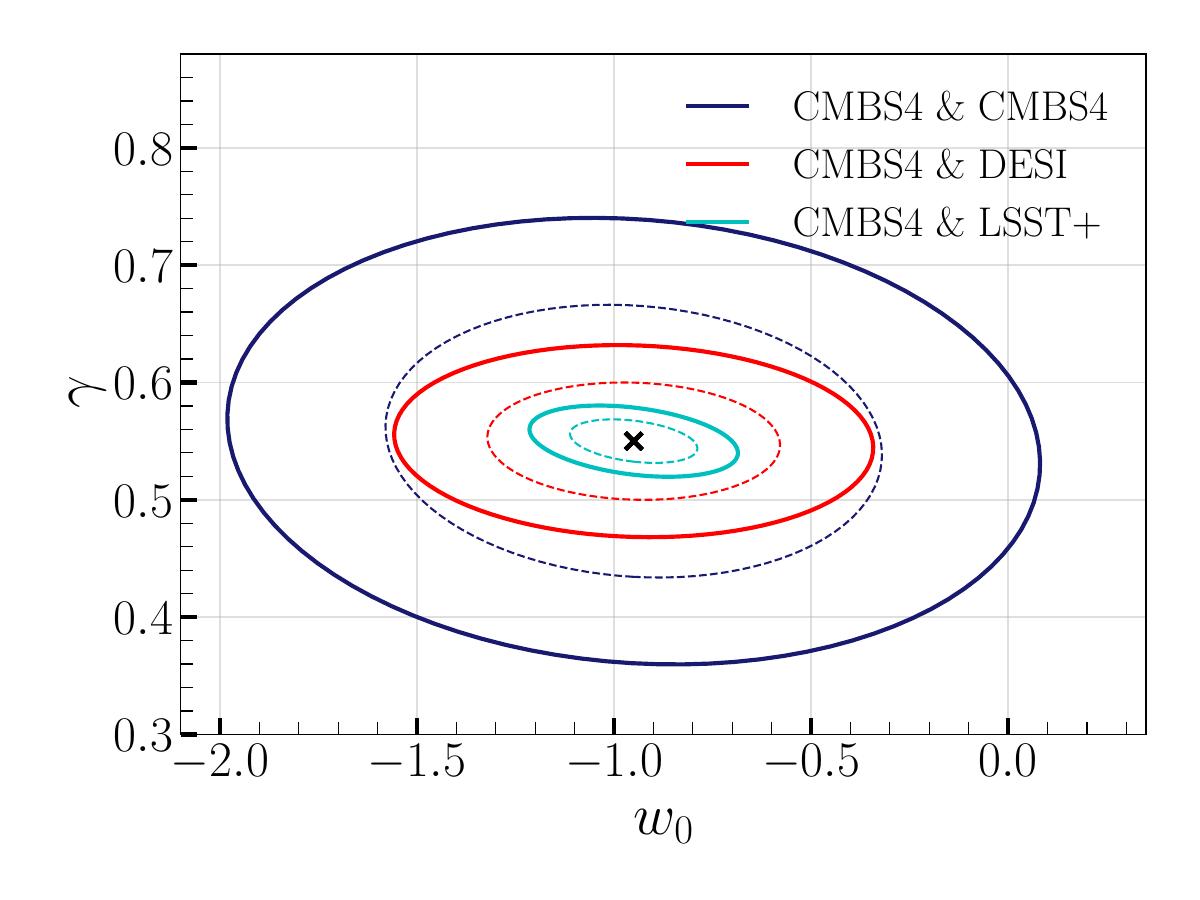}
			\caption{Planck (CMB)+CMB lensing prior}
			\label{fig:wo_gamma_planck_wn}
		\end{subfigure}%
		\begin{subfigure}[t]{0.51\textwidth}
			\centering
			\includegraphics[width=\linewidth]{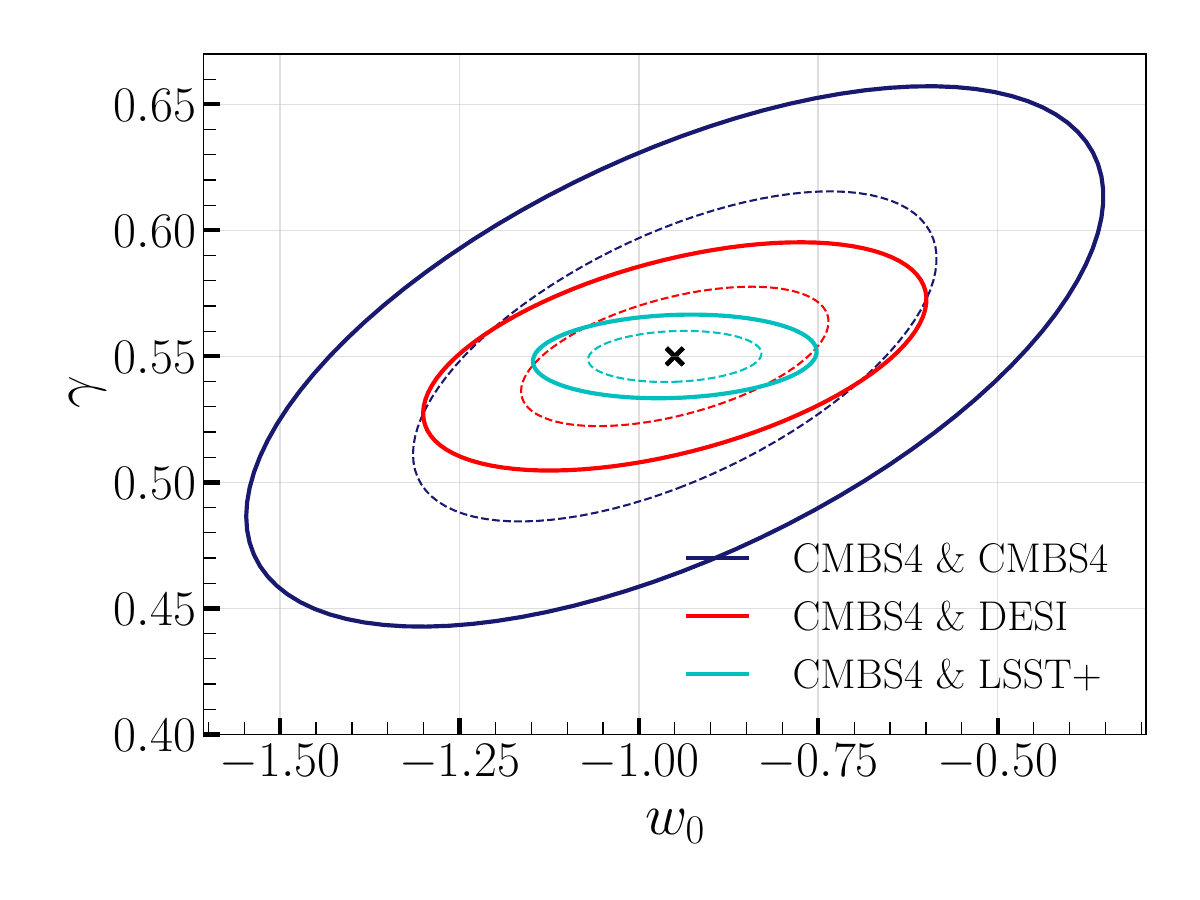}
			\caption{Planck (CMB)+CMB lensing+DESI BAO prior}
			\label{fig:wo_gamma_planck_BAO_wn}
		\end{subfigure}
		\begin{subfigure}[t]{0.51\textwidth}
			\hspace{-0.7cm}
			\centering
			\includegraphics[width=\linewidth]{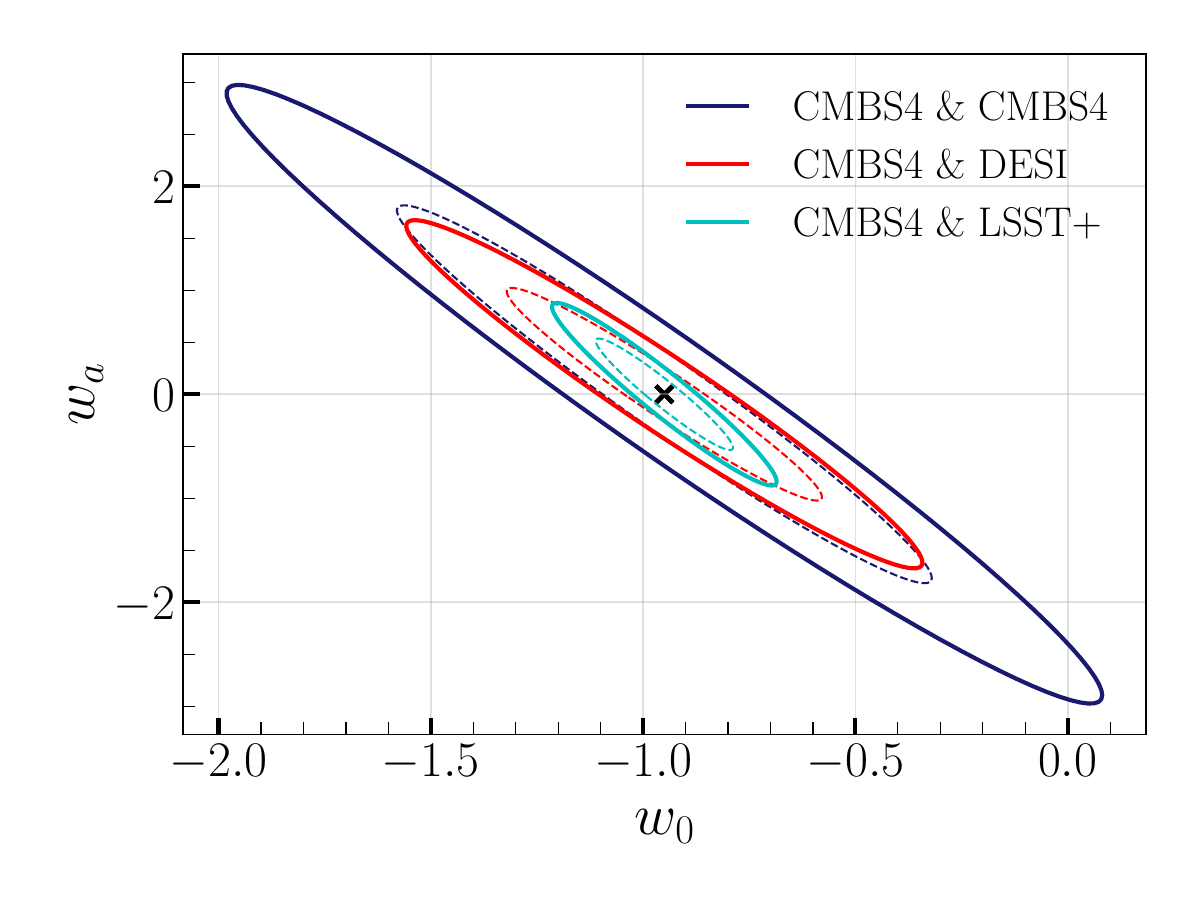}
			\caption{Planck (CMB)+CMB lensing prior}
			\label{fig:wo_wa_planck_wn}
		\end{subfigure}%
		\begin{subfigure}[t]{0.51\textwidth}
			\hspace{-0.7cm}
			\centering
			\includegraphics[width=\linewidth]{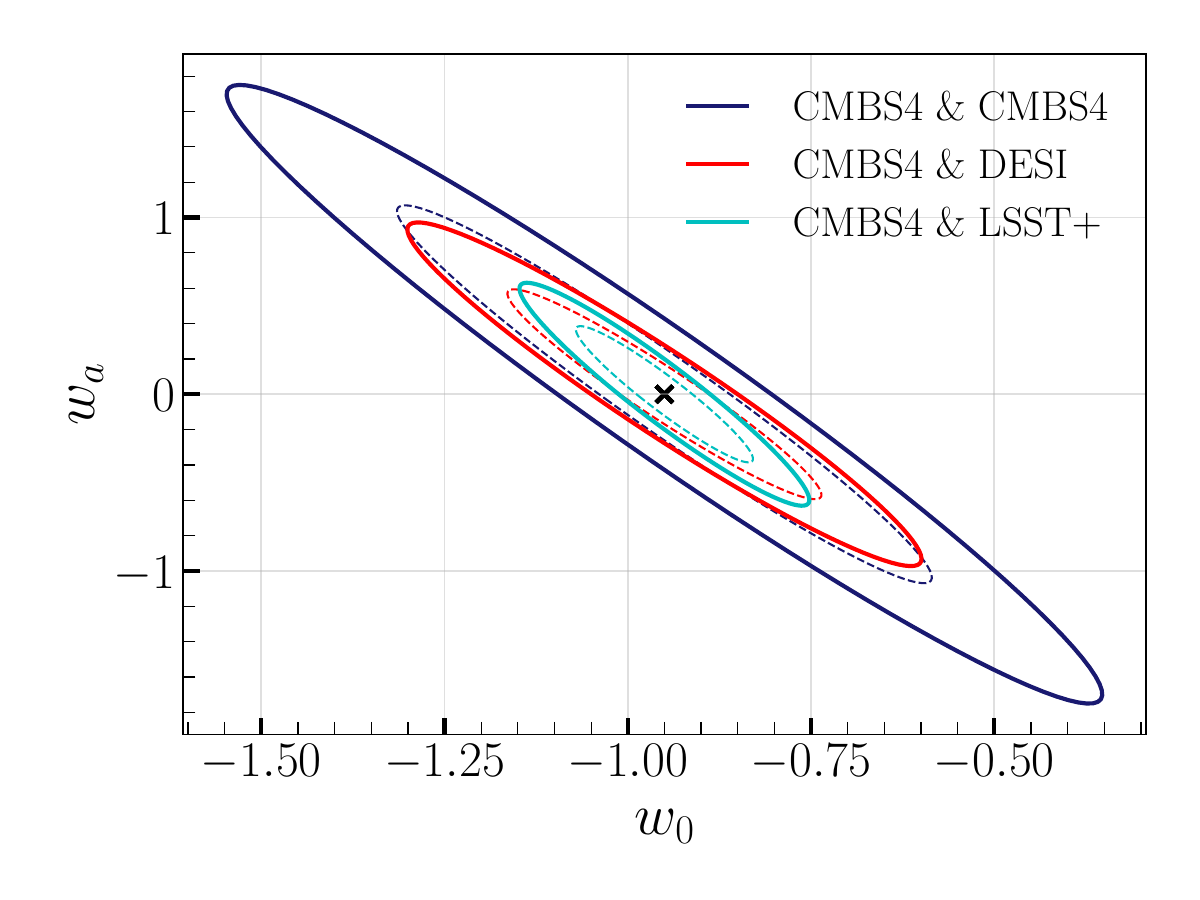}
			\caption{Planck (CMB)+CMB lensing+DESI BAO prior}
			\label{fig:wo_wa_planck_BAO_wn}
		\end{subfigure}	
		%
		%
		%
		%
		%
		%
		\caption{Same as figure \ref{fig:Conf_ellipse} assuming fully correlated Poisson noise between clusters and galaxies.}
		\label{fig:Conf_ellipse_wn}
	\end{figure}
	\section{The case of fully correlated shot noise in the cross-pairwise effect\label{App:Noise_case}}
	Although we expect uncorrelated Poisson noise for two independent tracers of the underlying dark matter field, we expect some correlation in real data. To see the effect of these correlations on our forecasts, we use the most pessimistic case of fully correlated Poisson noise. In this case the Poisson noise is given by,
	\begin{align}
		\mathrm{Poisson\;Noise}_\mathrm{\;cl-gal}=\frac{1}{\sqrt{n_\mathrm{gal}\;n_\mathrm{cl}}},
	\end{align}
	where $n_\mathrm{cl}$ and $n_\mathrm{gal}$ are the cluster and galaxy number density respectively. To include the Poisson \textit{shot noise} contribution we again have to just replace,
	\begin{align}
		b^{\mathrm{gal}}_1(\chi_i)	P(k,\chi_i)\rightarrow\left(b^{\mathrm{gal}}_1(\chi_i)P(k,\chi_i)+\frac{1}{\sqrt{n_\mathrm{gal}\;n_\mathrm{cl}}}\right),
	\end{align}
	in Eq.(\ref{cov_crosspair}).
	Therefore, in this case the statistical part of the covariance matrix is given by,
	\begin{align}
		&C_{T_{\mathrm{cl-gal}}}(r,r')=\frac{1}{V_s}\; \frac{1}{15\pi^2}\sum_{i,j}\;m_i\;m_j\,\tau_{\mathrm{eff}}(m_i,\chi_i)\tau_{\mathrm{eff}}(m_j,\chi_j)\,\bigg[
		1+2(\mathbf{\hat{n}_{i}}\cdot\mathbf{\hat{n}'_{j}})^2\bigg]\, [afH](\chi_i)\nonumber\\
		&\hspace{0.0cm}\,[afH](\chi_j)\int dk\,\left(b^{\mathrm{gal}}_1(\chi_i)P(k,\chi_i)+\frac{1}{\sqrt{n_\mathrm{gal}\;n_\mathrm{cl}}}\right)\left(b^{\mathrm{gal}}_1(\chi_j)P(k,\chi_j)+\frac{1}{\sqrt{n_\mathrm{gal}\;n_\mathrm{cl}}}\right)\;j_{1}(kr)\;j_{1}(kr')\nonumber\\
		&\hspace{0.0cm}\left[\left(1+b^{\mathrm{gal}}_1(\chi_i)b^{\mathrm{cl}}_1(m_i,\chi_i)\xi(x,\chi_i)\right)\left(1+b^{\mathrm{gal}}_1(\chi_j)b^{\mathrm{cl}}_1(m_j,\chi_j)\xi(x',\chi_j)\right)\right]^{-1}.
	\end{align}
	The confidence ellipses for this case is given in figure \ref{fig:Conf_ellipse_wn}. Since the galaxy number density is much larger than the cluster number density, the shot noise contribution in the cross-pairwise case is still much lower than the pairwise kSZ signal.  Hence, we still get an improvement in the forecast of cosmological parameters. We expect the reality to be closer to the uncorrelated Poisson noise case presented in the main text, with small degradation in error bars due to correlations \cite{2009_poisson_smith,2017_poisson_noise}.
	\section{Covariance matrices for pairwise and cross-pairwise kSZ measurements\label{App:Cov_mat}}
	{\color{black}We show the full covariance matrix for two representative cases, one each for the pairwise and the cross pairwise kSZ estimators. We note that the covariance matrix for the cross-pairwise kSZ effect is significantly more non-diagonal compared to the pairwise kSZ effect.}
	\begin{figure}[h]
		\hspace{-0.5cm}
		\begin{subfigure}{0.5\textwidth}
			\centering
			\includegraphics[width=1.05\linewidth]{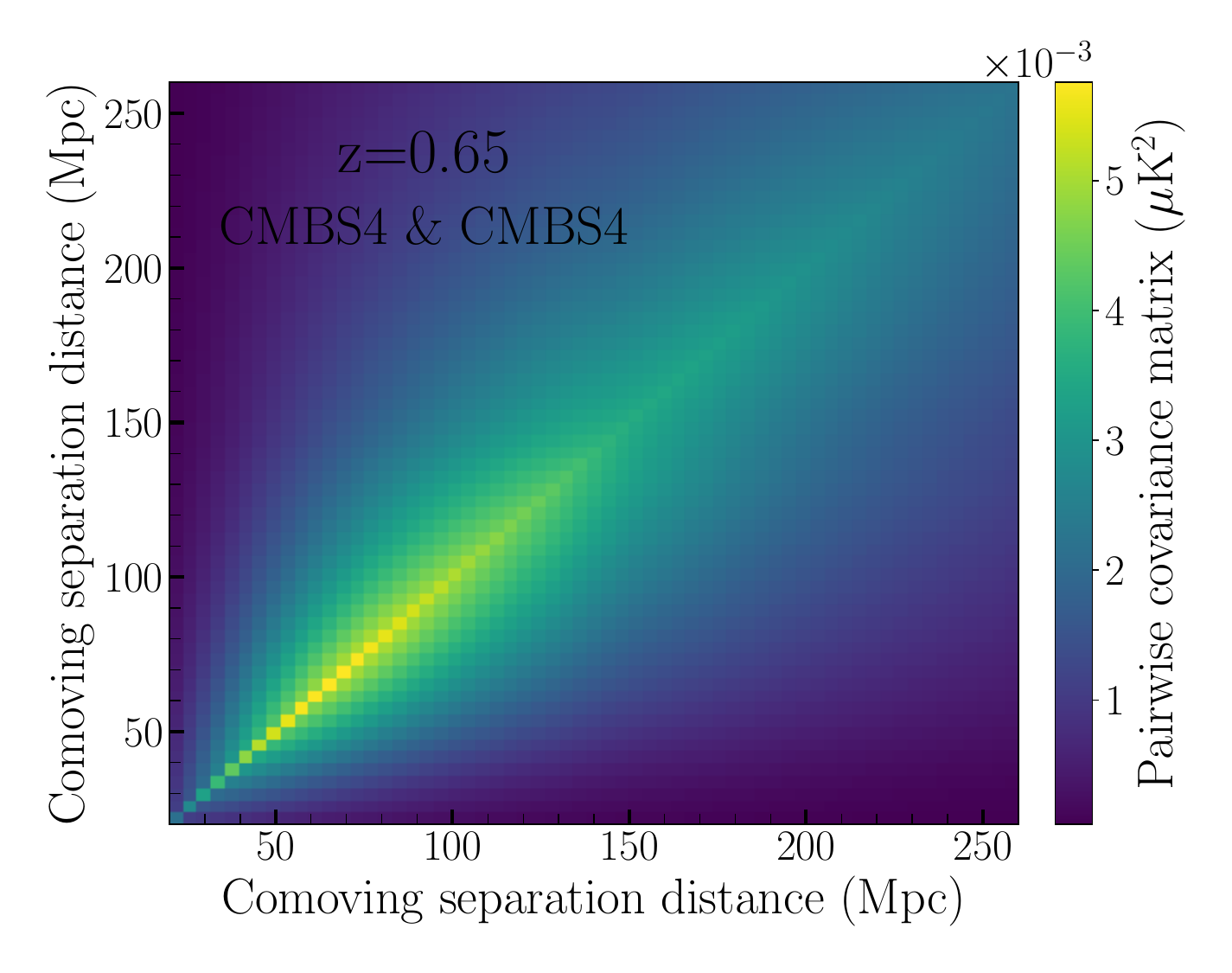}
			\caption{Pairwise covariance matrix}
			\label{fig:pairwisecovmat}
		\end{subfigure}%
		\hspace{0.1cm}
		\begin{subfigure}{0.5\textwidth}
			\centering
			\includegraphics[width=1.05\linewidth]{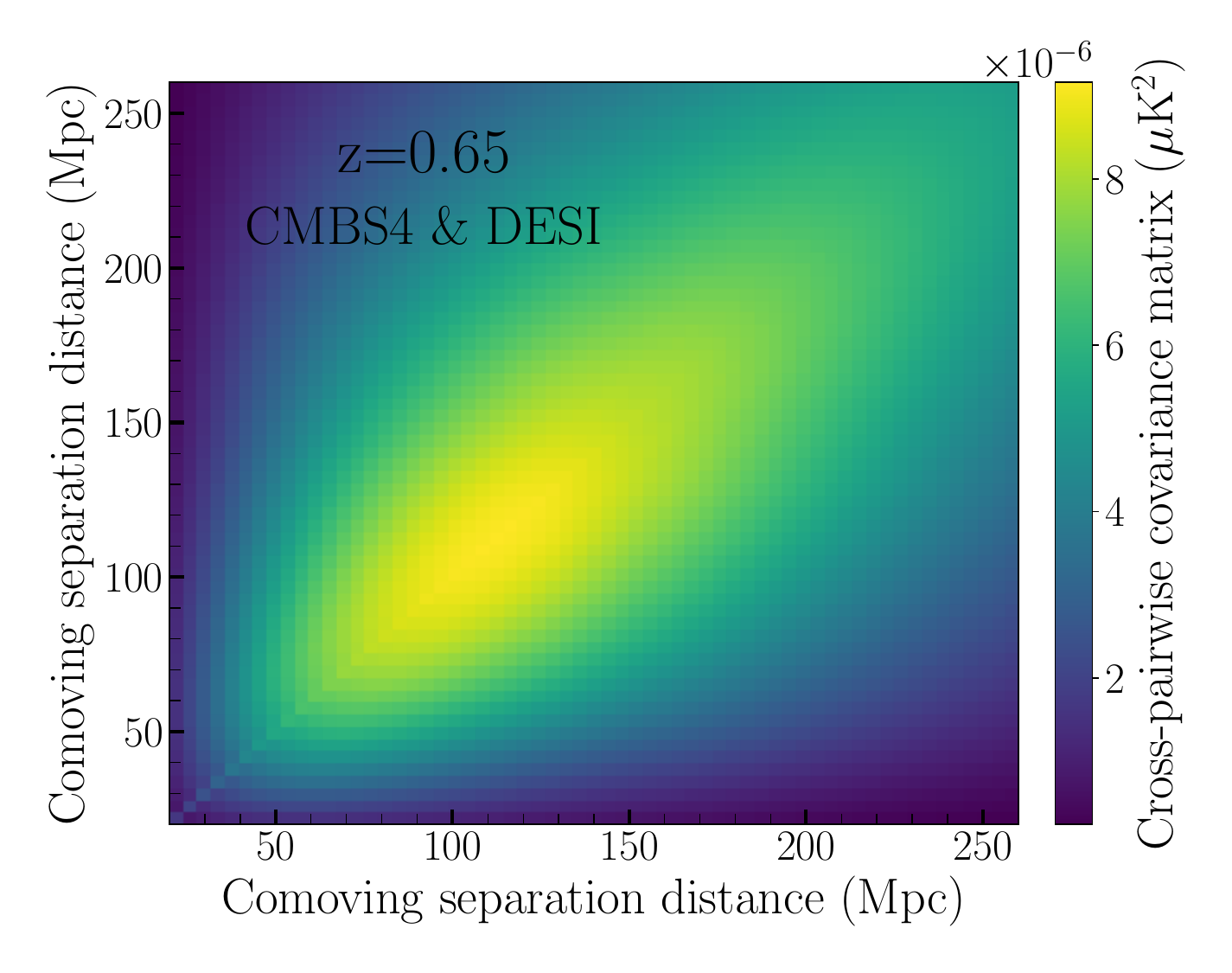}
			\caption{Cross-pairwise covariance matrix}
			\label{fig:crosspairwisecovmat}
		\end{subfigure}
		\caption{{\color{black}The pairwise and cross-pairwise covariance matrix at $z=0.65$. For the cross-pairwise case we show the covariance for luminous red galaxy sample of the DESI survey cross-paired with CMB-S4 clusters. }}
		\label{fig:pairwisecrosspairwisecovmat}
	\end{figure}
	
	\begin{flushleft}
		\justifying
		\bibliographystyle{unsrtads}
		\bibliography{references.bib}

\begin{thebibliography}{10}

\bibitem{1972_Sunyaev_Zeldovich}
R.~A. {Sunyaev} and Ya.~B. {Zeldovich}.
\newblock {The Observations of Relic Radiation as a Test of the Nature of X-Ray
  Radiation from the Clusters of Galaxies}.
\newblock {\em Comments on Astrophysics and Space Physics}, 4:173, November
  1972.
\newblock
  {\small[\href{https://ui.adsabs.harvard.edu/abs/1972CoASP...4..173S}{ADS}]}.

\bibitem{SZ_80}
R.~A. {Sunyaev} and Ya.~B. {Zeldovich}.
\newblock {The velocity of clusters of galaxies relative to the microwave
  background - The possibility of its measurement.}
\newblock {\em \mnras}, 190:413--420, February 1980.
\newblock \href {http://dx.doi.org/10.1093/mnras/190.3.413} {\path{[DOI]}},
  {\small[\href{https://ui.adsabs.harvard.edu/abs/1980MNRAS.190..413S}{ADS}]}.

\bibitem{zeldovich1969interaction}
Ya.~B. {Zeldovich} and R.~A. {Sunyaev}.
\newblock {The Interaction of Matter and Radiation in a Hot-Model Universe}.
\newblock {\em \apss}, 4(3):301--316, July 1969.
\newblock \href {http://dx.doi.org/10.1007/BF00661821} {\path{[DOI]}},
  {\small[\href{https://ui.adsabs.harvard.edu/abs/1969Ap&SS...4..301Z}{ADS}]}.

\bibitem{2002_Carlstorm}
John~E. {Carlstrom}, Gilbert~P. {Holder}, and Erik~D. {Reese}.
\newblock {Cosmology with the Sunyaev-Zel'dovich Effect}.
\newblock {\em \araa}, 40:643--680, January 2002.
\newblock \href {http://arxiv.org/abs/astro-ph/0208192}
  {\path{arXiv:astro-ph/0208192}}, \href
  {http://dx.doi.org/10.1146/annurev.astro.40.060401.093803} {\path{[DOI]}},
  {\small[\href{https://ui.adsabs.harvard.edu/abs/2002ARA&A..40..643C}{ADS}]}.

\bibitem{2014_renaux}
S{\'e}bastien {Renaux-Petel}, Christian {Fidler}, Cyril {Pitrou}, and Guido~W.
  {Pettinari}.
\newblock {Spectral distortions in the cosmic microwave background
  polarization}.
\newblock {\em \jcap}, 2014(3):033, March 2014.
\newblock \href {http://arxiv.org/abs/1312.4448} {\path{arXiv:1312.4448}},
  \href {http://dx.doi.org/10.1088/1475-7516/2014/03/033} {\path{[DOI]}},
  {\small[\href{https://ui.adsabs.harvard.edu/abs/2014JCAP...03..033R}{ADS}]}.

\bibitem{2022JCAP_Hotinli}
Selim~C. {Hotinli}, Gilbert~P. {Holder}, Matthew~C. {Johnson}, and Marc
  {Kamionkowski}.
\newblock {Cosmology from the kinetic polarized Sunyaev Zel'dovich effect}.
\newblock {\em \jcap}, 2022(10):026, October 2022.
\newblock \href {http://arxiv.org/abs/2204.12503} {\path{arXiv:2204.12503}},
  \href {http://dx.doi.org/10.1088/1475-7516/2022/10/026} {\path{[DOI]}},
  {\small[\href{https://ui.adsabs.harvard.edu/abs/2022JCAP...10..026H}{ADS}]}.

\bibitem{2022JCAP_Gon}
Aritra~Kumar {Gon} and Rishi {Khatri}.
\newblock {E and B modes of the CMB y-type distortions: polarised kinetic
  Sunyaev-Zeldovich effect from the reionisation and post-reionisation eras}.
\newblock {\em \jcap}, 2022(10):056, October 2022.
\newblock \href {http://arxiv.org/abs/2208.02270} {\path{arXiv:2208.02270}},
  \href {http://dx.doi.org/10.1088/1475-7516/2022/10/056} {\path{[DOI]}},
  {\small[\href{https://ui.adsabs.harvard.edu/abs/2022JCAP...10..056G}{ADS}]}.

\bibitem{2019_movinglens_hotinli}
Selim~C. {Hotinli}, Joel {Meyers}, Neal {Dalal}, Andrew~H. {Jaffe}, Matthew~C.
  {Johnson}, James~B. {Mertens}, Moritz {M{\"u}nchmeyer}, Kendrick~M. {Smith},
  and Alexander {van Engelen}.
\newblock {Transverse Velocities with the Moving Lens Effect}.
\newblock {\em \prl}, 123(6):061301, August 2019.
\newblock \href {http://arxiv.org/abs/1812.03167} {\path{arXiv:1812.03167}},
  \href {http://dx.doi.org/10.1103/PhysRevLett.123.061301} {\path{[DOI]}},
  {\small[\href{https://ui.adsabs.harvard.edu/abs/2019PhRvL.123f1301H}{ADS}]}.

\bibitem{2019_movlens_Elena}
Siavash {Yasini}, Nareg {Mirzatuny}, and Elena {Pierpaoli}.
\newblock {Pairwise Transverse Velocity Measurement with the Rees-Sciama
  Effect}.
\newblock {\em \apjl}, 873(2):L23, March 2019.
\newblock \href {http://arxiv.org/abs/1812.04241} {\path{arXiv:1812.04241}},
  \href {http://dx.doi.org/10.3847/2041-8213/ab0bfe} {\path{[DOI]}},
  {\small[\href{https://ui.adsabs.harvard.edu/abs/2019ApJ...873L..23Y}{ADS}]}.

\bibitem{1987_Kaiser_RSD}
Nick {Kaiser}.
\newblock {Clustering in real space and in redshift space}.
\newblock {\em \mnras}, 227:1--21, July 1987.
\newblock \href {http://dx.doi.org/10.1093/mnras/227.1.1} {\path{[DOI]}},
  {\small[\href{https://ui.adsabs.harvard.edu/abs/1987MNRAS.227....1K}{ADS}]}.

\bibitem{2009_Percival}
Will~J. {Percival} and Martin {White}.
\newblock {Testing cosmological structure formation using redshift-space
  distortions}.
\newblock {\em \mnras}, 393(1):297--308, February 2009.
\newblock \href {http://arxiv.org/abs/0808.0003} {\path{arXiv:0808.0003}},
  \href {http://dx.doi.org/10.1111/j.1365-2966.2008.14211.x} {\path{[DOI]}},
  {\small[\href{https://ui.adsabs.harvard.edu/abs/2009MNRAS.393..297P}{ADS}]}.

\bibitem{2013_RSD_Fabian}
Alvise {Raccanelli}, Daniele {Bertacca}, Davide {Pietrobon}, Fabian {Schmidt},
  Lado {Samushia}, Nicola {Bartolo}, Olivier {Dor{\'e}}, Sabino {Matarrese},
  and Will~J. {Percival}.
\newblock {Testing gravity using large-scale redshift-space distortions}.
\newblock {\em \mnras}, 436(1):89--100, November 2013.
\newblock \href {http://arxiv.org/abs/1207.0500} {\path{arXiv:1207.0500}},
  \href {http://dx.doi.org/10.1093/mnras/stt1517} {\path{[DOI]}},
  {\small[\href{https://ui.adsabs.harvard.edu/abs/2013MNRAS.436...89R}{ADS}]}.

\bibitem{2017_sdss12}
Shadab {Alam} et~al.
\newblock {The clustering of galaxies in the completed SDSS-III Baryon
  Oscillation Spectroscopic Survey: cosmological analysis of the DR12 galaxy
  sample}.
\newblock {\em \mnras}, 470(3):2617--2652, September 2017.
\newblock \href {http://arxiv.org/abs/1607.03155} {\path{arXiv:1607.03155}},
  \href {http://dx.doi.org/10.1093/mnras/stx721} {\path{[DOI]}},
  {\small[\href{https://ui.adsabs.harvard.edu/abs/2017MNRAS.470.2617A}{ADS}]}.

\bibitem{2013_Sayers}
J.~{Sayers}, T.~{Mroczkowski}, M.~{Zemcov}, P.~M. {Korngut}, J.~{Bock},
  E.~{Bulbul}, N.~G. {Czakon}, E.~{Egami}, S.~R. {Golwala}, P.~M. {Koch}, K.~Y.
  {Lin}, A.~{Mantz}, S.~M. {Molnar}, L.~{Moustakas}, E.~{Pierpaoli}, T.~D.
  {Rawle}, E.~D. {Reese}, M.~{Rex}, J.~A. {Shitanishi}, S.~{Siegel}, and
  K.~{Umetsu}.
\newblock {A Measurement of the Kinetic Sunyaev-Zel'dovich Signal Toward MACS
  J0717.5+3745}.
\newblock {\em \apj}, 778(1):52, November 2013.
\newblock \href {http://arxiv.org/abs/1312.3680} {\path{arXiv:1312.3680}},
  \href {http://dx.doi.org/10.1088/0004-637X/778/1/52} {\path{[DOI]}},
  {\small[\href{https://ui.adsabs.harvard.edu/abs/2013ApJ...778...52S}{ADS}]}.

\bibitem{2012_Hand}
Nick {Hand} et~al.
\newblock {Evidence of Galaxy Cluster Motions with the Kinematic
  Sunyaev-Zel'dovich Effect}.
\newblock {\em \prl}, 109(4):041101, July 2012.
\newblock \href {http://arxiv.org/abs/1203.4219} {\path{arXiv:1203.4219}},
  \href {http://dx.doi.org/10.1103/PhysRevLett.109.041101} {\path{[DOI]}},
  {\small[\href{https://ui.adsabs.harvard.edu/abs/2012PhRvL.109d1101H}{ADS}]}.

\bibitem{2016_DESYear1andSPT}
B.~{Soergel}, S.~{Flender}, others, {DES Collaboration}, and {SPT
  Collaboration}.
\newblock {Detection of the kinematic Sunyaev-Zel'dovich effect with DES Year 1
  and SPT}.
\newblock {\em \mnras}, 461(3):3172--3193, September 2016.
\newblock \href {http://arxiv.org/abs/1603.03904} {\path{arXiv:1603.03904}},
  \href {http://dx.doi.org/10.1093/mnras/stw1455} {\path{[DOI]}},
  {\small[\href{https://ui.adsabs.harvard.edu/abs/2016MNRAS.461.3172S}{ADS}]}.

\bibitem{2018_PlanckandBOSSdata}
Yi-Chao {Li}, Yin-Zhe {Ma}, Mathieu {Remazeilles}, and Kavilan {Moodley}.
\newblock {Measurement of the pairwise kinematic Sunyaev-Zeldovich effect with
  Planck and BOSS data}.
\newblock {\em \prd}, 97(2):023514, January 2018.
\newblock \href {http://arxiv.org/abs/1710.10876} {\path{arXiv:1710.10876}},
  \href {http://dx.doi.org/10.1103/PhysRevD.97.023514} {\path{[DOI]}},
  {\small[\href{https://ui.adsabs.harvard.edu/abs/2018PhRvD..97b3514L}{ADS}]}.

\bibitem{2021_PRD_AtacamaCosmologyTelescope}
V.~{Calafut} et~al.
\newblock {The Atacama Cosmology Telescope: Detection of the pairwise kinematic
  Sunyaev-Zel'dovich effect with SDSS DR15 galaxies}.
\newblock {\em \prd}, 104(4):043502, August 2021.
\newblock \href {http://arxiv.org/abs/2101.08374} {\path{arXiv:2101.08374}},
  \href {http://dx.doi.org/10.1103/PhysRevD.104.043502} {\path{[DOI]}},
  {\small[\href{https://ui.adsabs.harvard.edu/abs/2021PhRvD.104d3502C}{ADS}]}.

\bibitem{2022MNRAS_DESIgalaxyclustersandPlanck}
Ziyang {Chen}, Pengjie {Zhang}, Xiaohu {Yang}, and Yi~{Zheng}.
\newblock {Detection of pairwise kSZ effect with DESI galaxy clusters and
  Planck}.
\newblock {\em \mnras}, 510(4):5916--5928, March 2022.
\newblock \href {http://arxiv.org/abs/2109.04092} {\path{arXiv:2109.04092}},
  \href {http://dx.doi.org/10.1093/mnras/stab3604} {\path{[DOI]}},
  {\small[\href{https://ui.adsabs.harvard.edu/abs/2022MNRAS.510.5916C}{ADS}]}.

\bibitem{2023_SPT_3GandDESCollaborations}
E.~Schiappucci, others, SPT-3G, and DES Collaborations.
\newblock Measurement of the mean central optical depth of galaxy clusters via
  the pairwise kinematic sunyaev-zel'dovich effect with spt-3g and des.
\newblock {\em Phys. Rev. D}, 107:042004, Feb 2023.
\newblock URL: \url{https://link.aps.org/doi/10.1103/PhysRevD.107.042004},
  \href {http://dx.doi.org/10.1103/PhysRevD.107.042004} {\path{[DOI]}}.

\bibitem{1994ApJ_PerturbativeGrowth}
Somnath {Bharadwaj}.
\newblock {Perturbative Growth of Cosmological Clustering. I. Formalism}.
\newblock {\em \apj}, 428:419, June 1994.
\newblock \href {http://dx.doi.org/10.1086/174254} {\path{[DOI]}},
  {\small[\href{https://ui.adsabs.harvard.edu/abs/1994ApJ...428..419B}{ADS}]}.

\bibitem{1977ApJ_Davis}
M.~{Davis} and P.~J.~E. {Peebles}.
\newblock {On the integration of the BBGKY equations for the development of
  strongly nonlinear clustering in an expanding universe.}
\newblock {\em \apjs}, 34:425--450, August 1977.
\newblock \href {http://dx.doi.org/10.1086/190456} {\path{[DOI]}},
  {\small[\href{https://ui.adsabs.harvard.edu/abs/1977ApJS...34..425D}{ADS}]}.

\bibitem{1980_peebles_lss}
P.~J.~E. {Peebles}.
\newblock {\em {The large-scale structure of the universe}}.
\newblock 1980.
\newblock
  {\small[\href{https://ui.adsabs.harvard.edu/abs/1980lssu.book.....P}{ADS}]}.

\bibitem{1992ApJ_Fry}
Laddawan {Ruamsuwan} and J.~N. {Fry}.
\newblock {Stability of Scale-invariant Nonlinear Gravitational Clustering}.
\newblock {\em \apj}, 396:416, September 1992.
\newblock \href {http://dx.doi.org/10.1086/171729} {\path{[DOI]}},
  {\small[\href{https://ui.adsabs.harvard.edu/abs/1992ApJ...396..416R}{ADS}]}.

\bibitem{1999ApJ_DynamicalEstimator}
P.~G. {Ferreira}, R.~{Juszkiewicz}, H.~A. {Feldman}, M.~{Davis}, and A.~H.
  {Jaffe}.
\newblock {Streaming Velocities as a Dynamical Estimator of
  {\ensuremath{\Omega}}}.
\newblock {\em \apjl}, 515(1):L1--L4, April 1999.
\newblock \href {http://arxiv.org/abs/astro-ph/9812456}
  {\path{arXiv:astro-ph/9812456}}, \href {http://dx.doi.org/10.1086/311959}
  {\path{[DOI]}},
  {\small[\href{https://ui.adsabs.harvard.edu/abs/1999ApJ...515L...1F}{ADS}]}.

\bibitem{2016PRL_ProjectedFields:ANovelProbe}
J.~Colin {Hill}, Simone {Ferraro}, Nick {Battaglia}, Jia {Liu}, and David~N.
  {Spergel}.
\newblock {Kinematic Sunyaev-Zel'dovich Effect with Projected Fields: A Novel
  Probe of the Baryon Distribution with Planck, WMAP, and WISE Data}.
\newblock {\em \prl}, 117(5):051301, July 2016.
\newblock \href {http://arxiv.org/abs/1603.01608} {\path{arXiv:1603.01608}},
  \href {http://dx.doi.org/10.1103/PhysRevLett.117.051301} {\path{[DOI]}},
  {\small[\href{https://ui.adsabs.harvard.edu/abs/2016PhRvL.117e1301H}{ADS}]}.

\bibitem{2005_Spergel}
Simon {DeDeo}, David~N. {Spergel}, and Hy~{Trac}.
\newblock {The kinetic Sunyaev-Zel'dovitch effect as a dark energy probe}.
\newblock {\em arXiv e-prints}, pages astro--ph/0511060, November 2005.
\newblock \href {http://arxiv.org/abs/astro-ph/0511060}
  {\path{arXiv:astro-ph/0511060}}, \href
  {http://dx.doi.org/10.48550/arXiv.astro-ph/0511060} {\path{[DOI]}},
  {\small[\href{https://ui.adsabs.harvard.edu/abs/2005astro.ph.11060D}{ADS}]}.

\bibitem{2008_Bhattacharya}
Suman {Bhattacharya} and Arthur {Kosowsky}.
\newblock {Dark energy constraints from galaxy cluster peculiar velocities}.
\newblock {\em \prd}, 77(8):083004, April 2008.
\newblock \href {http://arxiv.org/abs/0712.0034} {\path{arXiv:0712.0034}},
  \href {http://dx.doi.org/10.1103/PhysRevD.77.083004} {\path{[DOI]}},
  {\small[\href{https://ui.adsabs.harvard.edu/abs/2008PhRvD..77h3004B}{ADS}]}.

\bibitem{2012_Philip}
Philip {Bull}, Timothy {Clifton}, and Pedro~G. {Ferreira}.
\newblock {Kinematic Sunyaev-Zel'dovich effect as a test of general radial
  inhomogeneity in Lema{\^\i}tre-Tolman-Bondi cosmology}.
\newblock {\em \prd}, 85(2):024002, January 2012.
\newblock \href {http://arxiv.org/abs/1108.2222} {\path{arXiv:1108.2222}},
  \href {http://dx.doi.org/10.1103/PhysRevD.85.024002} {\path{[DOI]}},
  {\small[\href{https://ui.adsabs.harvard.edu/abs/2012PhRvD..85b4002B}{ADS}]}.

\bibitem{2015ApJ_DarkEnergy}
Eva-Maria {Mueller}, Francesco {de Bernardis}, Rachel {Bean}, and Michael~D.
  {Niemack}.
\newblock {Constraints on Gravity and Dark Energy from the Pairwise Kinematic
  Sunyaev-Zel'dovich Effect}.
\newblock {\em \apj}, 808(1):47, July 2015.
\newblock \href {http://arxiv.org/abs/1408.6248} {\path{arXiv:1408.6248}},
  \href {http://dx.doi.org/10.1088/0004-637X/808/1/47} {\path{[DOI]}},
  {\small[\href{https://ui.adsabs.harvard.edu/abs/2015ApJ...808...47M}{ADS}]}.

\bibitem{2018_Madhavacheril}
Kendrick~M. {Smith}, Mathew~S. {Madhavacheril}, Moritz {M{\"u}nchmeyer}, Simone
  {Ferraro}, Utkarsh {Giri}, and Matthew~C. {Johnson}.
\newblock {KSZ tomography and the bispectrum}.
\newblock {\em arXiv e-prints}, page arXiv:1810.13423, October 2018.
\newblock \href {http://arxiv.org/abs/1810.13423} {\path{arXiv:1810.13423}},
  \href {http://dx.doi.org/10.48550/arXiv.1810.13423} {\path{[DOI]}},
  {\small[\href{https://ui.adsabs.harvard.edu/abs/2018arXiv181013423S}{ADS}]}.

\bibitem{2015PRD_massiveneutrinos}
Eva-Maria {Mueller}, Francesco {de Bernardis}, Rachel {Bean}, and Michael~D.
  {Niemack}.
\newblock {Constraints on massive neutrinos from the pairwise kinematic
  Sunyaev-Zel'dovich effect}.
\newblock {\em \prd}, 92(6):063501, September 2015.
\newblock \href {http://arxiv.org/abs/1412.0592} {\path{arXiv:1412.0592}},
  \href {http://dx.doi.org/10.1103/PhysRevD.92.063501} {\path{[DOI]}},
  {\small[\href{https://ui.adsabs.harvard.edu/abs/2015PhRvD..92f3501M}{ADS}]}.

\bibitem{2021_kszgalaxy_Kusiak}
Aleksandra {Kusiak}, Boris {Bolliet}, Simone {Ferraro}, J.~Colin {Hill}, and
  Alex {Krolewski}.
\newblock {Constraining the baryon abundance with the kinematic
  Sunyaev-Zel'dovich effect: Projected-field detection using P l a n c k , W M
  A P , and u n W I S E}.
\newblock {\em \prd}, 104(4):043518, August 2021.
\newblock \href {http://arxiv.org/abs/2102.01068} {\path{arXiv:2102.01068}},
  \href {http://dx.doi.org/10.1103/PhysRevD.104.043518} {\path{[DOI]}},
  {\small[\href{https://ui.adsabs.harvard.edu/abs/2021PhRvD.104d3518K}{ADS}]}.

\bibitem{2023_Gon}
Aritra~Kumar {Gon} and Rishi {Khatri}.
\newblock {The pairwise and cross-pairwise y-type polarised kinetic Sunyaev
  Zeldovich effect from transverse velocity of galaxy clusters}.
\newblock {\em \jcap}, 2023(11):072, November 2023.
\newblock \href {http://arxiv.org/abs/2308.01730} {\path{arXiv:2308.01730}},
  \href {http://dx.doi.org/10.1088/1475-7516/2023/11/072} {\path{[DOI]}},
  {\small[\href{https://ui.adsabs.harvard.edu/abs/2023JCAP...11..072G}{ADS}]}.

\bibitem{2009_poisson_smith}
Robert~E. {Smith}.
\newblock {Covariance of cross-correlations: towards efficient measures for
  large-scale structure}.
\newblock {\em \mnras}, 400(2):851--865, December 2009.
\newblock \href {http://arxiv.org/abs/0810.1960} {\path{arXiv:0810.1960}},
  \href {http://dx.doi.org/10.1111/j.1365-2966.2009.15490.x} {\path{[DOI]}},
  {\small[\href{https://ui.adsabs.harvard.edu/abs/2009MNRAS.400..851S}{ADS}]}.

\bibitem{2017_poisson_noise}
Kerstin {Paech}, Nico {Hamaus}, Ben {Hoyle}, Matteo {Costanzi}, Tommaso
  {Giannantonio}, Steffen {Hagstotz}, Georg {Sauerwein}, and Jochen {Weller}.
\newblock {Cross-correlation of galaxies and galaxy clusters in the Sloan
  Digital Sky Survey and the importance of non-Poissonian shot noise}.
\newblock {\em \mnras}, 470(3):2566--2577, September 2017.
\newblock \href {http://arxiv.org/abs/1612.02018} {\path{arXiv:1612.02018}},
  \href {http://dx.doi.org/10.1093/mnras/stx1354} {\path{[DOI]}},
  {\small[\href{https://ui.adsabs.harvard.edu/abs/2017MNRAS.470.2566P}{ADS}]}.

\bibitem{2001_CP}
Michel {Chevallier} and David {Polarski}.
\newblock {Accelerating Universes with Scaling Dark Matter}.
\newblock {\em International Journal of Modern Physics D}, 10(2):213--223,
  January 2001.
\newblock \href {http://arxiv.org/abs/gr-qc/0009008}
  {\path{arXiv:gr-qc/0009008}}, \href
  {http://dx.doi.org/10.1142/S0218271801000822} {\path{[DOI]}},
  {\small[\href{https://ui.adsabs.harvard.edu/abs/2001IJMPD..10..213C}{ADS}]}.

\bibitem{2003_Linder}
Eric~V. {Linder}.
\newblock {Exploring the Expansion History of the Universe}.
\newblock {\em \prl}, 90(9):091301, March 2003.
\newblock \href {http://arxiv.org/abs/astro-ph/0208512}
  {\path{arXiv:astro-ph/0208512}}, \href
  {http://dx.doi.org/10.1103/PhysRevLett.90.091301} {\path{[DOI]}},
  {\small[\href{https://ui.adsabs.harvard.edu/abs/2003PhRvL..90i1301L}{ADS}]}.

\bibitem{2019JCAP_Simons_Obs}
Peter {Ade} et~al.
\newblock {The Simons Observatory: science goals and forecasts}.
\newblock {\em \jcap}, 2019(2):056, February 2019.
\newblock \href {http://arxiv.org/abs/1808.07445} {\path{arXiv:1808.07445}},
  \href {http://dx.doi.org/10.1088/1475-7516/2019/02/056} {\path{[DOI]}},
  {\small[\href{https://ui.adsabs.harvard.edu/abs/2019JCAP...02..056A}{ADS}]}.

\bibitem{2016_cmbs4}
Kevork~N. {Abazajian}, Peter {Adshead}, {Ahmed}, et~al.
\newblock {CMB-S4 Science Book, First Edition}.
\newblock {\em arXiv e-prints}, page arXiv:1610.02743, October 2016.
\newblock \href {http://arxiv.org/abs/1610.02743} {\path{arXiv:1610.02743}},
  \href {http://dx.doi.org/10.48550/arXiv.1610.02743} {\path{[DOI]}},
  {\small[\href{https://ui.adsabs.harvard.edu/abs/2016arXiv161002743A}{ADS}]}.

\bibitem{2024_DESI}
{DESI Collaboration}.
\newblock {Validation of the Scientific Program for the Dark Energy
  Spectroscopic Instrument}.
\newblock {\em \aj}, 167(2):62, February 2024.
\newblock \href {http://arxiv.org/abs/2306.06307} {\path{arXiv:2306.06307}},
  \href {http://dx.doi.org/10.3847/1538-3881/ad0b08} {\path{[DOI]}},
  {\small[\href{https://ui.adsabs.harvard.edu/abs/2024AJ....167...62D}{ADS}]}.

\bibitem{2023_DESI_EDR}
{DESI Collaboration}.
\newblock {The Early Data Release of the Dark Energy Spectroscopic Instrument}.
\newblock {\em arXiv e-prints}, page arXiv:2306.06308, June 2023.
\newblock \href {http://arxiv.org/abs/2306.06308} {\path{arXiv:2306.06308}},
  \href {http://dx.doi.org/10.48550/arXiv.2306.06308} {\path{[DOI]}},
  {\small[\href{https://ui.adsabs.harvard.edu/abs/2023arXiv230606308D}{ADS}]}.

\bibitem{2009_LSST}
{LSST Science Collaboration} et~al.
\newblock {LSST Science Book, Version 2.0}.
\newblock {\em arXiv e-prints}, page arXiv:0912.0201, December 2009.
\newblock \href {http://arxiv.org/abs/0912.0201} {\path{arXiv:0912.0201}},
  \href {http://dx.doi.org/10.48550/arXiv.0912.0201} {\path{[DOI]}},
  {\small[\href{https://ui.adsabs.harvard.edu/abs/2009arXiv0912.0201L}{ADS}]}.

\bibitem{2012_erosita}
A.~{Merloni}, J.~{Predehl}, and the {German eROSITA Consortium}.
\newblock {eROSITA Science Book: Mapping the Structure of the Energetic
  Universe}.
\newblock {\em arXiv e-prints}, page arXiv:1209.3114, September 2012.
\newblock \href {http://arxiv.org/abs/1209.3114} {\path{arXiv:1209.3114}},
  \href {http://dx.doi.org/10.48550/arXiv.1209.3114} {\path{[DOI]}},
  {\small[\href{https://ui.adsabs.harvard.edu/abs/2012arXiv1209.3114M}{ADS}]}.

\bibitem{2017ApJ_Fender}
Samuel {Flender}, Daisuke {Nagai}, and Michael {McDonald}.
\newblock {Constraints on the Optical Depth of Galaxy Groups and Clusters}.
\newblock {\em \apj}, 837(2):124, March 2017.
\newblock \href {http://arxiv.org/abs/1610.08029} {\path{arXiv:1610.08029}},
  \href {http://dx.doi.org/10.3847/1538-4357/aa60bf} {\path{[DOI]}},
  {\small[\href{https://ui.adsabs.harvard.edu/abs/2017ApJ...837..124F}{ADS}]}.

\bibitem{2002PhR_Bernardeau}
F.~{Bernardeau}, S.~{Colombi}, E.~{Gazta{\~n}aga}, and R.~{Scoccimarro}.
\newblock {Large-scale structure of the Universe and cosmological perturbation
  theory}.
\newblock {\em \physrep}, 367(1-3):1--248, September 2002.
\newblock \href {http://arxiv.org/abs/astro-ph/0112551}
  {\path{arXiv:astro-ph/0112551}}, \href
  {http://dx.doi.org/10.1016/S0370-1573(02)00135-7} {\path{[DOI]}},
  {\small[\href{https://ui.adsabs.harvard.edu/abs/2002PhR...367....1B}{ADS}]}.

\bibitem{2004ApJ_Ma_CosmologicalKineticTheory}
Chung-Pei {Ma} and Edmund {Bertschinger}.
\newblock {A Cosmological Kinetic Theory for the Evolution of Cold Dark Matter
  Halos with Substructure: Quasi-Linear Theory}.
\newblock {\em \apj}, 612(1):28--49, September 2004.
\newblock \href {http://arxiv.org/abs/astro-ph/0311049}
  {\path{arXiv:astro-ph/0311049}}, \href {http://dx.doi.org/10.1086/421766}
  {\path{[DOI]}},
  {\small[\href{https://ui.adsabs.harvard.edu/abs/2004ApJ...612...28M}{ADS}]}.

\bibitem{1994ApJ_Bhuvnesh}
Bhuvnesh {Jain} and Edmund {Bertschinger}.
\newblock {Second-Order Power Spectrum and Nonlinear Evolution at High
  Redshift}.
\newblock {\em \apj}, 431:495, August 1994.
\newblock \href {http://arxiv.org/abs/astro-ph/9311070}
  {\path{arXiv:astro-ph/9311070}}, \href {http://dx.doi.org/10.1086/174502}
  {\path{[DOI]}},
  {\small[\href{https://ui.adsabs.harvard.edu/abs/1994ApJ...431..495J}{ADS}]}.

\bibitem{2007_growthrate_Linder}
Eric~V. {Linder} and Robert~N. {Cahn}.
\newblock {Parameterized beyond-Einstein growth}.
\newblock {\em Astroparticle Physics}, 28(4-5):481--488, December 2007.
\newblock \href {http://arxiv.org/abs/astro-ph/0701317}
  {\path{arXiv:astro-ph/0701317}}, \href
  {http://dx.doi.org/10.1016/j.astropartphys.2007.09.003} {\path{[DOI]}},
  {\small[\href{https://ui.adsabs.harvard.edu/abs/2007APh....28..481L}{ADS}]}.

\bibitem{1974ApJ_pH_estimator}
P.~J.~E. {Peebles} and M.~G. {Hauser}.
\newblock {Statistical Analysis of Catalogs of Extragalactic Objects. III. The
  Shane-Wirtanen and Zwicky Catalogs}.
\newblock {\em \apjs}, 28:19, November 1974.
\newblock \href {http://dx.doi.org/10.1086/190308} {\path{[DOI]}},
  {\small[\href{https://ui.adsabs.harvard.edu/abs/1974ApJS...28...19P}{ADS}]}.

\bibitem{2013A&A_estimator}
M.~{Vargas-Maga{\~n}a}, J.~E. {Bautista}, J.~Ch. {Hamilton}, N.~G. {Busca},
  {\'E}.~{Aubourg}, A.~{Labatie}, J.~M. {Le Goff}, S.~{Escoffier}, M.~{Manera},
  C.~K. {McBride}, D.~P. {Schneider}, and Ch. N.~A. {Willmer}.
\newblock {An optimized correlation function estimator for galaxy surveys}.
\newblock {\em \aap}, 554:A131, June 2013.
\newblock \href {http://arxiv.org/abs/1211.6211} {\path{arXiv:1211.6211}},
  \href {http://dx.doi.org/10.1051/0004-6361/201220790} {\path{[DOI]}},
  {\small[\href{https://ui.adsabs.harvard.edu/abs/2013A&A...554A.131V}{ADS}]}.

\bibitem{1999_Tegmark}
Daniel~J. {Eisenstein}, Wayne {Hu}, and Max {Tegmark}.
\newblock {Cosmic Complementarity: Joint Parameter Estimation from Cosmic
  Microwave Background Experiments and Redshift Surveys}.
\newblock {\em \apj}, 518(1):2--23, June 1999.
\newblock \href {http://arxiv.org/abs/astro-ph/9807130}
  {\path{arXiv:astro-ph/9807130}}, \href {http://dx.doi.org/10.1086/307261}
  {\path{[DOI]}},
  {\small[\href{https://ui.adsabs.harvard.edu/abs/1999ApJ...518....2E}{ADS}]}.

\bibitem{2009_Coe}
Dan {Coe}.
\newblock {Fisher Matrices and Confidence Ellipses: A Quick-Start Guide and
  Software}.
\newblock {\em arXiv e-prints}, page arXiv:0906.4123, June 2009.
\newblock \href {http://arxiv.org/abs/0906.4123} {\path{arXiv:0906.4123}},
  \href {http://dx.doi.org/10.48550/arXiv.0906.4123} {\path{[DOI]}},
  {\small[\href{https://ui.adsabs.harvard.edu/abs/2009arXiv0906.4123C}{ADS}]}.

\bibitem{2019_cmbs4_science_case}
Kevork {Abazajian} et~al.
\newblock {CMB-S4 Science Case, Reference Design, and Project Plan}.
\newblock {\em arXiv e-prints}, page arXiv:1907.04473, July 2019.
\newblock \href {http://arxiv.org/abs/1907.04473} {\path{arXiv:1907.04473}},
  \href {http://dx.doi.org/10.48550/arXiv.1907.04473} {\path{[DOI]}},
  {\small[\href{https://ui.adsabs.harvard.edu/abs/2019arXiv190704473A}{ADS}]}.

\bibitem{2022ApJ_raghunathan}
Srinivasan {Raghunathan}, {Whitehorn}, and others.
\newblock {Constraining Cluster Virialization Mechanism and Cosmology Using
  Thermal-SZ-selected Clusters from Future CMB Surveys}.
\newblock {\em \apj}, 926(2):172, February 2022.
\newblock \href {http://arxiv.org/abs/2107.10250} {\path{arXiv:2107.10250}},
  \href {http://dx.doi.org/10.3847/1538-4357/ac4712} {\path{[DOI]}},
  {\small[\href{https://ui.adsabs.harvard.edu/abs/2022ApJ...926..172R}{ADS}]}.

\bibitem{2022_Raghunathan}
Srinivasan {Raghunathan}.
\newblock {Assessing the Importance of Noise from Thermal Sunyaev-Zel'dovich
  Signals for CMB Cluster Surveys and Cluster Cosmology}.
\newblock {\em \apj}, 928(1):16, March 2022.
\newblock \href {http://arxiv.org/abs/2112.07656} {\path{arXiv:2112.07656}},
  \href {http://dx.doi.org/10.3847/1538-4357/ac510f} {\path{[DOI]}},
  {\small[\href{https://ui.adsabs.harvard.edu/abs/2022ApJ...928...16R}{ADS}]}.

\bibitem{2008_Tinker}
Jeremy {Tinker}, Andrey~V. {Kravtsov}, {Klypin}, et~al.
\newblock {Toward a Halo Mass Function for Precision Cosmology: The Limits of
  Universality}.
\newblock {\em \apj}, 688(2):709--728, December 2008.
\newblock \href {http://arxiv.org/abs/0803.2706} {\path{arXiv:0803.2706}},
  \href {http://dx.doi.org/10.1086/591439} {\path{[DOI]}},
  {\small[\href{https://ui.adsabs.harvard.edu/abs/2008ApJ...688..709T}{ADS}]}.

\bibitem{2020_Cobaya}
Jesus Torrado and Antony Lewis.
\newblock {Cobaya: Code for Bayesian Analysis of hierarchical physical models}.
\newblock {\em JCAP}, 05:057, 2021.
\newblock \href {http://arxiv.org/abs/2005.05290} {\path{arXiv:2005.05290}},
  \href {http://dx.doi.org/10.1088/1475-7516/2021/05/057} {\path{[DOI]}}.

\bibitem{2020_Planck_likelihoods}
{Planck Collaboration}.
\newblock {Planck 2018 results. V. CMB power spectra and likelihoods}.
\newblock {\em \aap}, 641:A5, September 2020.
\newblock \href {http://arxiv.org/abs/1907.12875} {\path{arXiv:1907.12875}},
  \href {http://dx.doi.org/10.1051/0004-6361/201936386} {\path{[DOI]}},
  {\small[\href{https://ui.adsabs.harvard.edu/abs/2020A&A...641A...5P}{ADS}]}.

\bibitem{2020_Planck_lensing}
{Planck Collaboration} .
\newblock {Planck 2018 results. VIII. Gravitational lensing}.
\newblock {\em \aap}, 641:A8, September 2020.
\newblock \href {http://arxiv.org/abs/1807.06210} {\path{arXiv:1807.06210}},
  \href {http://dx.doi.org/10.1051/0004-6361/201833886} {\path{[DOI]}},
  {\small[\href{https://ui.adsabs.harvard.edu/abs/2020A&A...641A...8P}{ADS}]}.

\bibitem{2024_DESI_BAO_Gal}
{DESI Collaboration} et~al.
\newblock {DESI 2024 III: Baryon Acoustic Oscillations from Galaxies and
  Quasars}.
\newblock {\em arXiv e-prints}, page arXiv:2404.03000, April 2024.
\newblock \href {http://arxiv.org/abs/2404.03000} {\path{arXiv:2404.03000}},
  \href {http://dx.doi.org/10.48550/arXiv.2404.03000} {\path{[DOI]}},
  {\small[\href{https://ui.adsabs.harvard.edu/abs/2024arXiv240403000D}{ADS}]}.

\bibitem{2024_DESI_BAO_Lya}
{DESI Collaboration} et~al.
\newblock {DESI 2024 IV: Baryon Acoustic Oscillations from the Lyman Alpha
  Forest}.
\newblock {\em arXiv e-prints}, page arXiv:2404.03001, April 2024.
\newblock \href {http://arxiv.org/abs/2404.03001} {\path{arXiv:2404.03001}},
  \href {http://dx.doi.org/10.48550/arXiv.2404.03001} {\path{[DOI]}},
  {\small[\href{https://ui.adsabs.harvard.edu/abs/2024arXiv240403001D}{ADS}]}.

\bibitem{2009_Albrecht}
Andreas {Albrecht}, Luca {Amendola}, Gary {Bernstein}, Douglas {Clowe}, Daniel
  {Eisenstein}, Luigi {Guzzo}, Christopher {Hirata}, Dragan {Huterer}, Robert
  {Kirshner}, Edward {Kolb}, and Robert {Nichol}.
\newblock {Findings of the Joint Dark Energy Mission Figure of Merit Science
  Working Group}.
\newblock {\em arXiv e-prints}, page arXiv:0901.0721, January 2009.
\newblock \href {http://arxiv.org/abs/0901.0721} {\path{arXiv:0901.0721}},
  \href {http://dx.doi.org/10.48550/arXiv.0901.0721} {\path{[DOI]}},
  {\small[\href{https://ui.adsabs.harvard.edu/abs/2009arXiv0901.0721A}{ADS}]}.

\bibitem{2013_fisher_Khatri}
Rishi {Khatri} and Rashid~A. {Sunyaev}.
\newblock {Forecasts for CMB {\ensuremath{\mu}} and i-type spectral distortion
  constraints on the primordial power spectrum on scales
  8lesssimklesssim{}10$^{4}$ Mpc$^{-1}$ with the future Pixie-like
  experiments}.
\newblock {\em \jcap}, 2013(6):026, June 2013.
\newblock \href {http://arxiv.org/abs/1303.7212} {\path{arXiv:1303.7212}},
  \href {http://dx.doi.org/10.1088/1475-7516/2013/06/026} {\path{[DOI]}},
  {\small[\href{https://ui.adsabs.harvard.edu/abs/2013JCAP...06..026K}{ADS}]}.

\bibitem{2013_fabian_photo_z}
Ryan {Keisler} and Fabian {Schmidt}.
\newblock {Prospects for Measuring the Relative Velocities of Galaxy Clusters
  in Photometric Surveys Using the Kinetic Sunyaev-Zel'dovich Effect}.
\newblock {\em \apjl}, 765(2):L32, March 2013.
\newblock \href {http://arxiv.org/abs/1211.0668} {\path{arXiv:1211.0668}},
  \href {http://dx.doi.org/10.1088/2041-8205/765/2/L32} {\path{[DOI]}},
  {\small[\href{https://ui.adsabs.harvard.edu/abs/2013ApJ...765L..32K}{ADS}]}.

\end{thebibliography}
	\end{flushleft}

		

\end{document}